\documentclass[preprint,3pt]{elsarticle}

\usepackage{xcolor}
\usepackage{graphicx}
\usepackage{amsmath}
\usepackage{amssymb}
\usepackage{amsfonts}
\usepackage{units}
\usepackage{bm}
\usepackage{hyperref}
\usepackage{enumerate}
\usepackage{longtable}
\usepackage{totcount}

\newtotcounter{citnum} 
\def\oldbibitem{} \let\oldbibitem=\bibitem
\def\bibitem{\stepcounter{citnum}\oldbibitem}

\newcommand{\dd}[1]{\mathrm{d}#1\,}

\renewcommand{\Re}{\mathop{\mathrm{Re}}}
\renewcommand{\Im}{\mathop{\mathrm{Im}}}

\DeclareMathOperator{\Tr}{Tr}
\DeclareMathOperator{\tr}{tr}

\newcommand{\sgn}{\mathop{\mathrm{sgn}}}

\renewcommand{\vec}[1]{\bm{#1}}

\definecolor{TTH-color}{named}{green}
\definecolor{FSB-color}{named}{magenta}
\definecolor{PV-color}{rgb}{0.97,0.57,0.11}
\definecolor{MS-color}{RGB}{128,0,128}

\definecolor{TTH-color2}{named}{red}
\definecolor{FSB-color2}{named}{blue}
\definecolor{PV-color2}{rgb}{0.87,0.47,0.01}
\definecolor{MS-color2}{RGB}{128,0,128}

\begin{document}

\title{Thermal, electric and spin transport in  superconductor/ferromagnetic-insulator structures}

\author[jyu]{Tero T Heikkil\"a}
\ead{Tero.T.Heikkila@jyu.fi}

\author[jyu]{Mikhail Silaev}

\author[jyu,nest]{Pauli Virtanen}

\author[csic,dipc]{F. Sebastian Bergeret}
\ead{fs.bergeret@csic.es}

\address[jyu]{University of Jyvaskyla, Department of Physics and Nanoscience Center, P.O. Box 35 (YFL), FI-40014 University of Jyv\"askyl\"a, Finland }

\address[nest]{NEST, Istituto Nanoscienze-CNR and Scuola Normale
  Superiore, I-56127 Pisa, Italy }

\address[csic]{Centro de Fisica de Materiales (CFM-MPC), Centro Mixto CSIC-UPV/EHU, Manuel de Lardizabal 5, E-20018 San Sebastian, Spain}

\address[dipc]{Donostia International Physics Center (DIPC), Manuel de Lardizabal 4, E-20018 San Sebastian, Spain}

\begin{abstract}
A ferromagnetic insulator (FI) attached to a conventional  superconductor (S) changes drastically the properties of the latter. Specifically, the  exchange field at the FI/S interface   leads to a splitting of the superconducting density of states.  If S is a superconducting  film,  thinner than the superconducting  coherence length,  the modification of the density of states occurs   over the whole sample. The coexistence of the  exchange splitting and superconducting correlations in S/FI structures leads to striking transport phenomena that  are of interest for applications  in thermoelectricity, superconducting spintronics and radiation sensors.  Here we review the most recent progress in understanding the transport properties of  FI/S structures  by presenting   a complete  theoretical framework based on the quasiclassical kinetic equations. We discuss the coupling between the electronic degrees of freedom, charge, spin and energy, under non-equilibrium conditions and its manifestation in thermoelectricity and spin-dependent transport. 

\end{abstract}

\maketitle

\tableofcontents

\clearpage

\section{Introduction\label{sec-introduction}}

The study of the interaction between ferromagnetism and conventional superconductivity in hybrid systems has attracted a great attention during the past decades \cite{RevModPhys.77.935,Bergeret2005,bergeret2018colloquium}.  In principle, these two states of matter are antagonistic: whereas the  ferromagnetic correlations  try to  align the spins of electrons, a conventional singlet Bardeen-Cooper-Schrieffer (BCS) \cite{bcs1957} superconductor  exhibits a condensate with electron pairs (Cooper pairs)  with opposite spins. This simple argument explains why usual ferromagnetic metals do not show transition to a conventional superconducting state at any temperature.  

Superconductivity and magnetism can coexist though  in  hybrid systems consisting of superconductor (S) and ferromagnet (F)  layers, where the two states couple through mutual proximity effects.  The interplay between 
them  leads to striking novel phenomena not present in either system alone.   For example the spatial oscillation of the superconducting correlations induced in F via the proximity effect leads to the $\pi$-junctions  in S/F/S Josephson systems \cite{buzdin1982critical,ryazanov2001coupling}. The exchange field in F also leads to  a triplet component of the superconducting condensate   \cite{RevModPhys.77.935,Bergeret2005,eschrig2015spin}.

The effects discussed above are caused by 
the  leakage of the superconducting condensate into the ferromagnet.  There is also a reciprocal effect, caused by the interaction of the condensate with the exchange field that can leak back into the S region over distances of the order of the superconducting coherence length $\xi_s$.  These triplet correlations induced in the superconductor  lead  to a finite magnetization
and a drastic change of the local density of states 
\cite{bergeret2004induced,bergeret2004spin,bergeret2005inverse,kharitonov2006oscillations}. 
If the superconductor is a thin film  with thickness smaller than $\xi_s$, this magnetic proximity effect may cause  an almost  homogeneous spin splitting of the BCS density of states (DOS) \cite{meservey1994spin}.  Such spin splitting plays a central role in this review.

In an all-metallic S/F system however it is difficult to achieve such a homogeneous spin splitting. The reason for this is the leakage of Cooper pairs from S to F. Usual metallic ferromagnets such as Fe, Co or Ni, have intrinsic  exchange fields much larger than the  characteristic superconducting energy ($\sim \Delta$, the BCS pair potential). For  thin S-films such strong exchange fields in F lead eventually to a full suppression of the superconducting state.  Indeed, the first observation of the $0$$-\pi$-transition  in S/F/S Josephson  could be achieved thanks to the use of ferromagnetic alloys with small exchange fields \cite{ryazanov2001coupling,kontos2002josephson}. 
However,  a sharp splitting of the DOS in a superconductor in all-metallic systems has not been observed.

A quite different situation emerges when instead of a metallic ferromagnet one uses a ferromagnetic insulator (FI)  adjacent to a thin S layer. The band gap of the FI prevents the penetration of the superconducting condensate  into the FI.  Electrons are reflected at the S/FI interface. Because the FI is magnetic, this  reflection is spin-dependent and  leads to the  creation of an effective exchange field in the superconductor \cite{PhysRevB.38.8823}. Superconductivity with a spin-splitting field has been observed in several spectroscopy experiments  on Al films  attached to europium chalcogenides, such as EuO, EuS and EuSe \cite{PhysRevLett.56.1746,Hao:1990,meservey1994spin}.

Interestingly, the spin splitting in the DOS observed in those experiments  resembles the spin splitting created by a strong  in-plane field applied to a thin superconducting film. This is caused by the fact that when a magnetic field is applied parallel to a thin S film, 
 the orbital diamagnetic effect is weak \cite{TinkhamBook} and the field  mainly  induces a paramagnetic response \cite{Chandrasekhar:1962,PhysRevLett.9.266,fulde_ferrell.1964,larkin_ovchinnikov.1964}. This effect was  explored experimentally already in the 1970s \cite{PhysRevLett.25.1270,meservey1975tunneling}. The spin-split DOS has been  utilized to determine the spin polarization of ferromagnets in S/F structures \cite{ PhysRevLett.26.192, tedrow1973spin,PhysRevB.16.4907, PhysRevB.22.1331,meservey1994spin}.

Even though spectral properties of S/FI have been intensively studied for a long time, only  recently it has been realized that  spin-split superconductors may find applications in spintronics, thermoelectricity, and sensors \cite{bergeret2018colloquium,de2018toward,Machon2013,giazotto2015,Hubler2012a,beckmann2016spin,heikkila2018,giazotto2015b}. To understand the physics underlying these applications one needs to consider nonequilibrium phenomena. Nonequilibrium effects  can extend over distances  larger than the coherence length and they  are related to the deviation of the electron distribution function from its equilibrium form. Such a deviation leads to an imbalance of the electron degrees of freedom:  charge, energy and  spin. 
From recent studies it has turned out that non-equilibrium effects in spin-split superconductors is a very rich field of research \cite{silaev2015long,bobkova2015,bobkova2016injection,Quay2013,Hubler2012a,kolendaup16}.

In this work  we review  the non-equilibrium properties of spin-split superconductors, focusing on thermal, electric and spin transport.  We have recently  published a Colloquium on this topic \cite{bergeret2018colloquium} in which  we gave a qualitative overview of the field and the experimental activity. Our main  purpose here  is different: We  mainly focus on the theoretical  aspects of the transport properties of S/FI structures and elaborate deeper on the thermoelectricity  in superconducting systems. 
We provide all the necessary theoretical tools for those readers interested in exploring   non-equilibrium effects in novel setups combining superconductors and magnetic materials. Those   readers seeking  more phenomenological approach are referred to the Colloquium mentioned above.

The review is organized as follows: In Sec.~\ref{sec-superwithh} we summarize the main  properties of S/FI structures and discuss the coexistence  between superconductivity and ferromagnetism in these systems. 
The main focus is on the spin-splitting induced in the S layer and its manifestation on the spectrum. We also  analyze the magnetic susceptibility of superconductors and briefly discuss other effects related to spin-splitting fields, including the Fulde-Ferrel-Larkin-Ovchnikov (FFLO) state,  triplet superconducting correlations and the cryptoferromagnetic state in the FI layer. 

In Sec. \ref{sec:quasiclassical_theory} we present the quasiclassical Green's function formalism, with a special focus on  the description of   non-equilibrium properties of diffusive S/FI systems, the role of inelastic scattering and transport through hybrid interfaces. In this section we introduce the concept of nonequilibrium modes which is essential for understanding the different ways of exciting a spin-split superconductor out of equilibrium. Especially the charge, spin and heat transport properties are determined by these modes and their coupling.  

On the basis of the quasiclassical formalism we analyze  in Sec.~\ref{spininjection} the transport properties of different FI/S structures. We focus in particular on spin injection, diffusion and relaxation, in different setups. 

In Sec.~\ref{sec:acdynamics} we study the non-equilibrium quasiparticle dynamics of a spin-split superconductor in an alternating radio frequency field, both in the linear and the non-linear regime.  For this we make use of time-dependent quasiclassical equations.  

Spin splitting the BCS spectrum leads to a strongly electron-hole asymmetric spin-resolved density of states. The symmetry is recovered by averaging over spin, but the electron-hole asymmetry shows up in the case of spin-polarized tunneling into spin-split superconductors. Section \ref{thermoel} discusses the transport properties of such spin-polarized contacts to spin-split superconductors, with a special emphasis on the thermoelectric response of such systems. We also review thermoelectricity in superconducting systems in general to set this finding into historical context. 

Finally, Sec.~\ref{sec:outlook} summarizes the main phenomena discussed in the review.

\section{Ferromagnetic insulator--superconductor structures}
\label{sec-superwithh}

At interfaces between a superconductor (S)  and a magnetic material (F) both   states of matter may coexist, however with certain modifications.  Such modifications extend over the characteristic  correlation lengths in each material.  
In S  this length is the coherence length, 
which in diffusive systems and at low temperatures is of the order of $\sqrt{D/\Delta_0}$, where $D$ is the diffusion coefficient and $\Delta_0$ the superconducting gap at zero temperature. 
If the ferromagnet is a metal,  the superconducting  condensate (Cooper pairs)  can penetrate into it.  
The superconducting correlation length  in usual ferromagnets as Ni, Co or Fe, is approximately given by $\sqrt{D/h}$, where $h$ is the intrinsic exchange field of the ferromagnet. Usually this length is very short (a few nanometers or even less)  and therefore one speaks about a short-range proximity effect. 
 Interestingly,  the  interaction of singlet electron pairs stemming from the S layer with the local exchange field leads to pairs in a triplet state that can get back into superconductor \cite{Bergeret2005}. Such leakage of pairs and the suppression of the density of Cooper pairs in singlet state leads to a suppression of superconductivity \cite{RevModPhys.77.935}.

If instead  of a metal, a ferromagnetic  insulator  (FI)  is placed adjacent to a  superconductor,  no leakage of pairs is possible. Conduction electrons are reflected at the FI/S interface and during  this process their spin interacts  via  exchange interaction  with the localized magnetic moment of the FI material.  If the latter is not too strong and the superconducting layer is thin enough, both magnetic and superconducting orders can coexist. In this case the superconductor  behaves as a ferromagnetic superconductor with a  density of states showing a spin splitting, similar to the one that an external in-plane magnetic field would induce in a thin S layer \cite{meservey1994spin}.

In this section we describe the main features of  superconductors with a spin-split density of states due to the proximity of a FI.   Such  artificial ferromagnetic superconductors are the  main focus of this review. They have been fabricated and measured in several experiments along the past decades \cite{PhysRevLett.56.1746,Hao:1990,PhysRevLett.106.247001,strambini2017revealing}.

\subsection{Induced spin splitting in a superconductor--ferromagnetic insulator structure\label{sec:S-FI}}

As mentioned above, a way of creating a superconductor with a spin-split density of states is via the exchange field induced by an adjacent ferromagnetic insulator.  
 This may lead to a spin splitting  of the order of the superconducting gap, even without an externally applied  magnetic field.  The first evidence of an exchange field induced in a superconductor via the magnetic proximity effect was observed in a thin Al film in contact with an EuO film \cite{PhysRevLett.56.1746} and  with EuS \cite{Hao:1990,Wolf2014,PhysRevLett.106.247001}.
The equivalent internal field of the Zeeman splitting in Al-EuS systems has been reported to be as large as 5 T \cite{Moodera_review}.  Another S-FI combination explored so far is GdN-NbN \cite{spin_filter_Blamire,muduli2017spin}. This system shows, however, a  weaker and less clear  spin splitting than the Al-europium chalcogenide combination, due to the influence of 
the sizable  spin-orbit interaction in Nb compared  to the Al-based devices.

 Because the electrons of the superconductor cannot penetrate into the insulating  FI, 
the superconducting properties are only modified by the induced spin-splitting field at the S/FI interface, 
 and not by the leakage of Cooper pairs into the FI. 
Moreover, FIs can also be used as  spin-filter barriers \cite{Moodera_review}, in some cases with a very high spin-filtering efficiency,
 and therefore they will play a crucial role for several of the applications discussed in subsequent sections  which require strong spin filtering.

The amplitude of the spin splitting in a FI/S structure depends on both the intrinsic properties of the superconductor, such as the amount of magnetic impurities and the strength of spin-orbit coupling, and on the quality of the S/FI interface. The latter is crucial for obtaining a large splitting, as shown for example in Ref.~\cite{Hao:1990}.

Large splitting and spin-filter efficiencies have been achieved in different FI/S combinations, such as 
in {EuO}/{Al}/AlO$_3$/Al \cite{PhysRevLett.56.1746},  Au/{EuS}/{Al} \cite{PhysRevLett.61.637}, {Al}/{EuS}/Al \cite{Hao:1990}, Ag/{EuSe}/{Al} \cite{Hao:1990}, {EuSe}/{Al}/AlO$_3$/Ag, {NbN}/{GdN}/{NbN} \cite{spin_filter_Blamire}, {NbN}/{GdN}/TiN \cite{PhysRevB.92.180510}. 
A summary of  these parameters in different FI-S structure is presented in Table I of Ref.~\cite{bergeret2018colloquium}. 

The spin splitting caused by a ferromagnetic insulator on an adjacent superconductor can be understood with the help of the   following  
model \cite{khusainov1996indirect,Kushainov_review}. The effective Hamiltonian describing a FI/S bilayer consists of the ferromagnetic coupling within the FI layer,
\begin{equation}
H_{\rm FI}=-\sum_{{\bf r,}{\bf r'}}J_{\bf r,r'}{\bf S_{r}S_{r'}}+H_{\rm an}\:,\label{eq:HFI} 
\end{equation}
where $J_{\bf r,r'}$ is the exchange coupling between the localized spins ${\bf S_{r}}$. The term $H_{\rm an}$ describes {the magnetic anisotropy whose specific form does not need to be fixed here.}

The S layer is described by the usual BCS Hamiltonian \cite{bcs1957,TinkhamBook}, whereas for the FI/S interface we consider a model of
  an ensemble of localized magnetic moments that  interact with the spin of the conduction electrons of the superconductor via exchange interaction,
\begin{equation}
H_{ex}=-J_{ex}\tilde{\sum}_{{\bf r}}\Psi_{\alpha}^{{\dagger}}({\bf r})\left({\bf S_{r}\sigma}\right)_{\alpha\beta}\Psi_{\beta}({\bf r)}\:.\label{eq:Hex-1}
\end{equation}
 Here, the symbol $\tilde\Sigma$ means that we consider only the magnetic moments ${\bf S_{r}}$ localized  at  the interface. $J_{ex}$ is an effective parameter which for example describes the s-d or s-f exchange interaction.
  We consider a ferromagnetic insulator with a Curie temperature much larger than the superconducting transition temperature. Thus we can assume that  the magnetization of the FI is only determined  by the Hamiltonian in Eq.~(\ref{eq:HFI}) and  not  affected by the superconducting state \cite{buzdin1988ferromagnetic,Bergeret2000}. 
With this assumption  Hamiltonian (\ref{eq:Hex-1}) describes  conduction  electrons interacting with an  effective  exchange field proportional  to the   local spin average  $\langle {\bf S_{r}}\rangle$ at the FI/S interface.  This average  can be computed by solving independently  the  magnetic Hamiltonian  (\ref{eq:HFI}) \cite{strambini2017revealing}.   

In the superconducting state the localized  exchange field leads to a modified   density of states characterized by the splitting of the coherent BCS peaks.  This modification of the spectral properties of the superconductor is non-local and survives over distances away from the FI/S interface of the order of the   coherence length  $\xi_{s}$ \cite{PhysRevB.38.8823,bergeret2004induced}.  
If  the thickness $d$ of the S film is  much smaller
than  $\xi_{s}$, the spin splitting  
can be assumed to be  homogeneous across the film. Thus the density of states is given  
by Eq.~(\ref{eq:split_Dos}) with an effective  exchange field  ${\bf h}_{eff}\approx J_{ex}\langle{\bf S_r}\rangle a/d$ \cite{deGennes1966coupling,khusainov1996indirect,PhysRevB.38.8823}, where $a$ is the characteristic distance between the localized spins.

In this review we mainly focus on  thin S films  and hence we adopt this approximation for the description of the uniform magnetic proximity effect when the film is adjacent to a FI layer. For an inhomogeneous magnetic configuration of the FI layer the spatial scale that determines  the effective splitting is the superconducting coherence length. Thus the splitting becomes observable  if the  ferromagnet  consists of magnetic domains with sizes larger than the superconducting coherence length.  This is  for example the case of  EuS films with magnetic  domains of micrometer size \cite{tischer1973ferromagnetic} that  explain the spin splitting observed in experiments   \cite{Hao:1990} on Al-EuS structures, where the Al layers are a few nanometers thick and with a coherence length  of the order of 100 nm.  In contrast, a  spatially fast changing magnetization averages to zero and results in a vanishing  effective splitting \cite{strambini2017revealing,aikebaier2018superconductivity}.

It also follows from this description that the strength of the spin splitting depends crucially on the quality of the FI/S interface. The growth of any non-magnetic oxide  between the FI and S layers drastically suppresses the effective exchange
 interaction $J_{ex}$ and hence reduces the spin splitting \cite{Hao:1990}.
In addition to the quality of the interfaces,   the spin splitting also depends on the intrinsic properties of the superconducting film. For example, magnetic disorder may lead to a strong suppression of superconductivity in the S film \cite{Abrikosov1961}.
But  even in the absence of magnetic disorder and magnetic impurities,  
spin-orbit coupling may lead  to a modifications in the  DOS. In particular, large spin-orbit scattering rates lead to a rounding of the BCS peaks and to a less sharp    spin splitting. This explains why splitting 
  has been observed in materials with relative  small atomic number $Z$, such as Al,  Be, V, but not in heavier materials such as Pb.   For a detailed description of the intrinsic properties of different superconductors in this respect, we refer the reader to the excellent review in Ref.~\cite{meservey1994spin}.

Spin splitting in a superconductor can also be achieved by applying an external magnetic field \cite{PhysRevLett.25.1270,PhysRevLett.26.192,PhysRevB.16.4907,PhysRevB.22.1331,Hao:1990,PhysRevLett.106.247001}. In this case, however, the magnetic field  also couples to the orbital motion of the electrons
and creates circulating currents, the Meissner effect, \cite{meissner33} that  try 
to expel the field  from the bulk (diamagnetic response).
By increasing the amplitude of the applied field superconductivity is gradually reduced.
We denote below this mechanism of suppression of superconductivity by the {\em orbital depairing} effect. 
This mechanism dominates in bulk samples or in thin films with the field applied perpendicular to the plane of the  film.  If the amplitude of the applied field reaches a critical value $H_{c}$, 
the created currents increase the free energy such that  the system undergoes a transition into
the normal state \cite{TinkhamBook}.
However,  in superconducting films with  thickness smaller than the 
London penetration length,  the critical value  $H_{c\parallel}$ for a magnetic field  applied 
in-plane   largely exceeds the critical value $H_{c\perp}$
of a perpendicular field \cite{TinkhamBook,Althinfilms}. In such a case the spin paramagnetic effect  (due to Zeeman effect) dominates with respect to the orbital one. The magnetic field  penetrates uniformly the film,  screening currents are relatively small and therefore 
$H_{c\parallel}$ is limited by the spin paramagnetic effect that tries to align the spin of the original singlet Cooper pairs, 
as demonstrated in Refs.~\cite{PhysRevLett.9.266,Chandrasekhar:1962}. At $T=0$, and in the absence of spin-orbit coupling and magnetic impurities, 
the critical field due to the paramagnetic effect is given by $H_{cP}=\Delta_{0}/(\sqrt{2}\mu_{B})$, where 
$\Delta_0$ is the superconducting gap at zero field and temperature and $\mu_{B}$
is the Bohr magneton. 
For conventional BCS superconductors with critical temperature ranging from 1 to 10 K,
 $H_{cP}$ can be of the order of  several Tesla and hence superconductivity exists within a large range of  field  strengths. In short,   the paramagnetic effect  in thin superconducting films  leads to a Zeeman splitting
  of the  density of states (DOS)  \cite{PhysRevLett.25.1270,PhysRevLett.26.192,PhysRevB.16.4907,PhysRevB.22.1331,Hao:1990,PhysRevLett.106.247001}, similar to the spin-splitting induced in S-FI structures.

\subsection{Spin-split density of states  in a superconductor\label{sec:dos_S-FI}}

\begin{figure}
  \centering
\includegraphics[width=8cm]{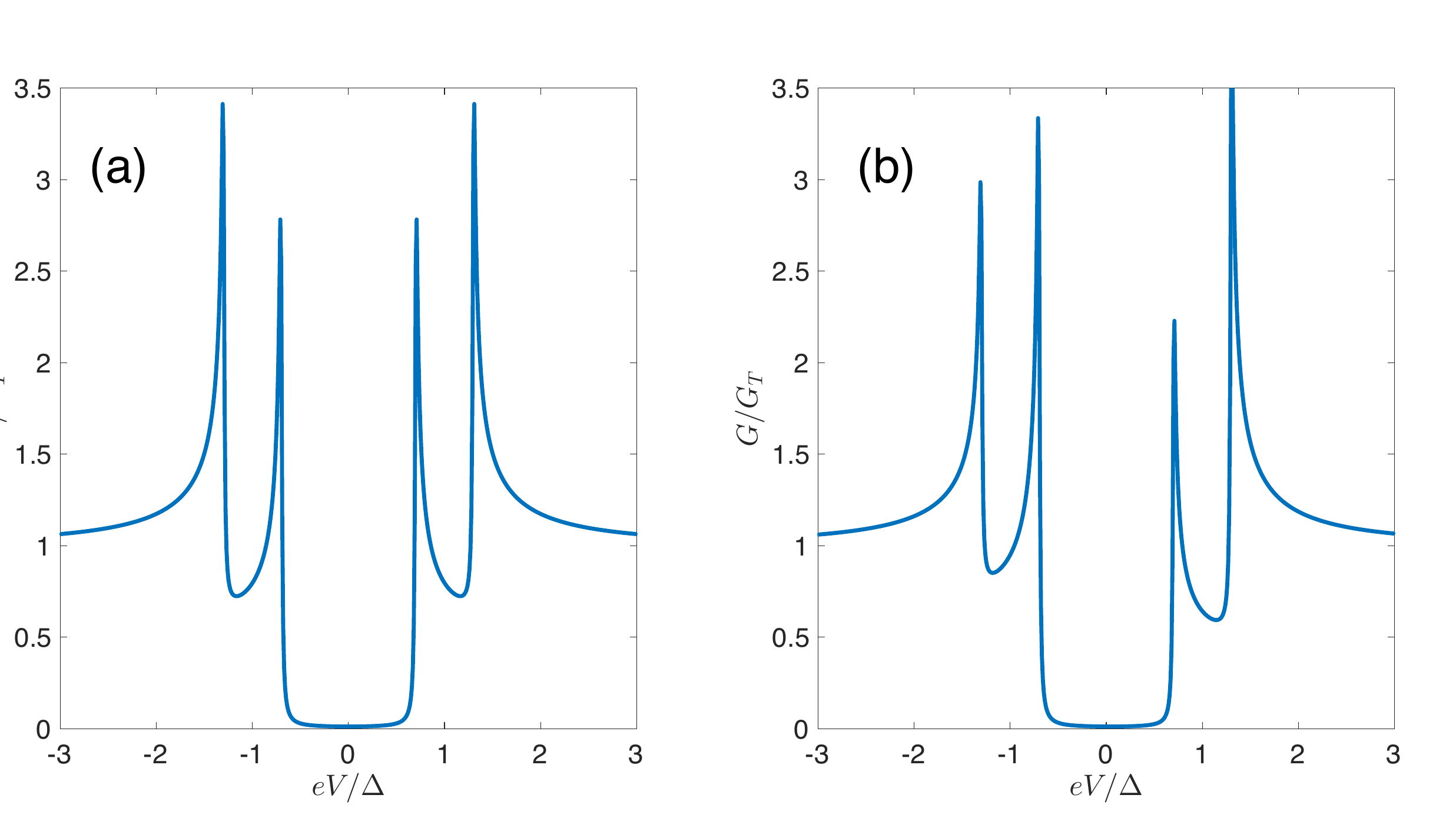}
\caption{Differential conductance of a FI-S-I-N obtained from Eq.~(\ref{eq:I_tun}) (a) when 
N is a non-magnetic metal; (b) and for a spin-polarized N.
 \label{fig:The-differential-conductance}
}
\end{figure}

 The splitting  discussed in the previous section can be detected  by measuring  the tunneling conductance of the  superconductor with a normal or a superconducting probe.  
For example,  if one uses a normal metal (N)  tunnel coupled to a superconducting film S, 
 the current through the  normal-insulator-superconductor (NIS)  junction
 is determined  by the tunneling expression \cite{giaver1961}
     \begin{align} 
 I(V)=\frac{G_T}{e}  
 \int_{-\infty}^{\infty} N(\varepsilon-V)N_{s}(\varepsilon)\left[n_{F}(\varepsilon-V)-n_{F}(\varepsilon)\right]d\varepsilon\;,\label{eq:I_tun}
\end{align}
where $N$ and $N_{s}$ are the reduced density of states of the normal metal
and the superconductor, respectively, $n_{F}$ the Fermi distribution function
and $G_T$ is the conductance of the insulating tunneling  barrier I. We first assume $G_T$ to have no spin polarization. Electronic transport occurs mainly at energies  
 close to the Fermi level where the density of states of the normal probe $N(\varepsilon)$ 
can be accurately approximated by a function independent of the energy.
In such a case, and at low temperatures the measured differential
conductance $G=dI/dV$ is proportional to the density of the
states of the superconductor. If the normal metal probe  is non-magnetic,
i.e., electrons are not spin-polarized, the measured
$G$ shows the spin-split peaks and  is symmetric with respect
to the voltage as sketched in  Fig. \ref{fig:The-differential-conductance}(a).

If the tunneling  barrier has a  spin-dependent transmission, i.e., a spin
filter, or a  ferromagnetic electrode with a non-vanishing spin polarization, 
the observed peaks are asymmetric \cite{PhysRevB.16.4907} [see Fig. \ref{fig:The-differential-conductance} (b)]. 
The asymmetry is proportional to the   spin polarization of the conduction electrons of the electrode or the spin-filter efficiency of the barrier. 
 Indeed spin-split superconductor-ferromagnet bilayers have  been used for determining the spin polarization of magnetic metals \cite{PhysRevB.16.4907, PhysRevB.22.1331,PhysRevLett.26.192}.

The normalized total density of states of the spin-split 
 superconductor, sketched in Fig.~\ref{fig:The-differential-conductance} (a), can be written  
 as the sum of the DOS $N_{\uparrow/\downarrow}$ of each spin species,\footnote{
  As conventional, here and below $\sqrt{\varepsilon^2-|\Delta|^2}$
  means $\sgn(\Re{}\varepsilon)\sqrt{\varepsilon^2-|\Delta|^2}$ in
  terms of the principal branch $\Re\sqrt{\cdot}\ge0$ of the square root.
}
\begin{equation}
N=\frac{1}{2}{\rm Re}\frac{\varepsilon+h_{\rm eff}}{\sqrt{\left(\varepsilon+h_{\rm eff}\right)^{2}-\Delta^{2}}}+\frac{1}{2}{\rm Re}\frac{\varepsilon-h_{\rm eff}}{\sqrt{\left(\varepsilon-h_{\rm eff}\right)^{2}-\Delta^{2}}}\;, \label{eq:split_Dos}
\end{equation}
where $\pm h_{\rm eff}$ is the effective spin-splitting
field.

Equation (\ref{eq:split_Dos}) for the DOS is a simplified description
of spin-split superconductors, which is accurate, as we see below, 
 only in certain limiting cases.  The  expression does not take into account  the effect of  magnetic impurities  or   spin-orbit coupling (SOC). 
For example,  magnetic impurities suppress superconductivity and
eventually lead to a gap-less situation \cite{Abrikosov1961}. 
On the other hand the SOC counteracts the magnetic impurities but it
may also  lead  to a broadening of the coherence peaks in the spectrum
\cite{maki1966effect,werthamer1966temperature,fulde1973high,meservey1975tunneling}. 
 Moreover, if the spin-splitting field $h_{\rm eff}$ in Eq.~(\ref{eq:split_Dos}) is due to an external magnetic
 field, also orbital depairing contributes to the pair breaking.
  We present a more quantitative analysis of these effects in the next section.

\subsection{Paramagnetic depairing mechanisms and the FFLO state}
\label{sec:paramagnetic}

When writing   the DOS of a superconductor in an exchange field, Eq. (\ref{eq:split_Dos}), 
we assume that the ground state of the S layer corresponds to a spatially homogeneous order parameter $\Delta$. The value of the order parameter is determined  from the self-consistency equation, written in terms of the Matsubara frequencies $\omega_n = (2n+1)\pi T$, \footnote{In this review we set $\hbar=k_B=1$ and hence
  the temperature, frequencies and inverse times have dimensions of energy,
  whereas the momentum has a dimension of inverse length.}
\begin{equation} \label{Eq:SelfConsistent}
\Delta \ln (T_{c0}/T) = 2 \pi T \sum_{\omega_n>0}^\infty \left( g_{01} - \frac{\Delta}{\omega_n}\right)
\end{equation}
and a free-energy functional discussed in Sec.~\ref{sec:freenergy}. Here $g_{01}$ is a singlet anomalous part of the Green's function described in more detail in Sec.~\ref{sec:quasiclassical_theory}. In the following we describe the behavior of $\Delta$ as a function of the exchange field $h$. For this, we denote $\Delta_0=1.76 T_{c0}$ the order parameter at $T=h=0$ without depairing effects. $T_{c0}$ is the corresponding critical temperature.
 In the absence of spin relaxation the results coincide with those of Ref.~\cite{maki_tsuneto.1964}: {For large values of $h$, the paramagnetic depairing drives the superconductor to the normal state.} There is a critical value of temperature, $T_0\approx 0.556T_{c0}$, above which the transition between the normal and superconducting states is of the second order. 
 For $T<T_0$,  the transition is of the first order, {and the self-consistency equation has three solutions, of which one is unstable. The stable solutions correspond to superconducting ($\Delta \neq 0$) and normal ($\Delta=0$) states. Below the critical field $h_c$, known as the Chandrasekhar-Clogston limit
 \cite{Chandrasekhar:1962,PhysRevLett.9.266}, the superconducting state is preferred, and above it the normal state. At $T=0$ this critical field is $h_c=\Delta_0/\sqrt{2}$. Finally, for $h>\Delta_0$, the superconducting solution does not exist at all.}

\begin{figure}
  \centering
 \includegraphics[width=8cm]{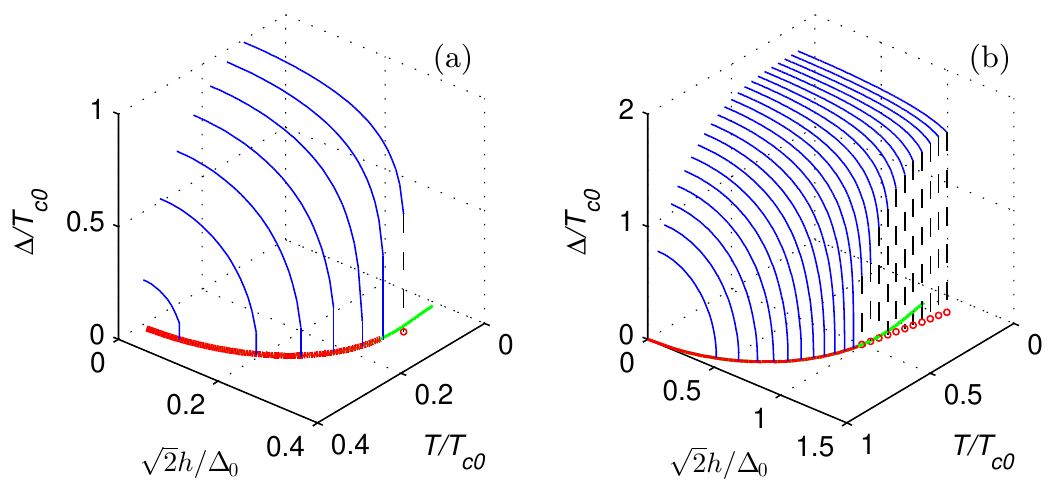}
 \caption{\label{Fig:opHT}
  Order parameter $\Delta$ as a function of the temperature $T$ and the exchange field $h$.
  Spin relaxation rate $\tau_{sn}=1/(\tau_{sf}^{-1}+\tau_{so}^{-1})$ is $1/(T_{c0}\tau_{sn})=0.96$.
  (a) Spin-flip relaxation $\beta=(\tau_{\rm so}-\tau_{\rm sf})/(\tau_{\rm so}+\tau_{\rm sf})=1$. (b) Spin-orbit relaxation $\beta=-1 $.
  The transition line $\Delta(h,T)=0$ is shown by the solid curve in the $(h,T)$
  plane. Its thick red part at $T>T_0$ shows the second-order transition line. 
  Its thin green part at $T<T_0$ shows the points on $T,h$ plane where the 
   unstable branch starts (not shown).
  The first-order transition is shown by the circles in the $(h,T)$ plane.
    }
 \end{figure}

  A non-zero spin relaxation rate leads to a quantitative modification of this behavior
\cite{bruno1973-mfs}. In Fig.~\ref{Fig:opHT} 
we show the computed $\Delta$ as a function of temperature and exchange field
for a  normal-state spin relaxation rate $1/T_{c0}\tau_{sn} =0.96$.
Panels (a) and (b) in Fig.~\ref{Fig:opHT} show the effects of 
spin-flip ($\beta=1$) and spin-orbit ($\beta=-1$) scattering, respectively.  
 The phase-transition line, $\Delta(T,h)=0$, is shown in Fig.~\ref{Fig:opHT} by the solid curve in the $(T,h)$ plane.  The red part on this curve corresponds to the second-order transition where $\Delta(h)$ continuously goes to zero  with increasing $h$.
This behavior changes to the abrupt first-order transition at $T<T_0$.  The first-order transition line
is shown by circles in the $(T,h)$ plane  and it is different from the $\Delta(T,h)=0$ curve shown by the green line at $T<T_0$.
Both relaxation mechanisms reduce the range of temperatures for which  the first-order transition takes place. In other words, the   threshold temperature $T_0$ between first- and second-order phase transitions  is reduced as compared to the case without relaxation.

In other respects the modification of the superconducting state 
strongly depends on the spin relaxation mechanism. 
Spin-flip scattering 
breaks the time-reversal symmetry and therefore
leads to a strong suppression of the superconducting gap
and $T_c$. On the other hand, spin-orbit interaction is time-reversal invariant 
and keeps the $T_c$ intact. 

\begin{figure}
  \centering
 \includegraphics[width=8cm]{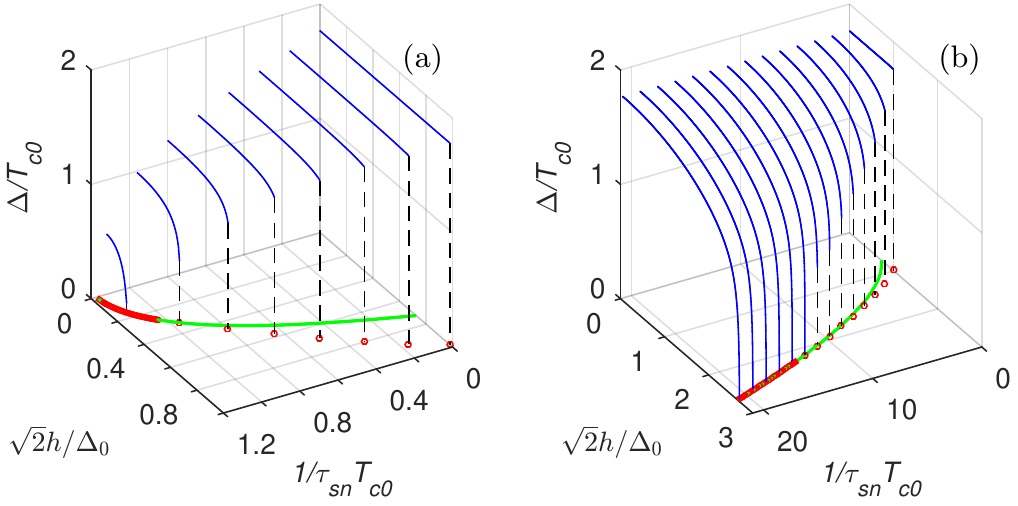}
 \caption{\label{Fig:opHtau}
  Order parameter $\Delta$ as a function of the exchange field $h$ and the normal-state spin 
  relaxation rate $1/\tau_{sn}$.
  The temperature is  $T=0.1 T_{c0}$.
  (a) Spin-flip relaxation $\beta =1$. (b) Spin-orbit relaxation $\beta=-1 $.
  The line of critical relaxation rates given by  $\Delta(h,\tau^{-1}_{sn})=0$ is shown by the solid curve in the $(\tau^{-1}_{sn},h)$
  plane. Its thick red part at $\tau^{-1}_{sn}> \tau_0^{-1}$ shows the second-order phase transition line. 
  Its thin green part at $\tau^{-1}_{sn}< \tau_0^{-1}$ corresponds to the 
  points where the  
   unstable branch starts (not shown).
  The first-order transition is shown by the circles in the $(\tau^{-1}_{sn},h)$ plane.
    }
 \end{figure}

\begin{figure}
  \centering
\includegraphics[width=8cm]{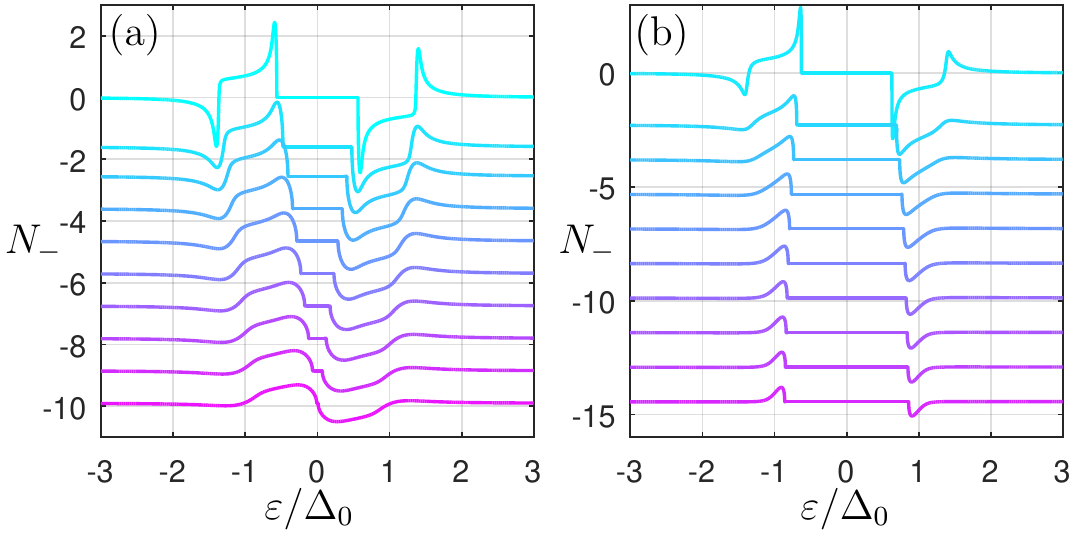}\caption{Calculated difference between spin-up and spin-down density of states $N_-=N_\uparrow - N_\downarrow$ of a thin superconducting film at $T=0.1T_c$,
 $h=0.4\Delta_0$ and different spin relaxation rates $\tau_{sn}$. 
 (a) Spin-flip  relaxation $\beta=1$,
curves from top to bottom correspond to an increasing
$(\tau_{sn}T_{c0})^{-1}$, varying equidistantly from $0.03$ by  $0.07$ steps.
(b) Spin-orbit relaxation $\beta=-1$, 
curves from top to bottom correspond to an increasing
$(\tau_{sn}T_{c0})^{-1}$, varying equidistantly from $1$ by  steps of $2$.
For clarity the curves are shifted along the vertical axis.
\label{fig:Calculated-density-of}}
\end{figure}

A striking effect of the spin-orbit  scattering is 
that it increases the critical field of the Chandrasekhar-Clogston limit \cite{bruno1973-mfs}.
This tendency is shown in Fig.~\ref{Fig:opHtau},
where we compare the dependencies of the superconducting gap on $h$ and relaxation rates $1/\tau_{sn}$ for (a) spin-flip and (b) spin-orbit relaxation, respectively. 
These two cases are characterized by the opposite behaviors of the critical field 
$h_c$ as a function of $1/\tau_{sn}$ --- it is suppressed by spin-flip scattering 
and enhanced by the spin-orbit one. 
{This effect can be understood from the comparison to the changes in the density of states caused by the two spin-relaxation mechanisms (see Fig.~\ref{fig:Calculated-density-of}): spin-flip scattering primarily lifts the gap in the density of states, whereas spin-orbit scattering acts to nullify the spin splitting without affecting the gap.}
In both cases there is a threshold value of the 
scattering rate $\tau^{-1}_0$ when the superconducting phase transition changes from the second order 
at $\tau^{-1}_{sn} > \tau^{-1}_0$ to the first order at $\tau^{-1}_{sn} < \tau^{-1}_0$. 

The assumption of a homogeneous order parameter does, however, not always correspond to the lowest-energy state of the system. In Refs.~\cite{fulde_ferrell.1964,larkin_ovchinnikov.1964} it was predicted that the exchange or Zeeman field can cause a transition to an inhomogeneous superconducting state, with a spatial periodical modulation of the order parameter at the scale of the coherence length. Such an FFLO state exists for temperatures below $T_0$ and only for exchange fields that satisfy $0.71\Delta_0<h<0.755\Delta_0$ when $T\rightarrow0$.

The FFLO state has not been observed in conventional superconductors. One of the main reasons for this is that it is very sensitive to disorder \cite{aslamazov1969-iie} and spin-orbit coupling and hence it is expected to occur only in extremely clean samples.  In the present review we mainly focus on diffusive systems and therefore we do  not  pay further attention to the FFLO state.  Moreover, the effective exchange fields we mostly consider are far below the range for the FFLO state to occur. 

The FFLO state  has been  also studied in the context of heavy-fermion superconductors \cite{gloos1993possible} and ultra-cold Fermi gases  \cite{koponen2007finite,korolyuk2010probing,kinnunenup17} especially in the context of the BCS-BEC crossover \cite{sheehy2006fflo,sheehy2007fflo}.

\subsection{Static spin susceptibility of conventional superconductors}

\label{sec:spin-susceptibility}

The spin splitting has a direct consequence on the 
paramagnetic spin susceptibility of the superconductor.
 In general, the  spin susceptibility $\chi$ can be found by calculating the 
spin polarization  
${\bm S} = - \chi {\bm h}$
generated by  a small spin-splitting field ${\bm h}$. 
In the superconducting state the susceptibility normalized with respect to that of the normal state 
$\chi_n$ can be expressed 
in terms of the difference between spin-up and spin-down DOS 
$N_- = N_\uparrow - N_\downarrow$ and the equilibrium distribution function $f_{\rm eq}=\tanh(\varepsilon/2T)$ \cite{abrikosov1961problem,Bergeret2005},
 \begin{equation}\label{Eq:SpinDens}
 \chi/\chi_n =1+ \frac{1}{4h} \int_{-\infty}^\infty 
 \tanh\left(\frac{\varepsilon}{2T}\right) N_- d\varepsilon .
 \end{equation}  
 Using Eq.~(\ref{eq:split_Dos}) for the spin-split DOS in the absence of magnetic depairing 
 processes one can see that $\int_{-\infty}^{\infty} N_- d\varepsilon = -4h $.
Therefore for $T\to 0$ and in the absence of spin-dependent scattering $\chi (T=0) =0$. 
 This is a consequence 
of the lack of polarizability of the condensate consisting of spin-singlet Cooper pairs.

However,  if  spin-dependent scattering is present, in the form of  spin-orbit and/or magnetic impurities causing spin flips, the superconducting condensate may exhibit  a nonzero paramagnetic susceptibility for $T\to 0$ \cite{abrikosov1961problem,yosida1958paramagnetic}.
 %
 
\begin{figure}
  \centering
 \includegraphics[width=8cm]{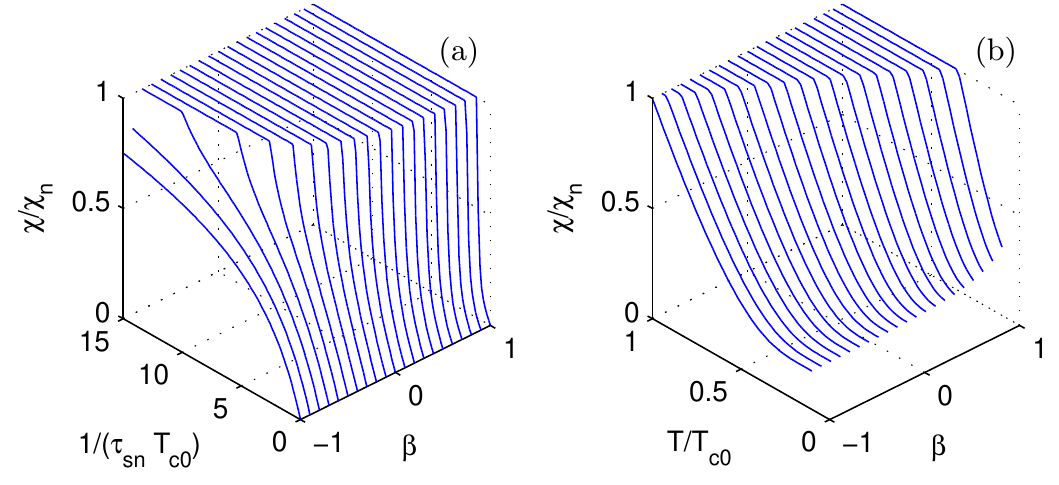}
 \caption{\label{Fig:SpinSusceptibilty}
 Paramagnetic susceptibility of a thin superconducting film for different values of $\beta$ as a function 
 of (a) the normal-state spin scattering rate $1/\tau_{sn}T_{c0}$ and (b) the temperature.
 In (a) $T=0.1 T_{c0}$ and in (b) $1/\tau_{sn} = 0.5 T_{c0}$, where $T_{c0}$
 is the critical temperature in the absence of spin scattering.  
   }
 \end{figure}

 A generalization of this result can be obtained by using the microscopic equations introduced in Sec.~\ref{sec:quasiclassical_theory} below. 
Here we are interested in the final results for the susceptibility  shown  in  Fig.~\ref{Fig:SpinSusceptibilty}. In particular, we show  the low-temperature dependence of the  spin susceptibility  as a function of  the normal-state spin scattering rate  {
 $1/\tau_{sn}=1/\tau_{so}+1/\tau_{sf}$ }
 and the parameter { $\beta=(\tau_{so}-\tau_{sf})/(\tau_{so}+\tau_{sf})$} determining the relative strength of spin-flip and spin-orbit scattering.
 From these plots one can see that both scattering mechanisms result in a nonzero susceptibility at low temperatures. Notice that  
  only the  spin-flip scattering  generates a strong suppression of the order parameter (see Fig.~\ref{Fig:opHT}). Therefore, at $T\ll T_{c0}$ the growth of $\chi$ as a function of the rate 
  $1/\tau_{sn}$ towards the normal-state value  is much faster when the spin-flip scattering dominates the 
  spin-relaxation in the normal state, {\it i.e.} when $\beta>0$.  
  The non-zero susceptibility explains the observed Knight shifts in ordinary superconductors \cite{meservey1978spinorbit}.

The static spin susceptibility (\ref{Eq:SpinDens}) characterizes the paramagnetic response of the superconductor to the external magnetic field. This expression cannot be used for calculating the non-equilibrium spin accumulation induced by the spin-dependent chemical potential shift 
$\pm\delta \mu$ when the distribution functions in different spin subbands are given by 
$f_{\uparrow}(E) = f_0(E+\delta\mu)$ and $f_{\downarrow}(E) = f_0(E-\delta\mu)$.
The theory of non-equilibrium spin states in superconductors is explained in detail in Secs.~\ref{sec:quasiclassical_theory} and \ref{spininjection}.
In particular, there we explicitly show [see the discussion after Eqs.~(\ref{Eq:ChPot0},\ref{Eq:ChPotZ})] that the  
 non-equilibrium spin accumulation generated by $\delta\mu$ is always exponentially suppressed at low temperatures
provided that there is a non-zero gap in the quasiparticle spectrum. 
This is the case  even in the presence of the strong spin-orbit interaction  [{\it cf.} Fig. \ref{fig:Calculated-density-of}(b)]
although the static spin susceptibility  (\ref{Eq:SpinDens}) is non-zero at $T=0$.

 \subsection{Other effects related to the Zeeman and exchange fields}
\label{subs:othereffects}
In the next sections we discuss effects related to non-equilibrium properties of spin-split superconductors in FI-S structures. 
It is however important to mention other effects that  we are not considering  but that may  appear
in the presence of Zeeman and exchange fields. 

\subsubsection{Fermi liquid effects}

The analysis in the previous subsections  assumes noninteracting electrons, which is a rather good approximation in metals.
Corrections due to interactions can be incorporated by renormalizing  certain parameters of the theory with the help of the powerful Landau's 
Fermi-liquid theory \cite{landau1957theory}.   In the context of the quasiclassical theory such effects have been studied in Ref.~\cite{Alexander1985}.
 In particular, and importantly for our work, the spin-splitting field amplitude $h\equiv|{\mathbf h}|$ entering Eq.~(\ref{eq:Usadel})
 is renormalized. According to the Fermi liquid theory it is given by \cite{Catelani2008Fermi,PhysRevLett.106.247001}:
 \begin{equation}
 \label{FL_h}
 h=2\mu_B\frac{H_{\rm ext}+H_{FI}}{1+G^0}\; , 
 \end{equation}
 where $H_{\rm ext}$ is the external magnetic field, $H_{FI}$ is the field induced by a FI, and $G^0$ is the effective antisymmetric Fermi-liquid parameter of Landau's theory. 
The value of $G^0$ in the superconducting state depends on temperature, is negligibly small at low temperatures ($T\ll T_c$), and nonzero for temperatures close to the 
superconducting critical temperature $T_c$ and above it.  In what follows we neglect the Fermi-liquid corrections by assuming that $G^0\ll1$. This is justified at the low temperatures where majority of our work concentrates. However, when describing equilibrium properties of spin-split superconductors at temperatures close to $T_c$, the Fermi-liquid corrections to the field $h$ in Eq.~(\ref{eq:Usadel}) should also be included for some superconductors. For example, for  Al films  $G^0\sim0.16-0.3$ \cite{PhysRevLett.52.1637,Catelani2008Fermi} close to $T_c$ and above, whereas in V it is negligibly small at all temperatures \cite{PhysRevB.40.8705}.


\subsubsection{Long-range triplet correlations}

In this review we mainly focus on the transport properties of the superconductor in S/F structures. 
For completeness, however, we briefly discuss in this section the appearance of 
 triplet correlations in ferromagnetic metals induced by the superconducting proximity effect \cite{bergeret2001long, Bergeret2005}.  Such correlations may induce non-dissipative spin-polarized 
 currents  that could be particularly useful to lower the  energy consumption in spintronic devices 
\cite{eschrig2011spin,linder2015superconducting,eschrig2015spin}.

Superconducting correlations can be induced in non-superconducting (N)  metals by means of  the proximity effect \cite{deGennes1964boundary}.
This can be seen as the "leakage" of Cooper pairs from the superconductor into the normal region. 
The microscopic origin of this "leakage" is the Andreev reflection \cite{andreev1964thermal}, where an incident electron from N is retro-reflected as a hole, resulting in the formation of a Cooper pair in S. An important consequence of the
 Andreev reflection  is that the correlation between the electron and the hole leads to a formation of a non-vanishing pair amplitude in N, within a  characteristic distance $\xi_N$ away from the S-N interface.  This distance depends on the intrinsic properties of N, such as the degree of disorder, spin-dependent fields, etc.   For example,  in a diffusive normal metal without magnetic impurities this length is of the order of the thermal length $\sqrt{D/T}$, where $D$ is the diffusion coefficient.

In a superconductor-ferromagnet (S-F) junction the situation is very different.  In usual ferromagnets the intrinsic exchange field $h$ is much larger than the superconducting characteristic energies. The proximity-induced superconducting correlations decay and oscillate on the length scale $\xi_F={\rm Re}\sqrt{D/h}$. The oscillation  of the condensate leads to the well-established  $0$-$\pi$  transition in S-F-S Josephson junctions predicted in Ref.~\cite{buzdin1982critical} and observed much later in experiments~\cite{ryazanov2001coupling,kontos2001inhomogeneous} on SFS structures with ferromagnetic alloys with a weak exchange field. In conventional  ferromagnets, such as Fe, Ni or Co, the effective exchange field is large  ($\sim 0.1-1$ eV) and hence $\xi_F$ is very short \cite{verbanck1994superconducting}. This explains why the Josephson effect in S-F-S junctions with transition metals has been only observed for thickness of the F layer smaller than 10 nm \cite{blum_PhysRevLett.89.187004,robinson_PhysRevLett.97.177003}.

The situation is very different if the exchange field is spatially non-homogeneous, as for example in a magnetic domain wall. In such a case, as shown in Ref.~\cite{bergeret2001long}, triplet components with equal spin components can generated. These are insensitive to the local exchange field and hence can penetrate the F metal over the thermal length. 
Such  long-range triplet component of the superconducting condensate  explains the  long-range Josephson currents observed 
in S-F-S junctions made for example  of half-metal \cite{keizer2006spin,anwar2012long}, ferromagnetic multilayers  \cite{khaire2010observation}, and ferromagnets with intrinsic inhomogeneous magnetization  \cite{robinson2010controlled}.

We emphasize that the physics of the triplet component  addressed 
in this section describes the  equilibrium state of ferromagnets in contact with superconducting electrodes. 
The main focus of the present review  is however the non-equilibrium properties of  superconductors
 in spin-splitting fields, either generated by an adjacent FI or by an external field.  As discussed in Sec.~\ref{sec:paramagnetic}, in order to preserve the superconducting state,  those fields have to be smaller than the superconducting gap, and hence all components of the condensate, singlet and triplets, vary over similar  length scales. This in particular means that no  distinction between short- and long-range has to be made. 

\subsubsection{Cryptoferromagnetic state}
The way we model the exchange field generated in a thin superconducting film  by an adjacent ferromagnetic--insulating layer, see Sec.~\ref{sec:S-FI}, is based on the assumption that  the magnetic ordering of the FI is the same in both the normal and superconducting states of the S layer.  This requires that the effective exchange constants $J$ and $J_{ex}$ in Eqs.~(\ref{eq:HFI})
 and (\ref{eq:Hex-1}), respectively, satisfy $J\gg J_{ex}$.  In this case the ferromagnetic order is robust against  the exchange interaction with the conduction electrons and the superconducting condensate.  Roughly speaking, this corresponds to the case when the Curie temperature $T_{\rm Curie}$ of the FI layer is larger than the effective exchange field, whose maximum value in the superconducting state is smaller than the  superconducting  order parameter $\Delta$. For example, in  EuS/Al junctions $T_{\rm Curie}\sim 16.6$ K and $\Delta_0\sim 0.25$ meV, thus $k_B T_{\rm Curie}/\Delta_0\sim 5.3$. In other FIs such as EuO or NbN, the Curie temperature is even larger and hence this ratio increases. Thus the approximation made in Sec.~\ref{sec:S-FI} is well justified.
 
In the case of a weak magnetic stiffness of the FI compared to the characteristic superconducting energy $\Delta$, one should take into account the effect of the superconducting condensate on the magnetic ordering mediated by the Hamiltonian in Eq.~(\ref{eq:Hex-1}). The competition between superconducting and magnetic ordering in such a case has been first considered in Ref.~\cite{anderson1959spin}. They considered a ferromagnetic state mediated via the Ruderman-Kittel-Kasuya-Yosida (RKKY) indirect exchange interaction. Above the superconducting critical temperature $T_c$ the ground state of the system is a homogeneous magnetic state. When the temperature is lowered below $T_c$ the ground state of the system corresponds to a spatially inhomogeneous  magnetic structure characterized by a   modulation with a wave vector of the order of $(\lambda_F^2\xi_s)^{1/3}$, where $\lambda_F$ is the Fermi wavelength and $\xi_s$ is the superconducting coherence length.   This state was called by Anderson and Suhl the cryptoferromagnetic state and may describe the situation of certain ternary rare earth compounds \cite{bulaevskii1985coexistence}.
 Such a magnetic  modulation  was also  predicted for thin metallic \cite{buzdin1988ferromagnetic,Bergeret2000} and insulating  \cite{Kushainov_review} ferromagnetic films on top of bulk superconductors. 

Although in this review we assume that the condition $k_BT_{\rm Curie}/\Delta_0\gg 1$ is satisfied, and hence no cryptoferromagnetic state is induced, it is important to keep in mind that such a state may exist in the case of FIs with weak magnetic stiffness.

\section{Quasiclassical theory of diffusive FI/S structures}
\label{sec:quasiclassical_theory}

In this section, we outline a microscopic theory useful for describing
nonequilibrium effects in spin-split superconductors.  We concentrate
on diffusive superconductors where scattering by non-magnetic
impurities results to a mean free path that is small compared to
other lengths involved in the problem, with the exception of the Fermi
wavelength $\lambda_F$. This condition is equivalent to the assumption that the
inverse of the elastic relaxation time $1/\tau$ is much larger than
all characteristic energies apart from the Fermi energy.

Within a quasiclassical theory framework, which operates on length scales large compared to the Fermi wavelength, the superconducting properties can be described by a diffusion-like equation \cite{usadel.1970}, now called the Usadel equation, which describes the behavior
of the Keldysh Green's function (GF),
\begin{equation}
 D\nabla\cdot(\check{g}\nabla\check{g})+[i \varepsilon\tau_{3}-i\vec{h}\cdot\vec{\sigma}\tau_{3}-\check{\Delta}-
 \check\Sigma,\check{g}]=0.\label{eq:Usadel}
\end{equation}
Here $D$ is the diffusion coefficient,
$\check{\Sigma}=\check{\Sigma}({\bm r},\varepsilon)$ is the general
self-energy term and $\check{g}({\bf r},\varepsilon)$ is the (momentum
isotropic part of the) quasiclassical Green's function, obtained by integrating
the microscopic Green's functions over the quasiparticle energy.  In
addition to \eqref{eq:Usadel}, the GF satisfies the quasiclassical
normalization condition
\begin{equation}
\check{g}^{2}=\check{1}\;.\label{normalization}
\end{equation}
The derivation of the Usadel equation~\eqref{eq:Usadel} is discussed
in several previous works
\cite{Eilenberger1968,usadel.1970,belzig1999quasiclassical,Bergeret2005},
and we do not present it here. For
the readers not familiar with the quasiclassical method it is
sufficient to understand that the knowledge of the structure of the GF
sheds light on the spectral and transport properties of
superconductors. In particular, the self-energies $\check{\Sigma}$
generate the collision integrals for the different scattering
processes of the kinetic transport theory, as described in more detail
in Sec.~\ref{sec:noneq-modes}.

In Eq.~(\ref{eq:Usadel}), 
the matrices $\tau_j$, $j=1,2,3$ are Pauli matrices in the Nambu (or
electron-hole) space.\footnote{In this review, when the
    matrices $\tau_i$ or $\sigma_i$, $i=1,2,3$ show up alone, 
    if required by the context, this
    should be understood in terms of an outer product of the
    $2\times2$ Pauli matrix in one (Nambu or spin) space and the
    $2\times 2$ identity matrix in the other space. See also
    \ref{app:paulimatrices}.}  The matrices
$\sigma_j$ ($j=0,1,2,3$), are the Pauli matrices in spin space,
with $\sigma_{0}$ denoting the unit matrix. The vector $\bm{h}$ denotes the spin-splitting field
generated either by an external field or the magnetic
proximity effect in a FI/S junction, and $\check\Delta=\Delta \tau_\uparrow + \Delta^* \tau_\downarrow$
is the singlet superconducting order parameter.

The Green functions and self-energies are functions of the spatial
coordinate and the energy. They are also matrices in the
Keldysh$\otimes$Nambu$\otimes$spin space.  The $4\times4$
Nambu$\otimes$spin matrix structure corresponds to the Nambu
bi-spinor, which we in this review choose as
\begin{equation} \label{Eq:SpinNambuBasis}
\Psi = \begin{pmatrix} \psi_\uparrow({\mathbf r},t) &
  \psi_\downarrow({\mathbf r},t) & -\psi_\downarrow^\dagger({\mathbf r},t) &
  \psi_\uparrow^\dagger({\mathbf r},t)\end{pmatrix}.
\end{equation}
The structure in the Keldysh
space can be represented via three independent components. 
Here we use the representation
\begin{equation}
{\check g}=\left(\begin{array}{cc}
{\hat g^{R}} & {\hat g^{K}}\\
0 & {\hat g^{A}}
\end{array}\right)\,, \label{eq:gmatrix}
\end{equation}
where the retarded (R), advanced (A) and Keldysh (K) components are
4$\times$4 matrices in the Nambu $\otimes$ spin space.

The set of equations is completed by expressions for the self-energy $\check{\Sigma}$
and a self-consistency relation for $\Delta$. The former is discussed in subsequent sections,
and the latter for conventional superconductors can be written as
\begin{equation}
  \Delta =\frac{\lambda}{16 i}\int_{-\Omega_D}^{\Omega_D} d\varepsilon\, {\rm Tr} [(\tau_1-i\tau_2) \hat g^K (\varepsilon) ]\, .
  \label{self_consistent}
\end{equation}
Here, $\lambda$ is the effective coupling constant and $\Omega_D$
is the Debye cutoff frequency.

Quite generally, the R/A components of the Green function are related to spectral properties
of the superconductors whereas K also contains information about the
quasiparticle kinetics. Indeed, it can be related to the electron
distribution function $\hat{f}$,
\begin{equation}
\hat{g}^{K}=\hat{g}^{R}\hat{f}-\hat{f}\hat{g}^{A}\;.\label{eq:K_param}
\end{equation}
In the most general case, the 4$\times$4 matrix distribution function $\hat{f}$ can be written as
\begin{equation}
  \hat{f}=f_{L}\hat{1}+f_{T}\tau_{3}+\sum_{j}(f_{Tj}\sigma_{j}+f_{Lj}\sigma_{j}\tau_{3})\;.\label{eq:distribution_function}
  \,
\end{equation}
where $f_{T/L,j}$ are real-valued.  Here, we generalize the notation
introduced in Ref.~\cite{schmid1975} to the spin-dependent case. The
different components of $\hat{f}$ carry specific information of the
nonequilibrium state of the electron system. 
In equilibrium at
temperature $T$ the distribution function is $\hat{f}=\hat{1}\tanh(E/2T)$.  We discuss the
physical interpretation of each component  in the following section.

\begin{figure}
  \centering
  \includegraphics[scale=1]{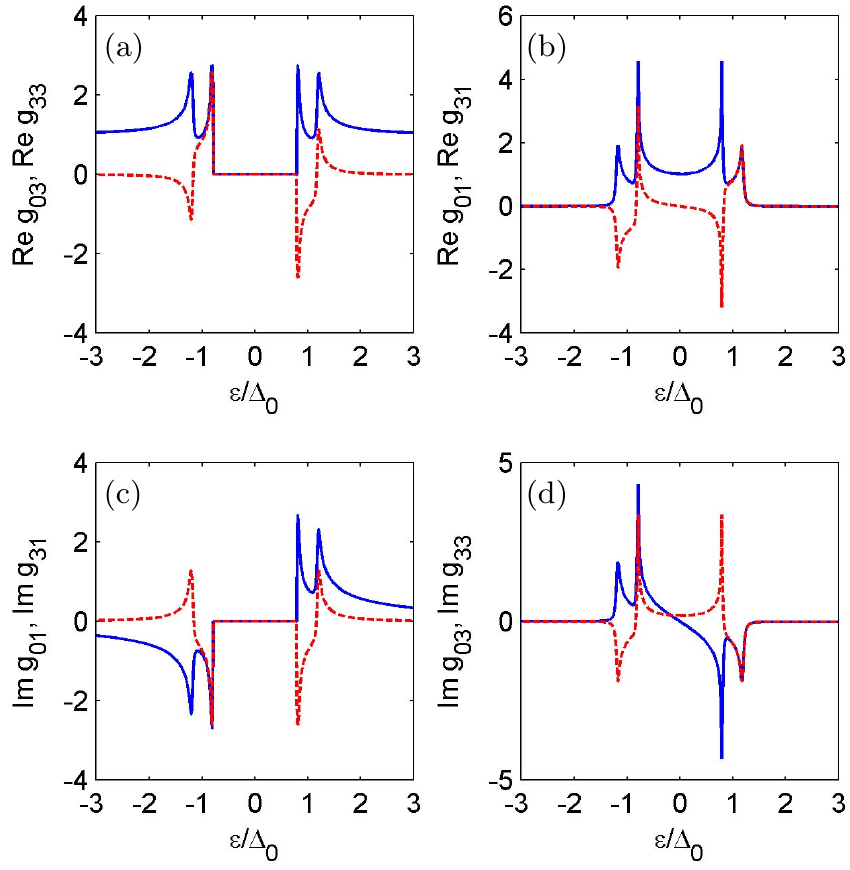}
  \caption{
    \label{fig:gR_Components_Hanle}
    Nonzero components of the retarded Green's function $g^{R}$ in a
    homogeneous spin-split superconductor. In all panels the blue
    solid curves show the spin-singlet amplitudes $g_{01}$, $g_{03}$
    and the red dashed lines show the spin-triplet ones $g_{31}$,
    $g_{33}$.  The curves are calculated for $h=0.2\Delta_0$,
    $\tau^{-1}_{sn} = 0.25 T_{c0}$ and $\beta=-0.9$.
  }
\end{figure}

For a homogeneous bulk superconductor, with no spatial dependence in
the quantities, we can write the general form of R and A functions as
\begin{equation}
\hat g^{R(A)}=\sum_{j=0}^{3}\left(\tau_{1}g_{j1}^{R(A)}+\tau_{3}g_{j3}^{R(A)}\right)\sigma_{j}\;.\label{eq:general_GR}
\end{equation}
The off-diagonal components, i.e., those proportional to $\tau_{1}$,
are characteristic of the superconducting state and describe the
anomalous GFs.  The component of the anomalous GFs proportional to
$\sigma_0$ describes the usual singlet correlations in the BCS theory.
On the other hand, the terms proportional to the Pauli matrices
$\sigma_{j}$, $j=1,2,3$, describe the three triplet components of the
condensate that appear in the presence of spin-dependent fields
\cite{bergeret2001long,Bergeret2005}.  
The density of states, normalized by its normal state value, can also
be written in terms of them:
\begin{equation}
N_+(\epsilon)=\frac{1}{2}\left(g^R_{03}-g^A_{03}\right)=N_\uparrow+N_\downarrow\; ,
\label{eq:dosplus}
\end{equation}
which for a bulk spin-split superconductor reduces to Eq.~\eqref{eq:split_Dos}. 
Accordingly, the difference of the DOS of the spin split bands is determined by
\begin{equation}
N_-(\epsilon)=\frac{1}{2}\left(g^R_{33}-g^A_{33}\right)=N_\uparrow-N_\downarrow\; .
\label{eq:dosminus}
\end{equation}
The latter quantity is shown in Fig.~\ref{fig:Calculated-density-of}. 
The real and imaginary parts of
the different components of the retarded Green's function are plotted
in Fig.~\ref{fig:gR_Components_Hanle} in the case of a homogeneous
magnetization pointing in the $z$ direction. In that case, the GF
components proportional to $\sigma_{1,2}$ vanish.

In what follows, we discuss the nonequilibrium
quasiparticle physics in more detail.

\subsection{Nonequilibrium quasiparticles}\label{sec:noneq-modes}

\begin{figure}
  \centering
  \includegraphics[width=14cm]{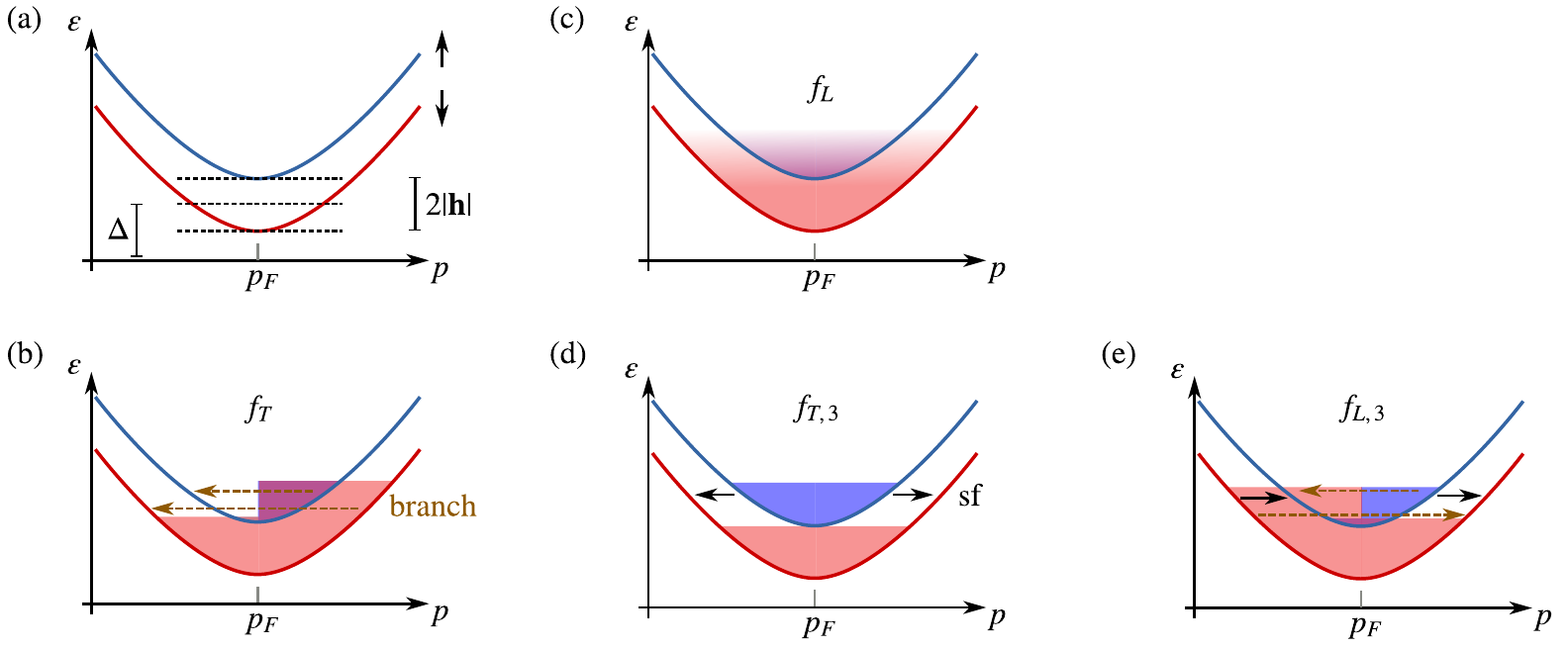}
  \caption{
    \label{fig:modes2}
    Schematic excitation spectrum for a spin-split superconductor, and
    illustration of the different nonequilibrium modes.
    (a)
    Schematic excitation spectrum.
    (b-e)
    Nonequilibrium population modes of the different branches.
    Relevant \emph{elastic} relaxation processes are also shown
    (spin-flip: black arrows, branch-imbalance relaxation: brown dashed arrows).
  }
\end{figure}

We call the different components $f_{T/Lj}(\varepsilon)$, $j=0,1,2,3$
of the distribution function the \emph{nonequilibrium modes}.  They
provide a description of the different ways in which the electron
distribution in the quasiclassical limit can deviate from equilibrium.

The excitation spectrum of a spin-split superconductor has distinct
electron- and hole-like branches, each with spin either up or
down. The exchange field $h$ splits the energies of the spin branches,
as illustrated in Fig.~\ref{fig:modes2}(a).  In a collinear situation
with a single spin quantization axis, the electron population can be
described with the four distribution function components $f_{T}$,
$f_{L}$, $f_{T3}$, and $f_{L3}$.

The mode $f_T\equiv{}f_{T0}$ [Fig.~\ref{fig:modes2}(b)] is well-known in
superconductor physics: it is the \emph{charge imbalance} (or ``branch
imbalance'') mode
\cite{PhysRevLett.28.1363,PhysRevB.6.1747,PhysRevLett.28.1366}, which
corresponds to an imbalance in the number of quasiparticles above and
below the Fermi surface.  This mode can be generated by charge
injection into the superconductor for example from a normal-state electrode,
and measured by observing the local potential $\mu$ of the
quasiparticles.  The mode $f_L$ is associated with the energy content
of the superconductor, and reflects local changes in the effective
temperature. It can generally be induced by any heating mechanism.

The spin imbalance mode $f_{T3}$ [Fig.~\ref{fig:modes2}(d)] is related to a
spin accumulation $\mu_z$, which can be nonzero also in the absence of
spin splitting of the spectrum. It can be induced for example by a
spin-polarized injection from a ferromagnetic electrode, even in the
normal state \cite{PhysRevLett.55.1790}.  Finally, the spin-energy
mode $f_{L3}$ encodes a nonequilibrium state with antisymmetric
differences in the electron-hole and spin-up/down distributions.
We are not aware of existing experiments probing $f_{L3}$
nonequilibrium.

The spin splitting of the superconductor spectrum also modifies how
the modes contribute to observables, which generally depend on
  the amount of quasiparticles on the different spectral branches.
The charge imbalance potential $\mu$ then acquires
contributions from both electron-hole antisymmetric modes
$f_T$ and $f_{L3}$. The other two modes, $f_L$ and $f_{T3}$, contribute
to the spin accumulation $\mu_z$.

To excite the modes $f_T$, $f_{T3}$ and $f_{L3}$, one only needs to
transfer quasiparticles between the different spectral branches in an
elastic process, i.e., between equal-energy states (horizontally
in Fig.~\ref{fig:modes2}). They can also relax back
to equilibrium due to elastic scattering processes (horizontal
  arrows in Fig.~\ref{fig:modes2}).  These relaxation
mechanisms depend on properties of the material, and also on the
superconducting spectrum, and are discussed in subsequent sections.

In contrast, relaxation of the $f_L$ mode generally requires inelastic
processes, which often are slow compared to elastic processes at low
temperatures. As a consequence, perturbations in $f_L$ can survive up
to longer times and propagate longer distances than those in the other
modes.  Because the $f_L$ mode also contributes to the spin accumulation,
this is crucial in understanding long-range spin signals observed in
spin-split superconductors, for example in Ref.~\cite{Hubler2012a},
which is discussed in Sec.~\ref{spininjection}.

\subsubsection{Kinetic equations}
\label{sec:Kin_eqs}
The different parts of the matrix distribution function $\hat f$ are
determined by a kinetic equation that describes the balance of
different transport and relaxation processes. In a diffusive conductor
in the steady-state limit, the kinetic equation can be obtained by
inserting Eq.~\eqref{eq:K_param} in the Usadel equation
\eqref{eq:Usadel} and evaluating its different matrix components. This
results into a set of diffusion-type equations of the generic tensor form
\begin{align}
  \nabla_k j_{kb}^a
  &=
  H^{ab}+R^{ab}+I_{\rm coll}^{ab}
  \,,
  \label{eq:kineticeq}
  \\
  j_{kb}^{a}
  &=
  \frac{1}{8}{\rm Tr}\tau_{b}\sigma_{a}(\check{g}\nabla_{k}\check{g})^{K}
  \,.
  \label{eq:general_current}
\end{align}
Here, $I_{\rm coll}^{ab}={\rm Tr} \tau_b \sigma_a [\check
  \Sigma,\check g]^K/8$ is the collision integral describing
scattering processes with self-energy $\check \Sigma$. The term
$H^{ab}={\rm Tr} \tau_b \sigma_a [-i{\mathbf h} \cdot {\mathbf \sigma}
  \tau_3,\hat g^K]/8$ describes the Hanle spin precession due to the
exchange field, and $R^{ab}={\rm Tr} \tau_b \sigma_a [\hat\Delta,\hat
  g^K]/8$ the quasiparticle branch conversion processes involving the
superconducting condensate. The detailed forms of the different
components of the kinetic equations in spin-split superconductors are
discussed in Sec.~\ref{sec:nl_detection}. Note that the definition of
the collision integrals here depends on the form of the diffusion
equation adopted; another definition is discussed in
Sec.~\ref{subs:linrespac}.

The above equations are formulated for a stationary situation where
$\Delta$ is time-independent, and here and below, we quite generally
only consider this case. The equations are written in a gauge where
the zero of the electric potential is taken to be the potential of the
superconducting condensate.  We also consider only length scales where
the metal can be assumed locally charge neutral.  The formulation
still allows describing normal-state electrodes at nonzero electrochemical
potential, by including distribution functions of the form
\cite{belzig1999quasiclassical}
\begin{align}
  f_{\mathrm{eq},L(T)}(\varepsilon)=\frac{1}{2}\left[\tanh\left(\frac{\varepsilon+eV}{2T}\right) +(-)
  \tanh\left(\frac{\varepsilon-eV}{2T}\right)\right] \,,
\end{align}
as boundary conditions. Superconducting electrodes however have $V=0$
as the quasiparticle and condensate potentials coincide at
equilibrium.

\subsubsection{Observables}

Local densities of charge, and excess spin and energy can be obtained from the Keldysh Green function.
These equations in the quasiclassical framework read for the charge imbalance and spin accumulation
\begin{eqnarray}
  \mu({\bf r},t) & = & -\int_{-\infty}^{\infty} \frac{d\varepsilon}{16}  \Tr \hat g^{K}(\varepsilon,{\bf r},t)\label{eq:def_mu}\\
  \mu_{sa}({\bf r},t) & = & \int_{-\infty}^{\infty} \frac{d\varepsilon}{16} \Tr \tau_{3}\sigma_{a}[ \hat g_{\rm eq}^{K}(\epsilon,\mathbf{r},t)- \hat  g^{K}(\varepsilon,\mathbf{r},t)]\; , \label{eq:def_mus}
\end{eqnarray}
and for the local energy and spin-energy accumulations:
\begin{eqnarray} \label{Eq:HeatDens}
  q({\bf r},t) & = & \int_{-\infty}^{\infty} \frac{d\varepsilon}{16}  \varepsilon \Tr \tau_3 [\hat  g_{\rm eq}^{K}(\varepsilon,{\bf r},t)- \hat  g^{K}(\varepsilon,\mathbf{r},t)]\label{eq:def_q}
  \\  \label{Eq:SpinEnergyDens}
  q_{sa}({\bf r},t) & = &
  \int_{-\infty}^{\infty} \frac{d\varepsilon}{16} \varepsilon \Tr\sigma_{a}[ \hat  g_{\rm eq}^{K}(\varepsilon,\mathbf{r},t)- \hat  g^{K}(\varepsilon,\mathbf{r},t)]. \label{eq:def_qs}\;
\end{eqnarray}
Here, polarization direction of the spin is encoded in the direction
components $a=1,2,3$. The energy content is written relative to a
system at equilibrium at the chemical potential $\mu_S=0$ of the
superconducting condensate.

The charge imbalance is related to the local charge density by
$\rho=-\nu_Fe^2\phi-e\nu_F\mu$, where $\phi$ is the electrostatic potential.
In the locally charge-neutral limit considered here, $\rho\approx0$ so that
$-e\phi\approx\mu$ and $\mu$ corresponds directly to the local electric potential
\cite{artemenko1979-efc,kopnin2001-ton}.

By using Eqs.~(\ref{eq:K_param}-\ref{eq:dosminus}) one can express  the charge 
and spin accumulations  in terms of the
distribution functions and the density of states,
\begin{align}
  \mu &= -\frac{1}{2}\int_{-\infty}^{\infty} d\varepsilon ( N_+ f_T+ N_-  f_{L3})
  \,,
  \label{Eq:ChPot0}
  \\
  \mu_z &= - \frac{1}{2}\int_{-\infty}^{\infty} d\varepsilon [ N_+ f_{T3}+ N_-(f_{L}-f_{\rm eq})]
  \,.
  \label{Eq:ChPotZ}
\end{align}
Similarly for the local energy and spin-energy content:
\begin{align}
  q &=
  \frac{1}{2}\int_{-\infty}^{\infty} d\varepsilon \varepsilon [ N_+ (f_L-f_{eq}) + N_-f_{T3}]
  \,,
  \label{Eq:HeatDensQuasi}
  \\
  q_{sa} &=  \frac{1}{2}\int_{-\infty}^{\infty} d\varepsilon \varepsilon [ N_- (f_{L}-f_{\rm eq}) + N_+f_{T3}]
  \,.
  \label{Eq:SpinEnergyDensQuasi}
\end{align}
The above equations apply to the collinear situation with a single spin quantization axis.

The observable current densities of charge, energy, spin, and spin-energy can
be obtained from the spectral current~\eqref{eq:general_current}.  Explicitly,
they read, (i) the charge current:
\begin{equation} 
  J_{k} = \frac{\sigma_{N}}{2e}\int_{-\infty}^\infty d\varepsilon\,j_{k3}^0,
  \label{eq:charge_current}
\end{equation}
(ii) the spin current density polarized in $a$-direction
\begin{equation}
J_{k}^{a}=\frac{\sigma_{N}}{2 e^2}\int_{-\infty}^\infty d\varepsilon \,j_{k0}^a,\label{eq:spin_current}
\end{equation}
(iii) the energy current density
\begin{equation}
J_{e,k}=\frac{\sigma_{N}}{2 e^2}\int_{-\infty}^\infty d\varepsilon \varepsilon \,j_{k0}^0,\label{eq:energy_current}
\end{equation}
and (iv) the spin energy current density
\begin{equation}
J_{e,k}^{a}=\frac{\sigma_{N}}{2 e^2}\int_{-\infty}^\infty d\varepsilon \varepsilon \,j_{k3}^a,\;.\label{eq:spinenergy_current}
\end{equation}
Here,  $\sigma_N=e^2 D \nu_F$ is the normal-state conductivity and
$\nu_F$ is the normal-state density of states at the Fermi level.

\subsection{Elastic relaxation mechanisms}
\label{sec:elastic-relaxation}

In addition to the impurity scattering that results in the momentum relaxation,
there are also other elastic scattering processes that, while
preserving the energy of the excitations, change their other quantum
numbers such as the spin. For example, in metals
and semiconductors there are two main types of
mechanisms that relax the spin: scattering from magnetic impurities
and spin-orbit coupling.  As discussed in Sec.~\ref{sec-superwithh},
these processes can modify substantially the spectral properties of a
superconductor.  In addition, the superconducting properties are
modified by the orbital effect due to external magnetic fields, which
for the quasiparticles can result to an electron-hole branch-mixing process.

Within Born approximation, the self-energies for these elastic
processes obtain the forms
\begin{align}
  \check{\Sigma}_{so}&=
  \frac{\vec{\sigma}\cdot\check{g}\vec{\sigma}}{8\tau_{so}}
  \,,\label{eq:SE-SO}
  \\
  \check{\Sigma}_{sf}&=
  \frac{\vec{\sigma}\cdot\tau_{3}\check{g}\tau_{3}\vec{\sigma}}  {8\tau_{sf}} 
  \,,
  \label{eq:SE-SF}
  \\
  \check{\Sigma}_{orb}&=
  \frac{\tau_{3}\check{g}\tau_{3}}{\tau_{orb}} .
  \label{eq:SE-ORB}
\end{align}
Here, $\tau_{so}$ and $\tau_{sf}$ are the two scattering times for
impurity spin-orbit and spin-flip scattering.  We  
parametrize these scattering rates in terms of the total spin
relaxation time,
$\tau_{\rm{}sn}^{-1}=\tau_{\rm so}^{-1}+\tau_{\rm sf}^{-1}$ and the
relative strength of the two scattering mechanisms
$\beta=(\tau_{\rm so}-\tau_{\rm sf})/(\tau_{\rm so}+\tau_{\rm sf})$.
Values with $\beta\approx1$ correspond to dominant spin-flip
scattering, whereas $\beta\approx-1$ to dominant spin-orbit
scattering.

The orbital self-energy above can be used to describe the orbital
depairing effect of (Meissner) screening currents induced by
an in-plane magnetic field in a thin film,
leading to the suppression of superconductivity \cite{deGennes:566105}.
It does not enter explicitly the kinetic
equation for the distribution function, but modifies indirectly its
energy dependence by affecting the spectral properties of the
superconductor, contributing to the relaxation of
charge imbalance \cite{schmid1975,nielsen1982pair}. In the case of a thin magnetic film with
thickness $d$ and an in-plane applied magnetic field, the orbital
depairing relaxation time is given by \cite{PhysRevB.54.9443}
$1/\tau_{orb}=De^{2}B^{2}d^{2}/6$.

In the case of spin-orbit coupling we can distinguish two types of
relaxation mechanisms according to the origin, intrinsic or extrinsic.
Intrinsic spin-orbit coupling occurs in systems without
inversion symmetry, either due to geometry constraints or the
crystal potential associated with the electronic band structure.
Momentum-dependent spin precession together with random momentum
relaxation due to impurities leads to the Dyakonov-Perel relaxation
\cite{dyakonov1972spin}. For example, for a diffusive 2D system with a
Rashba spin-orbit coupling $\alpha_R$, the Dyakonov-Perel relaxation
time is $\tau_\parallel=1/D(2m\alpha_R)^2$ for an in-plane
spin-polarization, and $\tau_\perp=\tau_\parallel/2$ for spins
perpendicular to the 2D system.
 
In contrast, the extrinsic spin-orbit effect originates from a random
impurity potential and hence leads to an isotropic spin relaxation
known as the Elliott-Yafet relaxation mechanism.  It is this component
that is described by the self-energy Eq.~\eqref{eq:SE-SO}.  A detailed
discussion of spin relaxation mechanisms originating from the
spin-orbit coupling can be found in the review~\cite{vzutic2004spintronics} and references therein.  Because we
here focus on centrosymmetric materials, we only consider the
extrinsic relaxation mechanism.  Moreover, we do not consider effects
related to the spin-charge coupling, such as the spin Hall effect
\cite{sinova2015spin} and the spin galvanic effects
\cite{edelstein1990spin}. Such effects enter the kinetic equations in
a higher order of the gradient expansion and can be neglected in the
leading order. The spin-charge coupling in diffusive  superconductors leads to
non-dissipative magnetoelectric effects and the appearance of an
anomalous phase in Josephson junctions, 
\cite{konschelle2015theory,bergeret2016manifestation,huang2018extrinsic}.

The parameters $\tau_{sf}$, $\beta$ of the spin-dependent scattering
are material specific. In Refs.~\cite{jedema:713,poli2008spin},
using spin injection experiments, the values $\tau_{\rm sn}
\approx 100$ ps and $\beta \approx 0.5$ were found for Al
for which then $\tau_{sf}<\tau_{so}$.
According to
Refs.~\cite{Abrikosov1962,meservey1978spinorbit}, the spin-orbit
scattering rate is the momentum relaxation rate times $(Z\alpha)^4$,
where $Z$ is the atomic number and $\alpha$ is the fine-structure
constant. Therefore, the spin-orbit scattering rate grows rapidly as
the atomic number grows. Indeed, in Nb
$\tau_{sn}$ is only 0.2 ps and
spin-orbit scattering dominates \cite{wakamura2014spin}.

\subsubsection{Nonequilibrium spin relaxation}

In this review we only consider centro-symmetric metals,  and therefore we only consider spin-relaxation caused by  impurities.
In this case, the collision integrals
entering the kinetic equation due to spin-dependent scattering are
given by
\begin{equation} \label{Eq:SpinCIgeneral}
  S_{ab} =
  \frac{1}{8}{\rm Tr}\left\{\tau_b \sigma_a \left[
      \check  \Sigma_{so} + \check  \Sigma_{sf} ,\check g\right]^K
  \right\}
  \,,
\end{equation}
where the components $a,b$ refer to the different nonequilibrium modes
and spin projections. The self-energies are defined in
Eqs.~\eqref{eq:SE-SO} and \eqref{eq:SE-SF}.  In the superconducting state,
the form of the spin-flip and spin-orbit contributions to the
collision integrals differ due to their different properties under
time reversal.

We now discuss  the relaxation of the nonequilibrium modes related
with the spin and the spin energy imbalances, i.e., the components
$f_{T3}$ [Fig.~\ref{fig:modes2}(d)] and $f_{L3}$
[Fig.~\ref{fig:modes2}(e)].  The relaxation of these
modes is described by the components ($b=0,a=3$) and ($b=3,a=3$) of
Eq.~\eqref{Eq:SpinCIgeneral}, respectively.

\begin{figure}
  \centering
  \includegraphics[width=10cm]{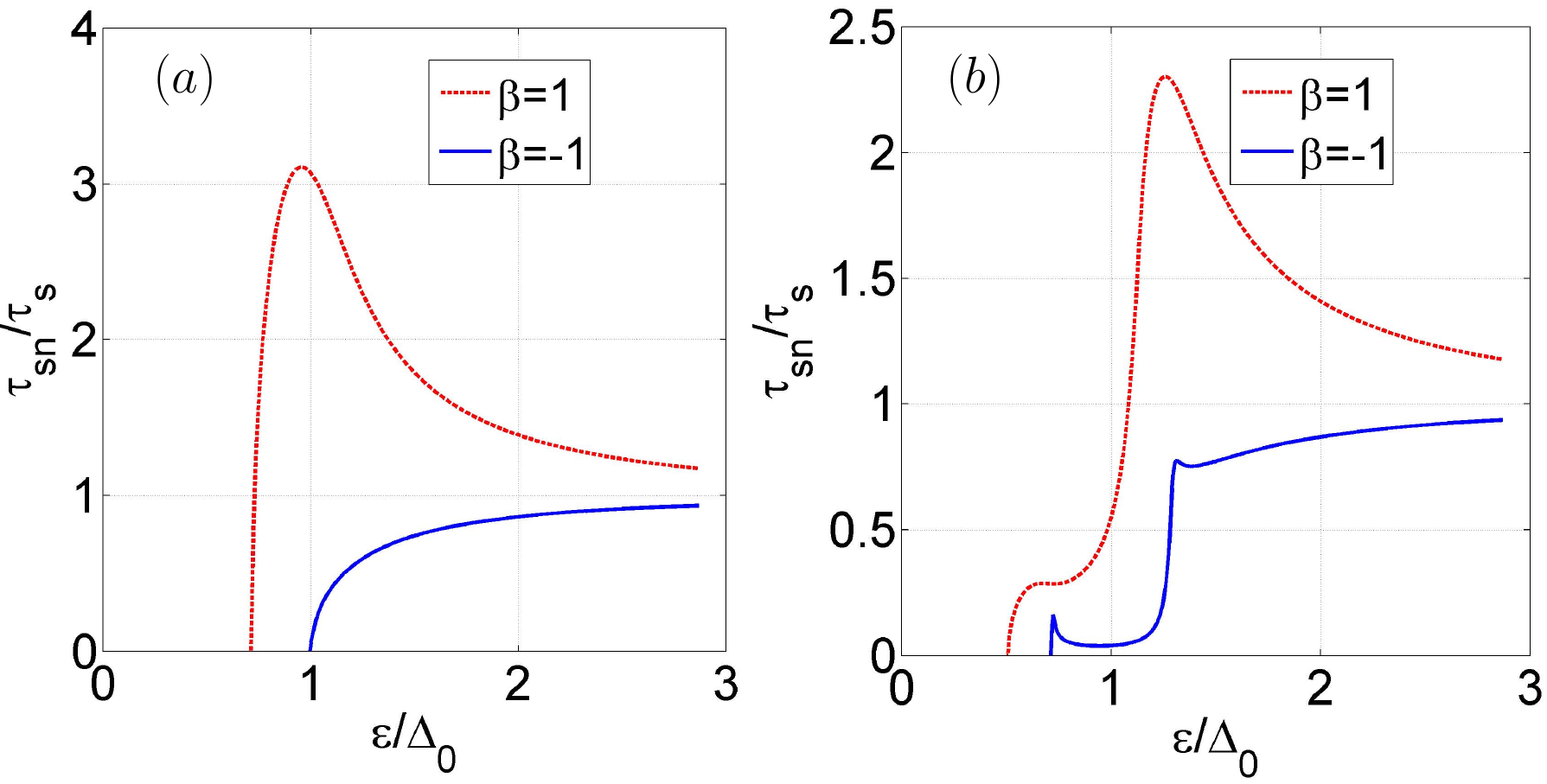}
  \caption{\label{Fig:SpinRelRate} Spin relaxation rates for (a) $h=0$
    and (b) $h=0.3\Delta_0$ and for spin-orbit ($\beta =- 1$) and
    spin-flip ($\beta = 1$) relaxation mechanisms.  We assume that the
    normal-state spin relaxation rate is
    $\tau_{sn}^{-1} =0.2T_{c0}$.
  }
\end{figure}

Specifically, the spin relaxation, shown schematically by the arrows
in Fig.~\ref{fig:modes2}(d), is described by the
collision integral $S_{03}^s = S_{T3} f_{T3}$, where
\begin{equation}
  \tau_{sn} S_{T3} =
  ({\rm Re}g_{03})^2  - ({\rm Re}g_{33})^2
  + \beta \left[ ({\rm Im}g_{01})^2 - ({\rm Im}g_{31})^2 \right] .
  \label{eq:ST3}
\end{equation}
In the normal state these processes are energy independent (within the
quasiclassical approximation), and have the rate $\tau_{sn}^{-1}$. In
the superconducting state, the spin-relaxation is described by
Eq.~\eqref{eq:ST3}, with the energy-dependent rate $\tau^{-1}_s =
S_{T3}(\varepsilon)/N_+$. Here the collision integral is normalized
with the spin-averaged density of states $N_+$ to counterbalance
effects due to the mere changes of the quasiparticle spectrum in the
superconducting state. The typical behavior of $\tau^{-1}_s
(\varepsilon)$ is shown in Fig.~\ref{Fig:SpinRelRate} for dominating
spin-flip and spin-orbit mechanisms for (a) zero and (b) nonzero spin
splitting.

In the presence of strong spin-orbit scattering ($\beta<0$), superconductivity
results to reduction of the spin relaxation rate $\tau_{s}^{-1}$.
Still, there is no significant change in the spin relaxation length
$\lambda_{so}$: \cite{morten2004spin} Qualitatively
$\lambda_{so}\sim{}v_g \tau_{s}$, and the reduction in the
quasiparticle momentum scattering cross section is mostly canceled by the
decrease in the group velocity
$v_g\sim{}v_F\sqrt{1-|\Delta|^2/\varepsilon^2}$.  In contrast, the
increase in $\tau_s^{-1}$ due to spin-flip scattering does result to a
reduced spin relaxation length: \cite{morten2004spin} this mechanism
is not related to the momentum scattering as the interaction with
magnetic impurities does not depend on the propagation direction and
the quasiparticle spin does not depend on energy.

In the presence of a Zeeman or exchange field, $h\neq 0$, this
situation changes drastically.  In this case the nonequilibrium spin
accumulation couples to the energy mode, $f_L$. This mode is robust
with respect to elastic spin relaxation processes, resulting to a
qualitative change in the observed spin relaxation. This effect, as
well as the detailed energy dependence of the spin relaxation lengths,
are discussed in detail in Sec.~\ref{sec:noneq}.

For the relaxation of the spin-energy mode $f_{L3}$, the situation is
more complicated. As sketched in
Fig.~\ref{fig:modes2}(e), the relaxation of this mode
involves transitions between electron and hole branches of the
quasiparticle spectrum.  As in the case of charge imbalance
relaxation, such inter-branch transitions involve the formation of
Cooper pairs via Andreev reflection.  This results in two
contributions to the relaxation of the spin-energy mode: One stems
from the ($j=3,k=3$) component of the collision integral
(\ref{Eq:SpinCIgeneral}) that describes transitions between spin-up
and spin-down subbands $S_{33}^s = S_{L3} f_{L3}$, where
\begin{align}
  \nonumber
  & \tau_{sn} S_{L3} = \\
  & 
  ({\rm Re}g_{03})^2 -({\rm Re}g_{33})^2 +
  \beta \left[ ({\rm Re}g_{31})^2 -
    ({\rm Re}g_{01})^2\right] .
  \label{eq:SL3}
\end{align}
The second contribution originates from the particle-hole relaxation
described by the non-diagonal self-energy in the Keldysh-Usadel
equation~(\ref{eq:Usadel}),
\begin{align}
R_T f_{L3}  = \frac{1}{8} {\rm Tr} (\tau_3\sigma_3 
[\check{\Delta},\check{g}^K] ) \\ \label{Eq:RT}
R_T = 2\Delta {\rm Re} g_{01}\; .
\end{align}
The relaxation of the spin-energy mode in the superconductor is given
by the sum $R_T+S_{L3}$.

Note that the particle-hole conversion processes responsible for the
spin energy relaxation also lead to the generation of charge imbalance
producing a coupling between spin and charge degrees of freedom. These
mechanisms are discussed in more detail in Sec.~\ref{sec:noneq}. At
large energies $\varepsilon \gg \Delta$ the charge and spin energy
nonequilibrium modes are decoupled. In this limit the charge
relaxation length tends to infinity because the coherence factor ${\rm
  Re} g_{01} \to 0$. At the same time, the spin-energy relaxation
length approaches its normal-state value equal to that of the spin
relaxation length $\Lambda_{sn} = \sqrt{D\tau_{sn}}$.

\subsection{Inelastic relaxation mechanisms}

The inelastic processes that are typically relevant in metallic systems,
the particle--phonon and particle--particle
collisions, are described by an inelastic self-energy
$\check\Sigma_{in} = \check\Sigma_{\rm eph}+ \check\Sigma_{\rm ee}$.  These
processes do not conserve the energies of colliding quasiparticles,
but conserve the total spin.  A model self-energy for inelastic
relaxation due to electron-phonon scattering is given by
\cite{eliashberg1972-iec}
\begin{align}\label{Eq:ElectronPhononSE}
 &  \check  \Sigma^{R/A/K}_{\rm eph} (\varepsilon) =  
  -i g_{\rm eph}  \int_{-\infty}^{\infty} d\omega \tilde{\Sigma}_{\rm eph}^{R/A/K}(\omega,\varepsilon+  \omega)
  \\
  \notag
 & \tilde{\Sigma}^{R(A)}_{\rm eph} =  D^K(\omega)\hat g^{R(A)} (\varepsilon+\omega) -D^{R(A)}(\omega)\hat g^K (\varepsilon+\omega) 
 \\
  \notag
 & \tilde{\Sigma}^K_{\rm eph} =
 D^K(\omega)\hat g^{K} (\varepsilon+\omega) -
 D^{RA}(\omega) \hat g^{RA} (\varepsilon+\omega), 
\end{align}
where $D^{R,A}=\pm i\omega |\omega|$, $D^{RA}=D^R-D^A$, $D^K(\omega) = D^{RA}(\omega)\coth(\omega/2T_{ph})$ are parts of Fermi surface
averages of the free phonon propagators, $T_{ph}$ is the phonon
temperature and we denote $X^{RA}= X^R - X^A $ for $X=D,\hat g$.
Inelastic particle-particle collision self-energies $\check \Sigma_{\rm ee}$ within the quasiclassical theory
have been discussed in Refs.~\cite{eliashberg1972-iec,serene1983-qat}.

\subsubsection{Electron-phonon coupling}

The relaxation due to the electron-phonon coupling is an important
limiting mechanism for some of the nonequilibrium effects discussed in
subsequent sections. Generally, it becomes weaker towards lower
temperatures as the phonon density of states decreases at low
energies.  As a consequence, the temperatures $T_{\rm qp}$ and $T_{\rm
  ph}$ can decouple and the two subsystems can be in different
temperatures, and eventually e-ph interaction is dominated by other
relaxation processes.  As the interaction is also sensitive to the
electronic spectrum, it is modified by superconductivity.
\cite{eliashberg1972-iec,kopnin2001-ton,kaplan1976} Below we
discuss what this results to in the spin-split case.

The collision integral for electron-phonon processes in a spin-split
superconductor was discussed in Ref.~\cite{grimaldi1997} and can be
obtained from the self-energies described in
Eq.~\eqref{Eq:ElectronPhononSE} \cite{virtanen2016stimulated}. The
overall collision integral entering the kinetic equations [Keldysh
  part of Eq.~\eqref{eq:Usadel}] is obtained from a commutator between
the self-energy and Green's function, $\check I_{\rm eph}=[\check
  \Sigma_{\rm eph},\check g]^K$.  This is still a Nambu-spin matrix,
its different matrix components describe the relaxation of the
different nonequilibrium modes. Electron-phonon interaction causes
inelastic scattering of quasiparticles, which means that energy flow
between quasiparticles and phonons becomes possible. That process is
described by the collision integral $I_L \equiv {\rm Tr}[\hat I_{\rm
    eph}]/8$. Also the other matrix components of the collision
integral are in general non-vanishing; for example the component ${\rm
  Tr}[\tau_3 \hat I_{\rm eph}]$ affecting $f_T$ is detailed in \cite{heikkila09}. However, since
electron-phonon interaction conserves charge and spin, the energy
integrals of ${\rm Tr}[\tau_3 \hat I_{\rm eph}]$ and ${\rm Tr}
[\sigma_3 \hat I_{\rm eph}]$ vanish (see \ref{app:ephonon}). The
remaining component ${\rm Tr}[\tau_3 \sigma_3 \hat I_{\rm eph}]$
describes the relaxation of the difference in the thermal energies of
electrons with opposite spins due to electron-phonon interaction
\cite{heikkila2010}. In the following, we only concentrate on energy
relaxation. Substituting the self-energy \eqref{Eq:ElectronPhononSE}
in the equation for the collision integral and assuming that $\check f=f_L \check 1$ we can write
\begin{equation}
\label{eq:ephcolli}
\begin{split}
I_L=&\frac{g_{eph}}{4}\int_{-\infty}^\infty d\omega\omega |\omega| {\rm Tr}[\hat g^{RA}(\varepsilon) \hat g^{RA}(\varepsilon+\omega)]\\&\times \bigg\{f_L(\varepsilon)f_L(\varepsilon+\omega)-1\\&-\coth\left(\frac{\omega}{2k_B T_{\rm ph}}\right)\left[f_L(\varepsilon)-f_L(\varepsilon + \omega)\right]\bigg\}.
\end{split}
\end{equation}
Typically this collision integral is used to calculate the total heat
current $\dot Q_{\rm eph}=2\nu_F \Omega \int d\varepsilon \varepsilon
I_L(\varepsilon)$ between the electron and phonon systems in a volume
$\Omega$. Assuming that the electron system resides in temperature $T_{\rm
  qp}$, we have $f_L=\tanh[\varepsilon/(2 k_B T_{\rm qp})]$ and thus
the heat current is \cite{timofeev09,maisi2013},
\begin{equation}
\begin{split}
&\dot Q_{\rm eph} = \frac{\Sigma \Omega}{96 \zeta(5) k_B^5}
\int_{-\infty}^\infty d\varepsilon \varepsilon \int_{-\infty}^\infty d\omega \omega |\omega|
 L_{\varepsilon,\varepsilon+\omega}\times \\&\bigg\{\coth\left(\frac{\omega}{2 k_B T_{\rm
        ph}}\right) \times \left[\tanh\left(\frac{\varepsilon+\omega}{2 k_B T_{\rm
        qp}}\right)-\tanh\left(\frac{\varepsilon}{2 k_B
      T_{\rm qp}}\right)\right] \\&+ \tanh\left(\frac{\varepsilon}{2 k_B T_{\rm qp}}\right)
\tanh\left(\frac{\varepsilon+\omega}{2 k_B T_{\rm qp}}\right)-1\bigg\}.
\end{split}
\label{eq:ephcoupling}
\end{equation}
Here $\Sigma=3072 \zeta(5) \nu_F k_B^5 g_{eph}$ describes the
materials dependent magnitude of the electron-phonon coupling (see
  tabulated values in \cite{giazotto2006}) typically used for heat
flow, and $\zeta(x)$ is the Riemann zeta function. The kernel
$L_{E,E'}$ depends on the spin-splitting field, as it is
\begin{equation}
\begin{split}
L_{\varepsilon,\varepsilon'}&=\frac{1}{2}\sum_{\sigma=\pm}
N_\sigma(\varepsilon)N_\sigma(\varepsilon')-F_\sigma(\varepsilon)F_\sigma(\varepsilon')\\
&=\frac{1}{2}\sum_{\sigma=\pm}  N_\sigma(\varepsilon)N_\sigma(\varepsilon')\left[1-\frac{\Delta^2(T_{\rm
      qp})}{(\varepsilon+\sigma h)(\varepsilon'+\sigma h)}\right],
\end{split}
\end{equation}
where $N_\sigma(\varepsilon)={\rm Re}(g_{03}+\sigma g_{33})/2$ is the
density of states for electrons with spin $\sigma$ and $F_\sigma={\rm
  Re}(g_{01}+\sigma g_{31})/2$ is the anomalous function for spin
$\sigma$ (see Eq.~\eqref{eq:general_GR} and below it). The latter formula applies in
the absence of spin-flip or spin-orbit scattering.

Strictly speaking, Eq.~\eqref{eq:ephcoupling} requires two conditions
to be met: (i) A well-defined local quasiparticle temperature that may
deviate from the phonon temperature. In other words, the
quasiparticles should first equilibrate between themselves. (ii) In
the presence of a spin-splitting field, also spin imbalance affects
the quasiparticle-phonon heat flow
\cite{grimaldi1997,virtanen2016stimulated}. Neglecting it therefore requires that
the spin relaxation rate exceeds the injection rate causing the
out-of-equilibrium situation. \footnote{Note that strictly speaking
  one should then include this spin relaxation into the kernel
  $L_{E,E'}$ as spin relaxation affects the spectral functions, see
  Fig.~\ref{fig:Calculated-density-of}. The relevant scales for the
  two effects differ, however, as the kernel is significantly modified only when
  $\hbar/\tau_{sn}$ is not very much weaker than $\Delta$.  This is
  further quantified in Sec.~\ref{subs:spinseebeck}.}

\begin{figure}
  \centering
  \includegraphics[width=8cm]{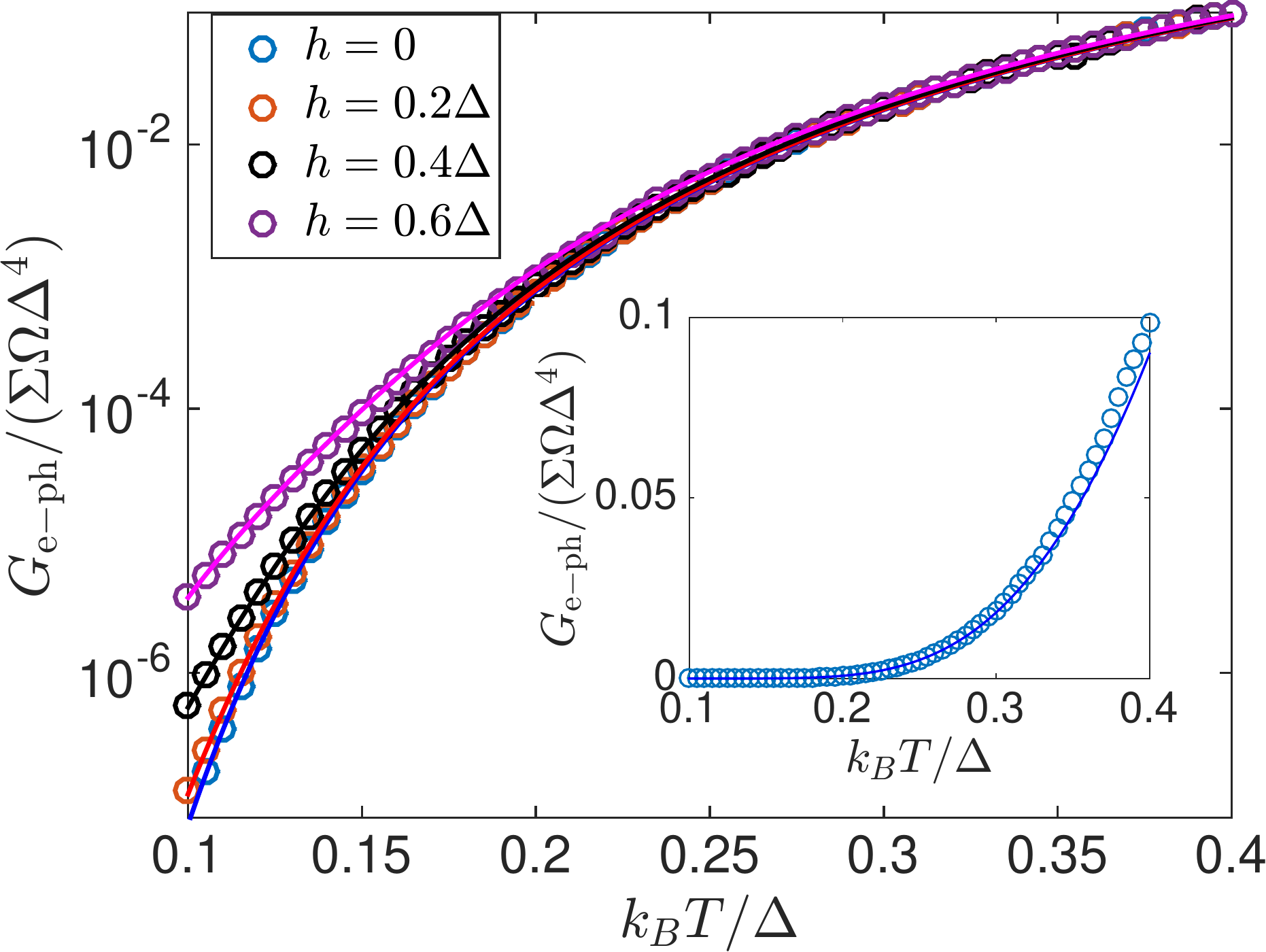}
  \caption{
    \label{fig:ephheatcond}
    Electron-phonon heat conductance in a spin-split
    superconductor. Circles show the numerical evaluation based on the
    linearization of Eq.~\eqref{eq:ephcoupling}, whereas the lines are
    from Eq.~\eqref{eq:sephconductance}. Note that the vertical axis
    in the main figure is in a log scale, whereas the inset shows a
    linear scale for $h=0$. On the linear scale the effects from
    non-vanishing $h$ are not visible.
  }
\end{figure}

In the general case Eq.~\eqref{eq:ephcoupling} needs to be evaluated
numerically. However, we may consider two limiting cases. In a normal
metal $\Delta=0$ implies $L_{\varepsilon,\varepsilon'}=1$. In that
case the heat current between the two systems becomes
\cite{wellstood1994}
\begin{equation}
\label{eq:normaleph}
\dot Q_{\rm eph} = \Sigma \Omega (T_{\rm qp}^5-T_{\rm ph}^5).
\end{equation}
Another tractable limit is that of linear response, where $T_{\rm qp}
= T_{\rm ph} + \Delta T$, $\Delta T \ll T_{\rm ph} \equiv T$. If
moreover $k_B T \ll \Delta-h$, $\dot Q_{\rm eph} = G_{\rm ph}\Delta T$
with \cite{heikkila2018}
\begin{equation}
\label{eq:sephconductance}
G_{\rm ph} \approx \frac{\Sigma \Omega T^4}{96\zeta(5)}  \left[f_1\left(\frac{1}{\tilde \Delta}\right) \cosh \tilde h e^{-\tilde \Delta} + \pi \tilde \Delta^5 f_2\left(\frac{1}{\tilde \Delta}\right) e^{-2\tilde \Delta}\right],
\end{equation}
where $\tilde \Delta=\Delta/k_B T$, $\tilde h=h/k_B T$, and $f_1(x)=
\sum_{n=0}^3 C_n x^n$, $f_2(x)=\sum_{n=0}^2 B_n x^n$, $C_0 \approx
440$, $C_1 \approx 500$, $C_2 \approx 1400$, $C_3 \approx 4700$,
$B_0=64$, $B_1=144$, $B_2=258$. The two terms in
Eq.~\eqref{eq:sephconductance} describe scattering and recombination
processes, respectively.  Equation \eqref{eq:sephconductance} is
compared to the exact result in Fig.~\ref{fig:ephheatcond}. The
recombination process, whose heat conductance is almost independent of
the spin splitting (as long as $h\ll \Delta$), dominates at high
temperatures, whereas the two become of the same order of magnitude
for $T \approx 0.1 \Delta$.  Note that the exponential suppression of
the electron-phonon heat conductance $\sim e^{-\Delta/T}$ at low
temperatures does not directly make it irrelevant in a superconductor,
because any effect related to quasiparticles contains such exponential
terms due to the gap in the superconducting density of
states. However, as lowering the temperature reduces the phase space
available for acoustic phonons, also the prefactor of the exponential
becomes small low temperatures. Thus, at very low temperatures
quasiparticles decouple from phonons, and other heat conduction
mechanisms become relevant.

\subsubsection{Particle-particle collisions}

The particle-particle collision integrals generally reflect symmetries
and conservation laws of the electron system. Below, we only point out
conservation laws that a consistent formulation of the collision
integrals must imply in a system involving collinear
magnetizations. The conservation of charge, spin, and overall energy
by spin-rotation symmetric scattering within the particle system can
be expressed as:
\begin{subequations}
\begin{align}
\int d\varepsilon \Tr[\tau_3 \sigma_0 \hat I_{\rm ee}]&=0\\
\int d\varepsilon \Tr[\tau_0 \sigma_3 \hat I_{\rm ee}]&=0\\
\int d\varepsilon\,\varepsilon \Tr[\tau_0 \sigma_0 \hat I_{\rm ee}]&=0.
\end{align}
\end{subequations}
The spin energy, involving the component $\varepsilon \tau_3 \sigma_3$
of the collision integral, is generally not conserved in
electron-electron collisions, because electrons with opposite spins
can exchange energy \cite{heikkila2010}.
The specific forms for $I_{\rm ee}$ for superfluids have been discussed in
\cite{eliashberg1972-iec,kopnin2001-ton,serene1983-qat}, however, typically
in the absence of spin splitting of the spectrum.

The above conservation laws are relevant in terms of studying what
happens in the limit of strong electron-electron scattering. In that
case, the usually assumed model is that the quasiparticle distribution
functions tend to equilibrium forms, but retaining some effective
temperature and a spin-dependent chemical potential. This {\it
  quasiequilibrium} limit (without spin accumulation) is often used in
analyzing heat transport in superconductors driven out of equilibrium
(for example, see \cite{muhonen12}).

\subsection{Description of hybrid interfaces}
\label{sec:hybrid_interfaces}

The spectral current density~(\ref{eq:general_current}) appearing in the kinetic 
equation is defined in the bulk of a superconductor (S) or a normal metal (N). In addition,
we need a description of interfaces between different materials in order
to model transport in hybrid
structures. At such interfaces, potentials and materials parameters
vary over atomic distances, and therefore they cannot directly be
described by quasiclassical equations that are valid only when changes
occur over distances much larger than $\lambda_F$. Rather, such
interfaces are included in the theoretical description via the
derivation of suitable boundary conditions for the Green's
functions. This was first done in Ref.~\cite{zaitsev1984quasiclassical}, and later 
extended in Refs.~\cite{Kupriyanov1988,PhysRevB.55.6015,nazarov1999}.  Magnetic
interfaces were discussed in Ref.~\cite{PhysRevB.38.8823} for the S/FI case and
extended to metallic interfaces and general cases in Refs.~\cite{PhysRevB.70.134510,PhysRevB.80.184511,Bergeret2012,Machon2013,eschrig2015general}.

To get a qualitative understanding of the phenomena related with spin-polarized tunneling into superconductors, we consider below low-transmissive interfaces between superconductors and normal or magnetic leads, using the description in Ref.~\cite{Bergeret2012}. Fixing the $z$ direction in the polarization direction of the interface, the transmission through it can be described by the tunneling matrix  $\hat{\Gamma}=t\tau_{3}+u\sigma_{3}$. It is defined
through the normalized transparencies satisfying $t^{2}+u^{2}=1$ and determining
the interface polarization via $2ut=P$. The possible non-collinearity between the interface and the bulk magnetizations
can be described by introducing rotation matrices in spin space. We assume that this interface connects a normal-metal electrode with bulk Green's function $\check g_N$ with a spin-split superconductor with GF $\check g$. The boundary condition gives an expression for the matrix current density through the interface in terms of $\hat \Gamma$, $\check g_N$ and $\check g$: 
\begin{equation}
\check{g}\nabla_{k}\check{g}=-\frac{1}{R_{\square}\sigma_{N}}\left[\hat{\Gamma}\check{g}_{N}\hat\Gamma^{\dagger},\check{g}\right]\label{eq:bc_gamma}\;,
\end{equation}
where $R_{\square}$ characterizes the spin-averaged barrier resistance per unit area. Its precise microscopic definition is given in Refs.~\cite{Bergeret2012,eschrig2015general}. In experiments it is usually used as a fitting parameter.
Equation \eqref{eq:bc_gamma}
is valid for an arbitrary spin polarization $P\in [-1,1]$ of the interface.  From this condition and Eqs.~(\ref{eq:charge_current}--\ref{eq:spinenergy_current})  we obtain the spectral current densities through spin-polarized barriers as
 \begin{equation}
 I_{ab} = -\frac{1}{16 e^2 R_{\square}} 
 {\rm Tr}\tau_{b}\sigma_{a}\left[\hat{\Gamma}\check{g}_{N}\hat\Gamma^{\dagger},\check{g}\right]^{K}\;.
 \label{eq:currents_at_interface}
 \end{equation}
The normal electrode Green's function $\check{g}_{N}$ is defined by $\hat g_{N}^{R(A)}=\pm\sigma_{0}\tau_{3}$
and $\hat g_{N}^{K}=2\tau_3\hat f^{(N)}$, where $f^{(N)}$ is the distribution function of the normal metal.
The boundary conditions for both spectral GF ($R$) and distribution functions ($K$) are then given by the 
continuity of the matrix current density, $n_k j_{kb}^{a} = I_{ab}$, where the tunneling current $I_{ab}$
and bulk current $j_{kb}^{a}$ are provided by Eqs.~(\ref{eq:currents_at_interface}) and (\ref{eq:general_current}), 
respectively, and $n_k$ are the components of the unit vector normal to the interface.
For small enough junctions, the approximate conservation of matrix current at short distances
implied by Eq.~\eqref{eq:Usadel} \cite{nazarov1999} allows neglecting the precise 3D structure of the junction.
Hence, in quasi-1D problems, the boundary condition can be integrated over the cross-sectional areas
of the wire $A$ and tunnel interface $A_T$ and written in the form $A j_{1D,b}^{a} = A_T I_{ab}$.

Low effective transmissivity of interfaces, which corresponds to a  large interface resistance (tunneling limit), also allows for further simplified approximations to be
made in the description of nonequilibrium effects.  In such a case,
changes in the spectrum due to the coupling can be neglected, and the probability for Andreev reflection \cite{andreev1964thermal} vanishes.  In the lowest order in transmissivity, we can also assume the electrodes to be in local thermal equilibrium. 
If, in addition, $P=0$, the charge current between two electrodes is
given by the usual tunneling expression
Eq.~(\ref{eq:I_tun}) with $G_T=A_T/R_\square$. In Sec.~\ref{spininjection} we describe the more general limit of an arbitrary $P$, and the different components of the spectral current.

It is worth emphasizing that the boundary condition, Eq.~\eqref{eq:bc_gamma}, when combined with the quasiclassical equations, allows for the description of effects which are beyond the quasiclassical limit and that involve strong ferromagnets with large spin polarization \cite{bergeret2012spin,silaev2017anomalous}.

\subsection{Free energy}
\label{sec:freenergy}

The quasiclassical approach can be used to describe the
collapse of superconductivity caused by an applied exchange field,
including the part of the parameter space where a first-order
phase transition occurs. The main physical features of this are
discussed in Sec.~\ref{sec:paramagnetic}.
In particular, a necessary condition for
the stability of a superconducting state is that the free energy
density of the superconducting state compared to the normal state is negative, $\Omega<0$. Below, we express
this condition in the quasiclassical framework, neglecting the FFLO
state.

The free energy density of a uniform superconductor relative to the
normal state can be written in terms of the quasiclassical Green
functions, \cite{Eilenberger1968,serene1983-qat,thuneberg1984-efp,vorontsov2003-tpt,rainer1976}: 
\begin{align}
  \label{eq:origina-qcl}
  \Omega[\hat{g},\hat{\Sigma}]
  &=
  -\frac{1}{2}\int_0^1\dd{\lambda}\Tr[\hat{\Sigma}(\hat{g}_\lambda[\hat{\Sigma}] - \hat{g})]
  + \Phi[\hat{g}]
  \,,
\end{align}
where the auxiliary Green function $\hat g_\lambda[\hat \Sigma]$ satisfies
\begin{align}
  0 &= [\hat{\Lambda} + \lambda\hat{\Sigma}, \hat{g}_\lambda[\hat{\Sigma}]]
  \,,
  \quad
  \hat{g}_\lambda[\hat{\Sigma}]^2 = 1
  \,.
  \label{eq:glambda-condition}
\end{align}
Here, $\hat{\Lambda}=(\omega_m+i\vec{h}\cdot\vec{\sigma})\hat{\tau}_3$, and
$\Tr=\pi{}TN(0)\sum_{\omega_n}\tr$ contains a Matsubara sum and Nambu
and spin traces. Variation of $\Omega$ vs. $\hat g$ produces the relation
$\delta\Phi[\hat{g}]=-\frac{1}{2}\Tr[\hat{\Sigma}_*\delta{}\hat{g}]$ between the
self-consistent self-energies $\hat{\Sigma}_*$ and the subtracted quasiclassical
$\Phi$-functional, with value $\Phi[\hat{g}_{*,N}]=0$ in the normal state.  
Variations and derivatives of $\hat{g}$ need to retain the condition $\hat{g}^2=1$,
and can generally be expressed in the form $\delta\hat{g}=\delta\hat{X}\hat{g}-\hat{g}\delta\hat{X}$ for some $\delta\hat{X}$.
This observation, and making use of Eq.~\eqref{eq:glambda-condition}, allows the evaluation of the $\lambda$ integral:
\begin{align}
  \Tr[\hat{\Sigma}(\hat{g}_\lambda[\hat{\Sigma}] - \hat{g})]
  =
  \frac{\mathrm{d}}{\mathrm{d}\lambda}
  \Tr[(\hat{\Lambda} + \lambda \hat{\Sigma})(\hat{g}_\lambda[\hat{\Sigma}] - \hat{g})]
  \,.
\end{align}
This enables writing the self-consistent free energy as
\begin{align}
  \Omega[\hat{g}_*,\hat{\Sigma}_*]
  &=
  \frac{1}{2}\Tr[\hat{\Lambda} (\hat{g}_{\lambda=0} - \hat{g}_{\lambda=1})]
  +
  \Phi[\hat{g}_{\lambda=1}]
  \,.
\end{align}
Note that in a spatially non-uniform situation, gradient terms would
complicate the $\lambda$-integration.  This
still allows for including the magnetic orbital effect, given the
approximation in Eq.~\eqref{eq:Usadel} where
Eq.~\eqref{eq:glambda-condition} retains its spatially uniform form
also in the presence of the vector potential \cite{maki1964-bst}.  The $\Phi$-functional
producing the elastic scattering self-energies discussed in preceding sections reads
\begin{align}
  \Phi
  &=
  \frac{1}{2}
  \Tr\{
  -\frac{1}{2}\Delta[\hat{g}] \hat{g}
  +
  \frac{1}{2\tau_{orb}}[1 - (\hat{\tau}_3{\hat{g}})^2]
  \\\notag&
  +
  \frac{1}{16\tau_{so}}[3 - (\vec{\sigma}{\hat{g}})^2]
  +
  \frac{1}{16\tau_{sf}}[3 - (\vec{\sigma}\hat{\tau}_3{\hat{g}})^2]
  \}
  \,,
  \\
  \Delta[\hat{g}]
  &=
  N(0)V
  \pi{}T
  \sum_{\omega_n}
  \frac{1}{2}
  (\hat{\tau}_+ \tr[\hat{\tau}_- \hat{g}] + \hat{\tau}_- \tr[\hat{\tau}_+ \hat{g}])
  \,.
\end{align}
For $\tau_{orb},\tau_{so},\tau_{sf}\gg\Delta^{-1}$, direct evaluation yields
\begin{align}
  \Omega
  &=
  2\pi{}TN(0)\Re\sum_{\omega_n>0}
  \frac{2w_n\sqrt{w_n^2+\Delta^2} - 2w_n^2 - \Delta^2}{\sqrt{w_n^2+\Delta^2}}
  \,,
\end{align}
where $w_n=\omega_n+ih$. At low temperatures $T/\Delta\ll1$,
$\Omega/N(0)\simeq{}-\frac{\Delta^2}{2}+h^2+\frac{\pi^2T^2}{3}$
\cite{maki_tsuneto.1964}, producing the Chandrasekhar--Clogston
result discussed in Sec.~\ref{sec:paramagnetic}.

\subsection{Linear response and generalized Onsager relations}
\label{sec:onsagersection}

A way of exciting the nonequilibrium modes  in superconductors is by injecting electrons from  
either a voltage or temperature biased  metallic lead. In this case, and in the language of  Eq.~(\ref{eq:distribution_function}),  the  distribution function in the normal metal is  given by  $\hat f^{(N)} = f_{L}^{(N)}+\tau_{3}f_{T}^{(N)}$, where $f_{L/T}^{(N)}=[f_{\rm eq}(\varepsilon - eV)\pm f_{\rm eq}(\varepsilon + eV)]/2$
 is the voltage-biased distribution function in the normal-metal electrode
 and $f_{\rm eq}(\varepsilon)=\tanh(\varepsilon/2T_N)$ is the equilibrium distribution corresponding to the normal-metal temperature $T_N$. 
 
 In a more general situation, all components  of  the distribution function,  Eq.~(\ref{eq:distribution_function}),
have to be taken into account.  The different components of the spectral current through a tunneling contact between a normal electrode 
and a superconductor can be  derived from the boundary condition  in Eq.~(\ref{eq:currents_at_interface}). 
In the case of collinear magnetizations the relevant components are the spectral charge 
$I_{30}$, energy $I_{00}$, spin $I_{03}$ and spin energy $I_{33}$ currents. 
They satisfy
 \begin{equation} \label{Eq:bcDistrFunc}
 \left( \begin{array}{cccc}
  I_{30} \\ I_{00} \\ I_{03} \\ I_{33}
  \end{array} \right)  =
   \kappa  \left(
    \begin{array}{cccc}
    N_+ & PN_- & PN_+ & N_-  \\
    PN_- & N_+ & N_- & PN_+ \\
    PN_+ & N_- & N_+ & PN_- \\
    N_- & PN_+ & PN_- & N_+ \\
   \end{array}
   \right)
   \left(\begin{array}{cccc}
   \tilde f_T \\ \tilde f_L  \\ \tilde f_{T3} \\ \tilde f_{L3}
   \end{array} \right) \; ,
   \end{equation} 
  %
   where the parameter $\kappa = 1/(R_{\square}e^2)$ describes the interface transparency
   and $\tilde{f}_k = f^{(S)}_k - f^{(N)}_k  $ are the differences of the various distribution functions 
   $\tilde{f}_k = f^{(S)}_k - f^{(N)}_k  $, with $k=T,L,T3,L3$,
   between the superconductor and normal-metal electrodes.  The response matrix is here described by the spin polarization $P$ and the energy-symmetric and energy-antisymmetric parts of the density of states, $N_\pm$ defined in Eqs.~(\ref{eq:dosplus}-\ref{eq:dosminus}).
            
  Expression  (\ref{Eq:bcDistrFunc}) is  valid for arbitrarily large deviations
  from the equilibrium state and 
  describes several intriguing effects which come into play in the
  nonequilibrium situation in junctions combining spin-split
  superconductors ($N_- \neq 0$) and
  a spin-polarized injector, modeled by a nonzero spin polarization $P
  \neq 0$. 
If both $N_-\neq 0$ and $P\neq 0$ simultaneously, all  the  nonequilibrium 
  modes are coupled. In turn, this coupling  results  in a rich variety of cross-couplings between different potentials and currents. 
    
  To illustrate the basic phenomena which can be expected in such a tunneling contact,
  let us consider the linear response limit.  The different components of the distribution function can be described by 
  spin-dependent temperatures\footnote{Note that the notion of spin-dependent temperatures in an out-of-equilibrium setup is questionable \cite{heikkila2010}. However, for linear response $\Delta T_s$ can be strictly defined to characterize the nonequilibrium mode via Eq.~\eqref{Eq:SpinDpendentPotentials}.} and voltages introduced according to Ref.~\cite{Bauer:2012fq}:
  \begin{equation} \label{Eq:SpinDpendentPotentials}
  \left( \begin{matrix}
  \Delta f_T \\ \Delta f_L \\ \Delta f_{T3} \\ \Delta f_{L3}
  \end{matrix} \right) 
   = 
  \frac{\partial f_{\rm eq}}{\partial \varepsilon}  
  \left(\begin{matrix} 
   eV \\ -\Delta T\varepsilon/T  \\ eV_s/2 \\ -\Delta T_s\varepsilon/2T
  \end{matrix} \right)\; , 
  \end{equation}
   where $V$ and $\Delta T$ are the voltage and the temperature bias and $V_s$ and $\Delta T_s$ the spin-dependent biases. 
   
   Integrating Eq.~(\ref{Eq:bcDistrFunc}) over energy  allows us to calculate the total  charge
     $I = e \int_{-\infty}^{\infty} I_{30} d\varepsilon$,  energy $\dot{Q} = \int_{-\infty}^{\infty} \varepsilon I_{00} d\varepsilon$,
   spin $I_s =\int_{-\infty}^{\infty} I_{03} d\varepsilon$, and  spin energy $\dot{Q}_s =\int_{-\infty}^{\infty} \varepsilon I_{33} d\varepsilon$ currents. 
   They are related to the generalized potentials via a $4\times 4$ Onsager matrix
   \footnote{In Eq.~(\ref{Eq:Onsager}) we assume that the biases are calculated as the shifts of the corresponding quantities in the normal metal with respect to the superconductor and the currents flow from the superconductor into the normal metal. }
   \begin{equation} \label{Eq:Onsager}
   \left( \begin{array}{ccc}
   I \\ \dot Q \\ I_s \\ \dot Q_s
  \end{array} \right)  
    = \left(
  \begin{array}{cccc}
    G & P\alpha & PG & \alpha  \\
    P\alpha & G_{\rm th}T & \alpha & PG_{\rm th}T \\
    PG & \alpha & G & P\alpha \\
    \alpha & PG_{\rm th}T & P\alpha & G_{\rm th}T \\
  \end{array}
  \right)
  \left(\begin{array}{ccc}
    V \\ - \Delta T/T   \\ V_s/2 \\ - \Delta T_s/2T
  \end{array} \right), 
  \end{equation}   
  where the conductance, heat conductance and thermoelectric coefficient are given by 
  \begin{align} \label{Eq:condG}
  G =  e^2\kappa \int_{-\infty}^{\infty} N_+ \frac{\partial f_{\rm eq}}{\partial \varepsilon} d\varepsilon \\ \label{Eq:condGQ}
  G_{\rm th} =  \frac{\kappa}{T}\int_{-\infty}^{\infty} N_+ \varepsilon^2 \frac{\partial f_{\rm eq}}{\partial \varepsilon}  d\varepsilon \\
  \label{Eq:condalpha}
  \alpha = -e\kappa\int_{-\infty}^{\infty} N_- \varepsilon \frac{\partial f_{\rm eq}}{\partial \varepsilon}  d\varepsilon,
  \end{align}
  respectively.\footnote{We denote the full thermoelectric
    coefficient by $\tilde \alpha$ and separate the dependence on the
    polarization $P$ in $\alpha$ to clarify the dependence on
    $P$. Thus, here $\tilde \alpha=P\alpha$.} In the absence of spin-orbit or spin-flip scattering, these integrals can be evaluated analytically in the limit $T \ll \Delta-h$. The results are in Eqs.~\eqref{eq:thermoelcoefsanalytically}.

Note that the four $2\times 2$ quadrants of the response matrix have different time-reversal symmetries: the quadrants on the diagonal are symmetric, whereas the off-diagonal quadrants coupling spin and charge are antisymmetric. This result follows from the fact that $G$ and $G_{\rm th}$ are symmetric whereas $\alpha$ and $P$ are antisymmetric in time reversal.

   From Eq.~(\ref{Eq:Onsager}) we can,
   for example,  study the  spin and charge 
   currents created at the SF interface in response to a temperature bias between the normal lead and the superconductor. These are the  
   thermospin and thermoelectric effects.
Interestingly, Eq.~(\ref{Eq:Onsager}) demonstrates also that the
charge current can be induced by the spin-dependent temperature bias
$\Delta T_s$ even without spin-filtering ($P=0$) but in the presence
of a spin-split DOS, $N_-\neq 0$. Qualitatively  the corresponding non-equilibrium mode $f_{L3}$ [Fig.~\ref{fig:modes2}(e)] can be interpreted as the spin-dependent particle-hole imbalance which produces electric signal due to  $N_-\neq 0$.  

   Let us focus on the case where  the nonequilibrium state is generated by applying a voltage
   and a temperature  bias to the injecting normal electrode in the
   absence of spin-dependent potentials, $\Delta T_s =0$ and 
   $V_s=0$. If the interface lacks spin polarization,  $P=0$,  then from  Eq.~(\ref{Eq:Onsager})
   it follows that  all currents  are  decoupled
   from each other, recovering standard expressions for charge and heat currents in terms of the local
   electrical and thermal conductances $I= G V$ and $\dot Q= - G_{\rm th} \Delta T/T$.  In addition, 
   the remaining block of the Onsager matrix leads to nonzero spin current $I_s = -\alpha \Delta T/T $  and spin heat current $\dot Q_s = \alpha V $.
   From this linear response analysis we can conclude that  in a setup consisting of a normal and a spin-split superconducting electrode,  a spin current can be generated even in the absence of  
   ferromagnetic electrodes.  It is worth mentioning that this thermospin effect leads to a pure  
    spin current  $I_s$ that does not carry electric charge and therefore 
    does not produce Joule heating and Ohmic losses. Clearly this situation is very  
    different from the one occurring in normal metal systems, 
   where spin currents can be generated only by injecting a 
   charge current from a ferromagnet --- a process corresponding to
     the relation $I_s=PGV$ obtained from Eq.~\eqref{Eq:Onsager}.
      
   If $P\neq 0$  all four types of currents are injected into the superconducting region
   just by applying the temperature bias $\Delta T$.  In particular a very large thermoelectric effect can be generated, 
  described by  $I = - P\alpha \Delta T/T$.   This effect is discussed in detail  in Sec.~\ref{thermoel}.

\subsection{Nonequilibrium effects on the superconducting order parameter}
\label{sec:selfconsistentgap}

Nonequilibrium also affects the superconducting state by modifying its order parameter. This can be seen by inserting the parametrization
(\ref{eq:K_param},\ref{eq:distribution_function}) of the
nonequilibrium Green's function in Eq.~\eqref{self_consistent}:
\begin{equation}
\Delta=\frac{\lambda}{2} \int_{-\Omega_D}^{\Omega_D} d\varepsilon {\rm
  Im}g_{01}^R f_L + {\rm Im} g_{31}^R f_{T3} + i ({\rm Re} g_{01}^R
f_T + {\rm Re} g_{31}^R f_{L3}).
\label{eq:gapeq}
\end{equation}
In equilibrium, only the first term contributes with
$f_L=f_{\rm eq}=\tanh[\varepsilon/(2T)]$. The imaginary terms affect the
phase of $\Delta$ and ensure charge current conservation in situations
where quasiparticle current is converted to supercurrent (described by
the $R^{ab}$ term in Eq.~\eqref{eq:kineticeq}). In the absence of spin
splitting, the nonequilibrium modifications in the size of the gap,
described by the first term, are
the most relevant. Often such modifications are
studied within the Rothwarf--Taylor phenomenological model \cite{rothwarf1967}
that neglects the energy dependence of the distribution functions, and
rather concentrates on the overall number of quasiparticles. This model was
derived under some simplifiying assumptions from the full energy
dependent kinetic equations in Ref.~\cite{chang1977}. It
has been used to study the gap suppression due to
nonequilibrium injection, see for example Ref.~\cite{yeh1978}. According to
that work, for small
changes around equilibrium, such models agree with the predictions of
the phenomenological models of Refs.~\cite{owen1972,parker1975}. For large
changes, the full kinetic equations with energy dependent distribution
functions should be used.  An extreme example is given in
Ref.~\cite{keizer2006} describing a voltage-driven transition of a
superconducting wire to the normal state. 

In the spin-dependent case we need to consider both the effects of
spin splitting and the spin-dependent distribution functions
describing spin accumulation. Disregarding the terms affecting mostly
only the phase of the order parameter, we can also write
Eq.~\eqref{eq:gapeq} as
\begin{equation}
\Delta=2\lambda \sum_{\sigma=\pm} \int_{-\Omega_D}^{\Omega_D} d\varepsilon {\rm
  Im}F_\sigma f_\sigma,
\label{eq:gapeq2}
\end{equation}
where $F_\sigma=(g_{01}^R+\sigma g_{31}^R)/2$ and
$f_\sigma=(f_L+\sigma f_{T3})/2$; the different signs of $\sigma=\pm$
representing the two spin directions. Ref.~\cite{takahashi1999} went a step
further, assuming that the spin accumulation in a superconductor can
be described in terms of a simple spin-dependent chemical potential
shift $\mu_s$, neglecting all other nonequilibrium effects. In that case
$f_\sigma=f_{\rm eq}(\varepsilon-\sigma \mu_s)$. As a result, by a
simple shift of the energy ($\varepsilon \mapsto \varepsilon + \sigma
\mu_s$) in Eq.~\eqref{eq:gapeq2}, $\mu_s$ shows up as a shift of the energy of the anomalous
function $F_\sigma$ and a small shift of the cutoff $\Omega_D$. When
$\Omega_D \gg \mu_s$, the latter effect can be disregarded, and the
net effect of the spin accumulation is the same as that of the spin
splitting field, explained in Sec.~\ref{sec-superwithh}, eventually leading to
the suppression of superconductivity. Moreover, in the presence of
both spin splitting and spin accumulation, this model shows how in the
special case $h=\mu_s$ spin accumulation can actually lead to the
recovery of superconductivity suppressed by spin splitting
\cite{bobkova2011}. 

This physics can be probed in a FISIF system where a superconducting
island or layer is placed between two ferromagnetic electrodes. In
this case, the current induced suppression of the superconducting gap
should be larger when the magnetizations of the two ferromagnets are
antiparallel than when they are parallel \cite{takahashi1999}. Only in
the previous case the spin accumulation builds up. This effect was
measured in Ref.~\cite{yang2010extremely}. They indeed found a stronger
suppression of superconductivity in the antiparallel
configuration. However, they found that the required spin relaxation
time for fitting the results to the above theory is much longer than that
expected from the normal-state measurements. This discrepancy may
result from the somewhat simplified model for the spin accumulation described above. 

The suppression of superconducting properties due to spin injection
has been measured in the case of high-temperature superconductors
\cite{vasko1997,koller1998,yeh1999,gim2001}. However, these features
are typically attributed to the non-conventional character of
superconductivity, and are therefore outside the scope of this review.

Besides the superconducting gap, in principle also the induced
spin-splitting field can obtain nonequilibrium corrections in the presence
of spin injection. The theory for such effects was outlined already in
Ref.~\cite{Alexander1985}, but to our knowledge such effects have not been
thoroughly examined in spin-split superconductors.

\subsection{Overall strategy to explore the transport and spectral
  properties of hybrid superconducting systems}

Spectral and transport properties of a diffusive metal or
superconductor of mesoscopic size attached to electrodes can be fully
described by using the theoretical framework presented in this section.
Specifically, one should solve the boundary problem defined by the
Keldysh-Usadel equation (\ref{eq:Usadel}), the normalization condition
for $\check{g}$, Eq.~(\ref{normalization}), and the boundary
conditions at the interfaces with the electrodes,
Eq.~(\ref{eq:bc_gamma}).  Once the Green's functions are determined
one can compute the currents and potentials from
Eqs.~(\ref{eq:def_mu}-\ref{eq:spinenergy_current}).

If the system under consideration is in the normal state, the
equations for the retarded and advanced GFs are decoupled from each
other and from the Keldysh one. The solution of the spectral equations
within the quasiclassical approach is trivial
and leads to $\hat{g}^{R(A)}=\pm \tau_3$. Thus the problem reduces to solving the
equation for the distribution functions.

If the system is a superconductor attached to leads, the equations for
the retarded and advanced GFs are coupled to the Keldysh component
through the self-consistency equation for the superconducting order
parameter, Eq.~(\ref{self_consistent}).  This complication can be
overcome if the superconductor and the electrodes are coupled via
tunneling junctions. In such a case the spectral properties of the
superconductor remain unchanged within the leading order in the
interface transmission. This means that retarded and advanced GFs
coincide with those of the homogeneous superconductor. In
Sec.~\ref{spininjection}, we concentrate on the case where the
normal-state tunnel junction conductance satisfies
$G_T R_{\ell_{E}} \ll 1$, where $R_{\ell_{E}}$ is the normal-state
resistance of a wire with length equal to the energy relaxation length
(due to electron-phonon or electron-electron scattering) $\ell_E$ at
the energy around the amplitude of the superconducting gap. On the
other hand, in Sec.~\ref{thermoel}, we mostly consider the cooling of
the superconducting island with a starting temperature much below
$T_c$. In these cases we can disregard the nonequilibrium effects on
the energy gap, and can rather use the self-consistent gap that is
calculated from the equilibrium version of self-consistency equation,
i.e., Eq.~\eqref{eq:gapeq}, with $f_L=\tanh[\varepsilon/(2 k_B T)]$.

In most of the discussed examples of this review we follow this
approach and use the self-consistent $\Delta$ calculated in the
decoupled spin-split superconductor.  From this we obtain the spectral
functions of the spin-split superconductor that yield the coefficients
of the kinetic equation for the nonequilibrium distribution
functions. The latter is obtained from the Keldysh component of
Eq.~(\ref{eq:Usadel}).  Only in Sec.~\ref{sec:ninlin_ac}, where we
study the effect of an ac field on the spectrum of a spin-split
superconductor, we compute the self-consistent gap using the
nonequilibrium distribution function (see
Fig.~\ref{fig:Delta-vs-tausn}).

Inspection of the different components in Keldysh space of
Eq.~(\ref{eq:Usadel}) provides a first insight about the characteristic
lengths involved in the different situations.  In a superconductor
changes of the spectral properties due, for example, to the magnetic
proximity effect with a ferromagnetic insulator, occur over the
superconducting coherence length $\xi_s$. On the other hand, this
$\xi_s$ depends on the temperature, and on the concentration of
magnetic and spin-orbit impurities via the self-consistent $\Delta$.

Different length scales govern the decay of nonequilibrium components
of the distribution function generated, for example, {by injecting a current through a contact with an electrode.} 
In this case the
characteristic lengths depend on the nature of the excited mode
(charge, energy or spin) and the type of relaxation process in the
system.  The calculation of these characteristic lengths is one of the
main goals of Sec.~\ref{spininjection}.  As we show there, the spin
splitting plays a crucial role in determining these length scales,
since it couples the different nonequilibrium modes, and hence changes
the range over which non-local spin and charge signals can be
detected.

\section{Transport properties of ferromagnetic insulator--superconductor heterostructures}
\label{spininjection}

Electrical injection of spins into a superconductor was first studied in Ref.~\cite{Aronov1976}.   The injection of an  electric current from a F electrode into a superconductor creates not only spin, but also   charge imbalance.  Moreover, 
 such imbalances  relax away from the injection point over distances different from those in the normal state. 

 In the absence of a spin-splitting field in the superconductor the spin imbalance is decoupled from the other modes. However,
in spin-split superconductors charge, spin and energy modes couple with each other. The goal of this section is to describe this coupling, the characteristic relaxation lengths of the different nonequilibrium modes and   effects that  occur  in superconducting structures as a consequence of  such coupling.

We start  by reviewing experiments on charge and spin injection in superconductors and 
by summarizing the main theoretical works on spin injection in superconductors with no spin splitting. 

Further on, with the help of kinetic equations derived in Sec.~\ref{sec:Kin_eqs} we describe how the spin splitting  affects transport properties in S/FI structures, {and discuss several possible experimental situations both with collinear and non-collinear spin configurations}. 

\subsection{Detection of spin and charge imbalance: non-local transport measurements\label{sec:nl_detection}}

\begin{figure}
  \centering
\includegraphics[width=8cm]{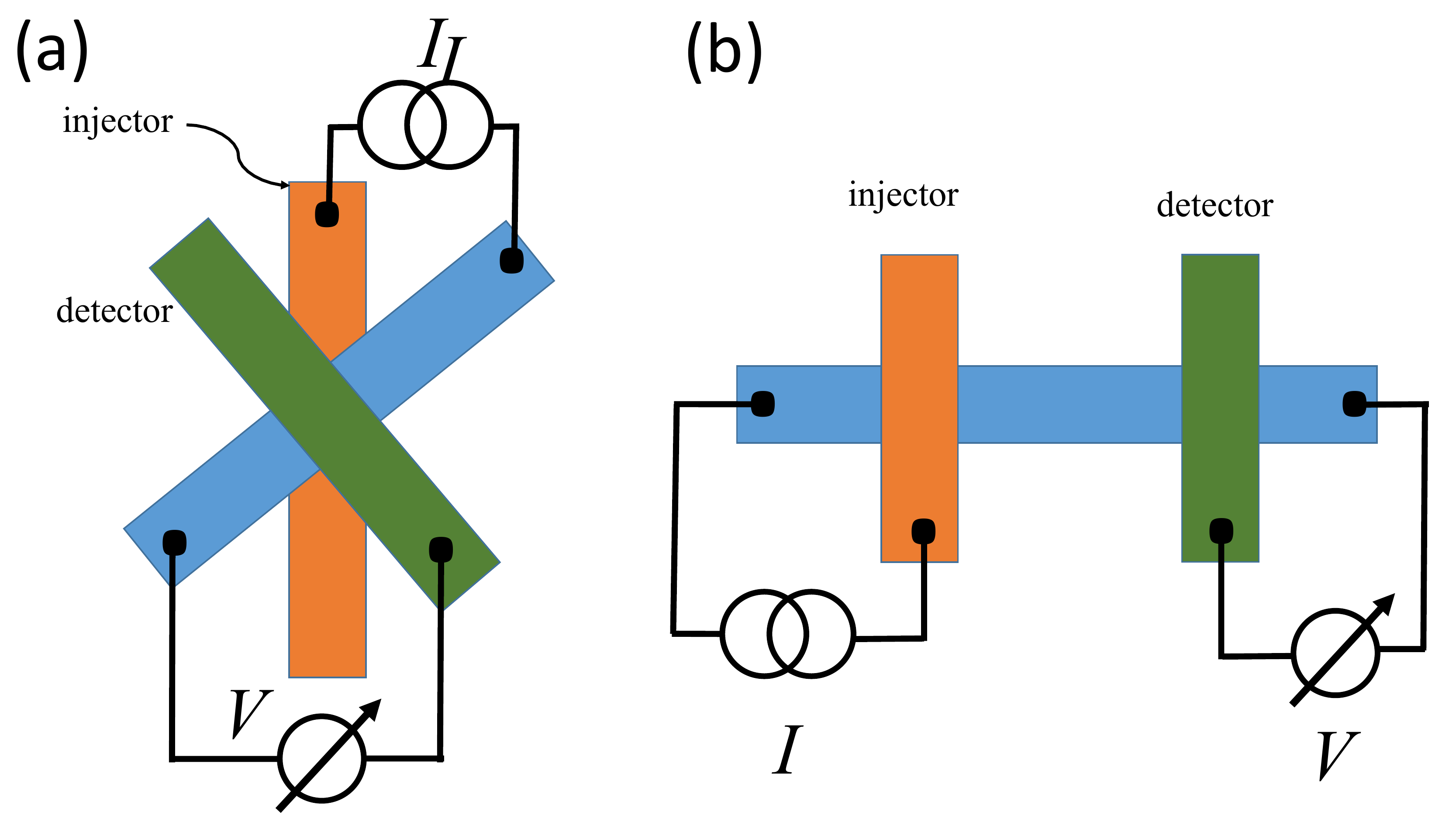}
\caption{Non-local detection of charge and spin imbalances via transport measurements. Sketch of (a) 
 the  original setup \cite{PhysRevLett.28.1363} for charge imbalance measurement in a superconductor (blue) and (b) a
lateral structure used in several more recent experiments for both spin and charge imbalance detection. 
 \label{fig:experiemnts_setups}}
\end{figure}

Nonequilibrium states in superconducting systems can be experimentally accessed using the non-local transport measurement setup 
invented in Ref.~\cite{PhysRevLett.28.1363}.   The geometry used in that experiment was a vertical structure such as the one  sketched in Fig.~\ref{fig:experiemnts_setups}(a). Injection of a current from a normal metal (injector) into the superconductor (blue stripe) generates the charge imbalance mode shown schematically in Fig.~\ref{fig:modes2}(b). 
Qualitatively this mode can be understood as the difference between the chemical potentials of the 
quasiparticles and the {superconducting} condensate. This relative shift of chemical potentials  induces an electric signal {(voltage or current), depending on the measurement setting} at the detector electrode [green stripe in Fig.~\ref{fig:experiemnts_setups}(a)]. 
First experiments demonstrating  the charge imbalance   induced in superconductors by current injection were performed at temperatures close to the critical temperature $T_c$ by using  vertical samples with large cross sections, such as the one depicted in  Fig.~\ref{fig:experiemnts_setups}(a).  More recent experiments have  allowed for non-local transport measurements
at lower temperatures using  lateral structures such as the one sketched in Fig.~\ref{fig:experiemnts_setups}(b)  
\cite{beckmann2004evidence,hubler2010charge,wolf2014charge,PhysRevB.87.024517,Quay2013,poli2008spin}.   Such  structures allow for an accurate measurement of the charge imbalance and nonequilibrium energy mode and their spatial dependence by placing  contacts (detectors)  at different distances from the injector.   For example, in an aluminum wire at 100 mK the charge and energy modes decay over 5 and 10 $\mu$m, respectively, as reported in Ref.~\cite{PhysRevB.83.104509}.

Substituting the normal injector by a ferromagnet, the injected current becomes spin polarized and the  charge injection is 
accompanied by a  spin injection  into the superconductor. 
%
%
%
The nonequilibrium charge and spin relax over different lengths.   
When the superconductor shown by the blue wire  in Fig.~\ref{fig:experiemnts_setups}b  is in the normal state,  charge accumulation is negligible  at the detector and only a spin imbalance contributes to the nonlocal signal.  The nonequilibrium spin density induced at the interface is polarized in the direction of the injector's magnetization and it diffuses into the normal wire over the spin-relaxation length, which can be  several hundreds of  nanometers.  The spin accumulation can be detected  by measuring the non-local voltage between a ferromagnetic detector and the N wire.
 
 First measurement of the  nonequilibrium spin accumulation was reported   in Ref.~\cite{PhysRevLett.55.1790} on a single-crystal aluminum bar at temperatures below 77 K. 
This pioneering experiment did not only demonstrate the spin accumulation associated with the injection from a ferromagnetic electrode, but also the coherent spin precession, the Hanle effect, that occurs when an external magnetic field non-collinear with the injector magnetization is applied. Later these effects have been reported at room temperature in  Al-based nanostructures in Refs.~\cite{jedema:345,jedema2003spin}. Since then, electrical injection of spins have been used in several experiments on metallic spintronics devices, for example for the direct detection of the spin Hall effect in  metallic structures  \cite{valenzuela2006direct}, electronic spin transport in graphene \cite{tombros2007electronic}, and modulation of the spin density in metal/ferromagnetic insulator bilayers \cite{Villamor2015, PhysRevB.91.100404}.

In the superconducting state  the situation drastically changes, as reported in several experiments   
on lateral superconducting structures with  ferromagnetic electrodes.
In contrast to the charge imbalance, which has been well 
understood since the 1980s,  spin injection and accumulation
in superconductors is a more recent research line.  
In 1990 Kivelson~\cite{Kivelson1990} suggested   that charge and spin should exhibit different relaxation times in superconductors, leading to the possibility of separating charge and spin transport. Also 
 Ref.~\cite{PhysRevB.52.3632} studied the spin injection
from a ferromagnet to a bulk superconductor and found theoretically that, while
the charge imbalance survives only within the field penetration length
from the surface, the spin imbalance may also exist in the bulk.
However, early experiments measuring spin diffusion length in Nb  did not show any evidence of
such a long-range spin signal  \cite{Johnson1994, gu2002direct}. Thus the possibility of having strongly different
relaxation scales for spin accumulation and  charge imbalance in superconducting state  was not confirmed at that time.

More recently, clearer  insight into  the spin and charge modes  has been obtained in experiments using   
lateral  nanostructures  with ferromagnetic injectors and detectors \cite{beckmann2004evidence,PhysRevB.71.144513,cadden2007charge,Hubler2012a,Kolenda2016,PhysRevB.87.024517,Wolf2014,poli2008spin,Quay2013,yang2010extremely}.
First, the decrease of spin relaxation length in the superconducting state as compared to the normal one has been observed \cite{poli2008spin} by measuring the non-local spin-dependent resistances.  
Later, by applying the external spin-splitting field to the superconducting wire it has been possible 
to prove experimentally the charge-spin separation \cite{beckmann2016spin}.  {A detailed overview of the recent experiments on charge, energy and spin imbalance in superconductors can be found in the recent topical reviews  in Refs.~\cite{beckmann2016spin,quay2017up}.}

In order to describe theoretically the different modes excited in  experiments on lateral S/F structures, it
is convenient to use the generalized quasiclassical model introduced in Sec.~\ref{sec:noneq-modes}.
 First theoretical works on spin injection into mesoscopic superconductors
\cite{morten2004spin,morten2005spin} addressed the question of 
how spin relaxation  changes in
the superconducting state and how these changes  depend 
on the spin relaxation mechanism. These works pointed out that in the presence of spin-polarized injected 
currents additional spin-resolved components of the distribution function may appear. 
The main conclusion of Refs.~\cite{morten2004spin,morten2005spin}
was that the  spin-relaxation length changes significantly in the superconducting state 
and   depends on the energy of the injected quasiparticles. 
In particular, it has been shown that  for electrons with energy close to the superconducting gap 
the spin relaxation length can decrease  in agreement with the experiment \cite{poli2008spin}. However,  two important  features  observed in subsequent studies of spin transport in superconductors with spin-splitting field could not be understood in terms of that  theory. One intriguing observation  was the drastic increase of the spin accumulation length in the superconducting state as compared to the normal one \cite{Quay2013,Hubler2012a,PhysRevB.87.024517}.
In addition, it has been observed that 
spin accumulation in such a setup can be created by the current injection from a non-magnetic electrode \cite{PhysRevB.87.024517}.

As we discuss in this section, 
to explain these two observations we need to use the kinetic theory which takes into account the modification of the quasiparticle spectrum due to the spin splitting. If it is caused by the Zeeman effect from the external magnetic field, it is also important to take into account the orbital depairing effect which leads to the suppression of  superconductivity. 
Recent works showed that such a modification of the spectrum of the superconductor leads to 
an intriguing coupling between the nonequilibrium modes in a superconductor
\cite{silaev2015spin,silaev2015long,virtanen2016stimulated,bobkova2016injection,bobkova2015,krishtop2015-nst}. 
The effect of this coupling  between the modes on the transport properties of a multi-terminal superconducting structure can be 
theoretically explored with the  quasiclassical formalism developed in Sec.~\ref{sec:quasiclassical_theory}. 
Here we review these theory works  and provide  a quantitative explanation of the long-range spin accumulation detected 
in multi-terminal superconducting devices.

 \subsection{Nonequilibrium properties of  a superconductor with spin splitting \label{sec:noneq}}
 
 \begin{figure}
  \centering
 \includegraphics[width=8cm]{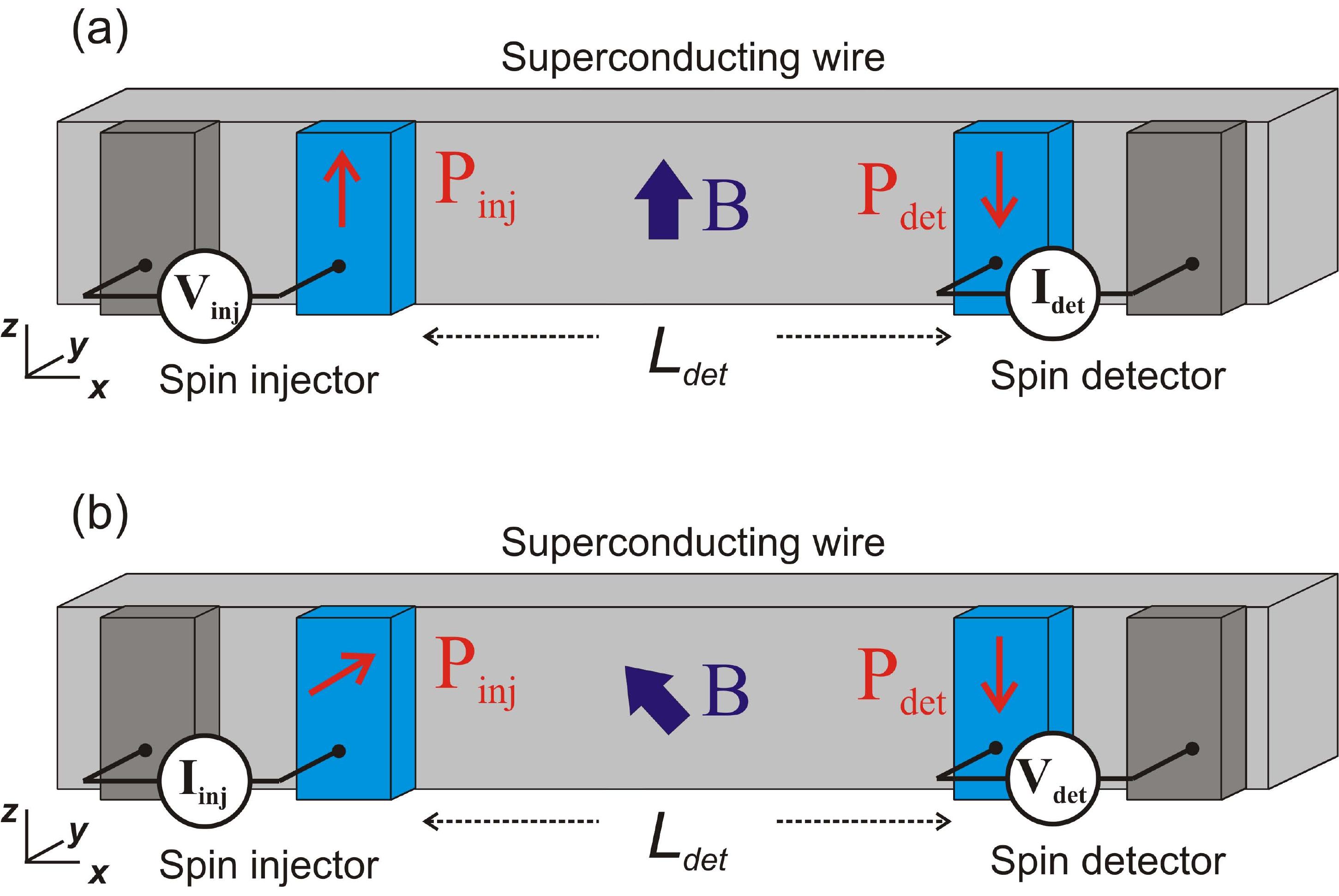}
 \caption{\label{Fig:SpinValveSketch}
 Schematic view of the
 setup for nonlocal transport measurements. (a) Collinear configuration where
 the polarizations of the magnetic contacts are collinear with the
 magnetic field, ${\bm P_{\rm inj}}\parallel {\bm P_{\rm det}}\parallel {\bm B}$.
 The non-local conductance $g_{nl} = d I_{\rm det}/d V_{\rm inj}$ is measured.   
 (b)  General non-collinear case with the magnetic field 
  having finite transverse component with respect to the polarizations 
 ${\bm P_{\rm inj}} , {\bm P_{\rm det}}$. The non-local voltage $V_{\rm det}$ is measured in the absence of the current in the detector electrode, $I_{\rm det} =0$. 
The latter detection scheme enhances the spin signal in the superconducting state, see discussion in the text. 
   }
 \end{figure}
 Spin accumulation and  transport  in superconductors has been considered in a number of theory papers \cite{Takahashi2003,morten2004spin,morten2005spin,chevallier2018, silaev2015spin,silaev2015long,virtanen2016stimulated,bobkova2016injection,bobkova2015,krishtop2015-nst}. Here we follow the approach developed in \cite{silaev2015long} that enables the description of all non-equilibrium modes, the effect of the spin-splitting field, and the various relaxation mechanisms.
As discussed in Sec.~\ref{sec:noneq-modes}, nonequilibrium conditions lead to the excitation of charge, energy and spin modes in superconducting systems. 
These modes are related to the different components of the distribution function introduced in Eq.~(\ref{eq:distribution_function}).
 A  prototypical experimental geometry 
to  reveal the different nonequilibrium modes in superconductors  
 is the lateral structure shown in Fig.~\ref{Fig:SpinValveSketch}. 
 It consists of a mesoscopic superconductor attached to  four electrodes.
 One of them serves as an injector for the current, 
 whereas another one  is used as a detector. {The other two electrodes serve as separate charge current sinks, allowing for such a non-local measurement.}
 
  We assume first that both electrodes, injector and detector, are  ferromagnetic and that their magnetizations  are collinear as in Fig.\ref{Fig:SpinValveSketch}a
    \[
    {\bm P_{\rm inj}}\parallel {\bm B}\parallel {\bm P_{\rm det}}\; .
    \]
The tunnelling current at the detector can be expressed through the 
charge imbalance $\mu$ and spin accumulation $\mu_z$ (\ref{Eq:ChPot0},\ref{Eq:ChPotZ})
  \begin{equation}\label{Eq:ZeroCurrentYGen}
  I_{\rm det}= (\mu  +  P_{\rm det} \mu_z)/R_{\rm det},
    \end{equation}
where 
%
$R_{\rm det}=R_\square/A$ is the detector interface resistance in the
normal state, and $A$ is the cross-sectional area of the detector. 
Here we consider  collinear magnetizations
 along the $z$-axis and therefore  only the spin components $f_{T3}$ and $f_{L3}$    enter these expressions. A non-collinear case is discussed in Sec.~\ref{sec:spinhanle}. 

If  the current through the detector
 is measured at zero bias $V_{\rm det}=0$, then   the nonlocal differential conductance
 is defined as 
 \begin{equation}
  g_{nl}= \frac{d I_{\rm det}}{d V_{\rm inj}} \; .\label{eq:gnl0}
  \end{equation}
 This quantity provides information about the modes that, being  excited at the injector via the injected current, 
 reach the detector by diffusing in the superconducting wire.  
The nonlocal conductance $g_{nl}$  depends on the distance $L_{\rm det}$ between the injector and the detector. This dependence reveals 
 the characteristic  length scale  over which the modes relax {and in practice it is measured by using several detectors at different distances to the injector}.

According to Eqs.~(\ref{Eq:ChPot0},\ref{Eq:ChPotZ})  the non-local current is determined, in the most general case,  by the contributions of  all four non-equilibrium modes shown schematically in Fig.~\ref{fig:modes2}. 
Here the quantities $\mu$ and $\mu_z$ characterize arbitrary nonequilibrium states of the superconducting wire, including those that cannot be described by the effective potentials $V$, $\Delta T$, $V_s$ and $\Delta T_s$ introduced in Sec.~\ref{sec:onsagersection}. However, if we restrict the analysis to   linear response then  Eq.~(\ref{Eq:ZeroCurrentYGen}) reduces to the expression for the current obtained from Eq.~(\ref{Eq:Onsager}) with $P=P_{\rm det}$,  $\kappa = A/(e^2R_{\rm det})$,  $\mu =  R_{\rm det} (\alpha \Delta T_s /2T - G V ) $ and $\mu_z = R_{\rm det} (\alpha \Delta T /T - G V_s/2 )$. 

When the distance between the injector and detector electrodes is much larger than the charge imbalance relaxation length, the contribution of $\mu$ to the detected current can be neglected  \cite{bobkova2015,krishtop2015-nst}.
In the absence of spin splitting $h=0$ ($N_-=0$), the spin-dependent part of $I_{\rm det}$ is determined only by the $f_{T3}$ mode related to the spin-dependent chemical potential shift \cite{Takahashi2003,morten2004spin,morten2005spin} (see Eq.~\eqref{Eq:SpinDpendentPotentials}).
In this case in the linear response regime
and at low temperatures $|e V_s|, T \ll \Delta$,
the spin accumulation generated by a spin-dependent chemical potential shift $\delta\mu = e V_s/2$ is exponentially small.  This can be seen from the first term on the r.h.s of   Eq.~(\ref{Eq:ChPotZ}), by noticing that  if the energy gap $\Delta$ is non-zero then $N_{+}(|\varepsilon|<\Delta) =0$. 
 In this case the nonequilibrium spin state is described by 
$f_{T3} = (\partial f_{\rm eq}/\partial\varepsilon) e V_s/2 $ and hence  the spin accumulation is given by  $\mu_z = - (eV_s/2)\int_{\Delta}^{\infty} d\varepsilon  N_+ (\partial f_{\rm eq}/\partial\varepsilon)$
 so that $\mu_z \propto e^{-\Delta/T}$
at $T\ll \Delta$ . 

The non-trivial behaviour of spin accumulation $\mu_z$ 
is determined by the
   second term in the  r.h.s  of Eq.~(\ref{Eq:ChPotZ}). {It}  
 appears when the spin splitting becomes non-zero, i.e., $h\neq 0$,  $N_-\neq 0$. 
This term 
allows for the spin-charge separation which   explains  experiments in  \cite{Quay2013,Hubler2012a,PhysRevB.87.024517}. 
 On the qualitative level the physics related to this term can be described as follows. 
 When $N_-\neq 0$ the spin accumulation has a contribution from  energy nonequilibrium mode 
 $f_L$ shown schematically in Fig.~\ref{fig:modes2}. This mode, once excited, can only relax via inelastic processes, especially mediated by the electron-phonon interaction.  
At low temperatures in metals the electron-phonon relaxation  can be much slower than the spin decay in the normal state.    
This mechanism explains the long-range spin accumulation observed in experiments. Since the nonequilibrium mode $f_L$ can be generated in particular by the temperature rise in the quasiparticle system, the  long-range non-local spin signals can be explained in terms of the thermo-spin effect.
Its  essence  is the generation of  
 spin accumulation by heating up  quasiparticles in the superconductor. 
 In practice  this effective heating can be achieved with the help of the voltage-biased tunnel junction
 which can inject the nonequilibrium quasiparticles 
with energies larger than the gap into the superconducting wire \cite{silaev2015spin,silaev2015long,virtanen2016stimulated,bobkova2016injection,bobkova2015,krishtop2015-nst}. 
Such a mechanism does not require injector electrode to be  spin-polarized. Thus the spin accumulation can be generated  by an injected
current even from a non-ferromagnetic electrode. 
This can be seen  from the second term in the  r.h.s. of  Eq.~(\ref{Eq:ChPotZ}),  which shows that  $\mu_z$ can be non-zero  
 even if the injection occurs from a non-magnetic lead, provided that  $N_-\neq 0$, i.e., if the superconductor 
  shows a spin-split spectrum \cite{PhysRevB.87.024517}.
  
  In addition, with the help of the qualitative picture described above one can understand the  antisymmetric shape of the non-local spin signal in $g_{nl}$ with respect to $V_{\rm inj}$ observed in the experiments \cite{Quay2013,Hubler2012a,PhysRevB.87.024517}. 
  The origin of such a  $g_{nl}(V_{\rm inj})$ dependence is again the thermo-spin effect 
  in the superconductor with $N_-\neq 0$.
  The spin accumulation generated in this way is an even function of the bias voltage  at the injector electrode $\mu_z(V_{\rm inj})=\mu_z(-V_{\rm inj})$ 
  because the effective  heating of quasiparticles  is not sensitive to the sign of  
   $V_{\rm inj}$. 
  Hence the non-local spin signal 
  (\ref{eq:gnl0}),
that is a derivative of $\mu_z$, is an odd function  $ g_{nl}( V_{\rm inj}) = - g_{nl} (- V_{\rm inj}) $. This is also what has  been observed in the experiments  \cite{wolf2014charge}.

\subsubsection{Relaxation of the nonequilibrium modes}
 To provide  a  quantitative description of the effects discussed in the previous section, we now  calculate the potentials  $\mu$,  $\mu_z$ and the non-local conductance 
  by using  the kinetic equations for superconductors with 
   spin-split subbands introduced  in Sec.~\ref{sec:Kin_eqs}. 

 Our starting point is   the general Usadel equation (\ref{eq:Usadel}).  
 By assuming that the  transparencies of the interfaces between the superconductor and the electrodes are small we 
  can neglect the changes of the spectral properties of the superconductor due to the proximity effect. 
  This means that the retarded and advanced  GFs correspond to the homogeneous  superconducting state ({\it cf.} Sec.~\ref{sec-superwithh}), 
  and we only have to focus  on the calculation of the components of the distribution function, Eq.~(\ref{eq:distribution_function}). In this discussion we disregard inelastic relaxation, assuming the corresponding scattering length to far exceed the spin relaxation length. We thus set $\check \Sigma_{in}=0$.

 In the collinear situation considered here  only the four components of the distribution function entering Eqs.~(\ref{Eq:ChPot0}-\ref{Eq:ChPotZ}) are finite. 
 In a homogeneous superconductor, for example in the absence of the supercurrent they are pairwise coupled.
The expressions for different currents can be obtained by combining Eqs.~(\ref{eq:K_param},\ref{eq:distribution_function},\ref{eq:kineticeq},\ref{eq:general_current}).
It is instructive to represent these expressions in a matrix form 
\begin{equation} \label{Eq:CurrentsNoPhaseGradient}
    \begin{pmatrix} {\bm j_e} \\ {\bm j_s} \\ {\bm j_c} \\ {\bm j_{se}} \end{pmatrix} = \begin{pmatrix} D_L \nabla & D_{T3}
 \nabla & 0 & 0 \\ D_{T3} \nabla & D_L \nabla & 0 & 0 \\ 0 & 0 & D_T \nabla & D_{L3} \nabla \\
 0 & 0 & D_{L3} \nabla & D_T \nabla \end{pmatrix} \begin{pmatrix} f_L \\ f_{T3} \\ f_{T} \\ f_{L3} \end{pmatrix}.\end{equation}
Here, following the notation in Eq.~\eqref{eq:kineticeq}, ${\bm j_c} \cdot \bm e_k = D j_{k3}^0$, ${\bm j_{se}} \cdot \bm e_k = D j_{k3}^3$, ${\bm j_e} \cdot \bm e_k = D j_{k0}^0$ and ${\bm j_s} \cdot \bm e_k = Dj_{k0}^3$ 
 where $\bm e_k$ is the unit vector in direction $k=x,y,z$. The {components $f_L$ and $f_{T3}$ determine the spectral energy ${\bm {j_e}}$ and spin ${\bm j_s}$ currents, whereas the} components $f_T$ and $f_{L3}$ determine  the spectral charge ${\bm {j}_c}$ and spin-heat ${\bm {j}_{se}}$ currents 
 . 
  The spectral coefficients appearing here  are defined in terms of the spectral GFs discussed in Eq.~\eqref{eq:general_GR}, and the diffusion coefficient $D$
  \begin{align}
  & {\mathcal{D}}_{T} = \frac{D}{2}\left( 1+|g_{01}|^2 + |g_{03}|^2 + |g_{31}|^2 + |g_{33}|^2 \right)
  \\
  & {\mathcal{D}}_{L3} = D{\rm Re} \left(g_{03}g_{33}^* + g_{01}g_{31}^* \right) 
  \\
  & {\mathcal{D}}_{L} = \frac{D}{2}\left(1-|g_{01}|^2 + |g_{03}|^2 - |g_{31}|^2 + |g_{33}|^2 \right) 
  \\
 & {\mathcal{D}}_{T3} = D{\rm Re} 
 \left(g_{03}g_{33}^* - g_{01}g_{31}^*\right) .
   \end{align}
   According to Eq.~(\ref{eq:Usadel})  the modes $f_T$ and $f_{L3}$ satisfy the  diffusion equations
  \begin{eqnarray}\label{Eq:fTKinC}
  {\bm {\nabla\cdot j_c}} &=&
  R_{T} f_{T}  + R_{L3} f_{L3}  \\
  \label{Eq:fTKinCS}
  {\bm {\nabla\cdot j_{se}}}
  &=& (R_{T} + S_{L3}) f_{L3}  + R_{L3} f_{T}\;  ,
  \end{eqnarray}
  where $ R_{T}=2\Delta{\rm Re} g_{01}$,  $R_{L3}=2\Delta{\rm Re} g_{31}$ {describe the coupling of the quasiparticles to the superconducting condensate}, and the collision 
  integral {for spin relaxation} $S_{L3}$ is given by Eq.~(\ref{eq:SL3}).  {They are hence of the form of spectral current conservation equations, except for the source/sink terms provided by the collision integrals.}
  
  The kinetic equations (\ref{Eq:fTKinC},\ref{Eq:fTKinCS}) are supplemented by the boundary conditions at the injector electrode
  given by the second and the third rows on the matrix of Eq.~(\ref{Eq:bcDistrFunc}).  
     
\begin{figure}
  \centering
  \includegraphics[scale=0.9]{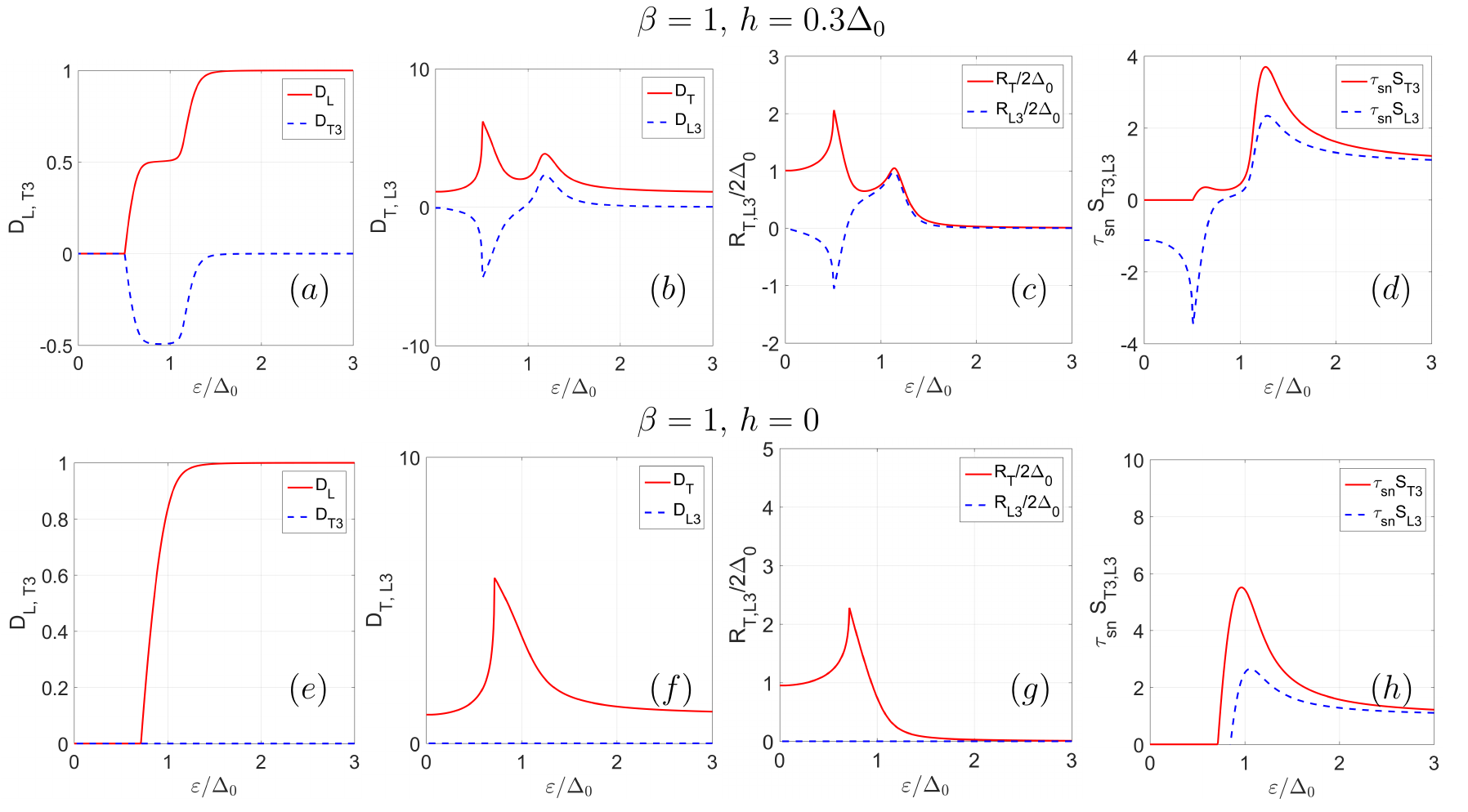} 
  \caption{\label{fig:DiffCoeff}
    Kinetic coefficients   in the presence of Zeeman splitting for  $\beta = 1$, $h=0.3\Delta_0$ (upper panel) and in the absence of Zeeman splitting $h=0$ for the same parameters  (lower panel).
 The  normal state spin relaxation rate $(\tau_{sn}\Delta_{0})^{-1} = 0.1 $, temperature  $T=0.1T_{c0}$.
 (a,e) Diffusion coefficients for energy and spin imbalance modes, (b,f) diffusion coefficients for spin energy and charge imbalance modes,
 (c,g) Andreev reflection-related coefficients $R_{T,L3}$, and (d,h)
 spin relaxation $S_{L3,T3}$ coefficients. 
  }
\end{figure}

 On the other  hand, according to Eq.(\ref{Eq:CurrentsNoPhaseGradient}) the second pair of modes,  $f_L,f_{T3}$, 
  determines the energy and  spin currents 
  ${\bm j_e} $ and ${\bm j_s}$, respectively.  
     They satisfy  the diffusion equations
    \begin{eqnarray}\label{Eq:fLKinE}
     {\bm {\nabla\cdot j_e}}  &=&  0\\
    \label{Eq:fLKinS}
     {\bm {\nabla\cdot j_s}} &=& S_{T3} f_{T3}.
 \end{eqnarray}
Here
 the spin-relaxation term $S_{T3}$ is given by  Eq.~(\ref{eq:ST3}).  
   Typical energy dependencies of the kinetic coefficients in 
    Eqs.(\ref{Eq:fTKinC}, \ref{Eq:fTKinCS}, \ref{Eq:fLKinE}) in the presence of spin splitting and spin-flip relaxation ($\beta=1$) are shown in Fig.~\ref{fig:DiffCoeff}.

 \paragraph*{Charge and spin-energy modes}
 
  The two systems of diffusion equations (\ref{Eq:fTKinC}-\ref{Eq:fTKinCS},\ref{Eq:fLKinE}-\ref{Eq:fLKinS})  
 describe the coupled transport of spin, charge and heat in superconductors.  

\begin{figure}
  \centering
 \includegraphics[width=8cm]{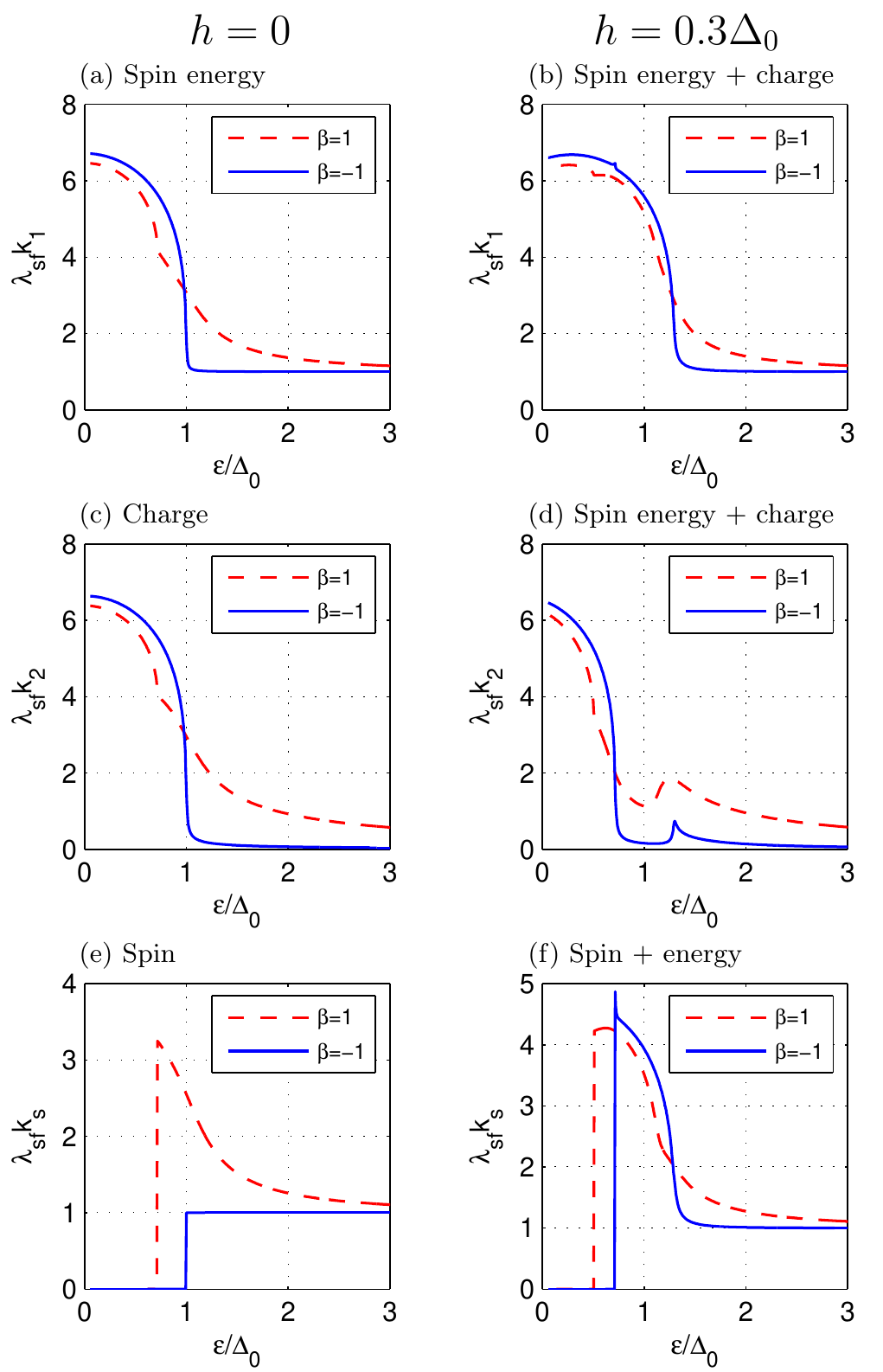}
 \caption{\label{Fig:ScalesCollinear}
 Energy dependence of the inverse length scales $k_{1}$, $k_{2}$ and
 $k_{s}$ for $h=0$ (left panels) and $h=0.3 \Delta_{0}$ (right panels). In each plot 
 $\beta = -1; 1$ corresponding to dominating spin-orbit and spin-flip scattering, respectively. The temperature is $T=0.1 T_{c0}$, 
 normal-state spin relaxation rate $(\tau_{sn}\Delta_{0})^{-1} = 0.1 $.  
  }
 \end{figure}

   We start by solving the system  (\ref{Eq:fTKinC},\ref{Eq:fTKinCS}) describing the 
 coupled charge and spin energy modes. The solution 
 for $f_T$ and  $f_{L3}$ is a superposition of two exponentially decaying functions
  \begin{equation}\label{Eq:SolFT}
\begin{pmatrix} f_T \\ f_{L3} \end{pmatrix}
  = \sum_{j=1,2}
  A_{j} e^{-k_{j}x} 
 \begin{pmatrix}
  R_{T} - {\mathcal{D}}_{L} k_{j}^2   
  \\   
  {\mathcal{D}}_{T} k_{j}^2-R_{L3} 
  \end{pmatrix},  
 \end{equation}
 where the coefficients $A_{i}$ have to be determined from the boundary conditions. 
 The inverse lengths $k_{1,2}$ are energy dependent. This dependence is shown in Figs.~\ref{Fig:ScalesCollinear}(a-d).
 Notice that both the charge and the spin-heat imbalance relaxation
 are nonvanishing for all energies, below and above the gap,  due to the magnetic pair breaking effects \cite{schmid1975,nielsen1982pair}.  
 In the absence of spin splitting, $h=0$, $R_{L3}={\cal D}_{L3}=0$, and the charge and spin-energy modes are  decoupled [{\it cf.} Eqs. (\ref{Eq:fTKinC},\ref{Eq:fTKinCS})].  Then 
 $k_{1}$ and $k_{2}$ can be ascribed to the spin energy and charge imbalance relaxation, respectively.
  The non-zero Zeeman splitting $h\neq 0$ leads to the coupling  of $f_T$ and $f_{L3}$,  at low energies.   At high energies $\varepsilon \gg \Delta$ though,  when the superconducting correlations become negligible,  the decoupled behavior is restored. This can be seen in the asymptotic behavior for $\varepsilon\rightarrow\infty$  shown in Figs.~\ref{Fig:ScalesCollinear}(b,d). In this limit   $k_{2}\to 0$  corresponding  to a vanishing charge relaxation, whereas   $k_{1}\to \lambda_{sn}^{-1}$  which is the spin-energy relaxation length in  the normal state.

\paragraph*{Spin and energy modes}   
 We now analyze the system of equations (\ref{Eq:fLKinE}-\ref{Eq:fLKinS}) that describes the coupling between the  spin and energy modes, i.e., the components $f_{L}$ and $f_{T3}$ of the distribution function.
 The solution of the system  can be written as the sum 
of two qualitatively different terms
 \begin{equation}\label{Eq:SolFL}
 \begin{pmatrix} f_L\\ f_{T3} \end{pmatrix} =
 B_1\begin{pmatrix} {\mathcal{D}}_{T3} \\ - {\mathcal{D}}_L \end{pmatrix}
 e^{-k_{s}x}+
 B_2 \begin{pmatrix} x-L \\ 0 \end{pmatrix} + \begin{pmatrix} f_{\rm eq}\\0\end{pmatrix},
 \end{equation}
 with coefficients $B_j$ determined by the boundary conditions. 
  The first term in (\ref{Eq:SolFL})
 describes a decay of the (spectral) spin imbalance with a
 characteristic length scale
 $ k_{s}= \sqrt{S_{T3}{\mathcal{D}}_{L}/ ( {\mathcal{D}}_{L}^2-{\mathcal{D}}_{T3}^2)}$.
 In the absence of a spin-splitting field and depairing
 mechanisms the expression for $k_{se}$ yields  the energy dependent spin-relaxation length in superconductors  obtained in Ref.~\cite{morten2005spin}: 
 $k^{-1}_{s} = \lambda_{sn}\sqrt{\varepsilon^2-\Delta^2 }/\sqrt{\varepsilon^2 +\beta \Delta^2}$ .  This length is strongly renormalized by superconductivity if the  spin-flip scattering rate is nonzero, $\beta\neq -1$.  A more accurate calculation of $k_{se} (\varepsilon)$, including the modification of spectral functions in the superconductor due to the spin relaxation, is  shown in Fig.~\ref{Fig:ScalesCollinear}(c). We see that  $k_s$ is constant for $\beta= -1 $ and has a pronounced peak near the self-consistent gap edge for $\beta = 1$. Note that the spectral gap is reduced by spin-flip scattering to the values smaller than $\Delta_0$. 
 Comparing Figs.~\ref{Fig:ScalesCollinear}(c,e) one can see that the 
relaxation lengths of charge imbalance and spin accumulation are very different even in the absence of a spin-splitting field.   For example at subgap energies   $k_{s}<k_{2}$.
In contrast,  for quasiparticles with energies
$\varepsilon\gg \Delta$ the charge relaxation vanishes whereas the spin relaxation remains non-zero.  Hence particle injection at voltages $V_{inj}$  well above the gap   leads at large distances from the injector mainly to a charge imbalance.
 
It is worth noticing that in  the presence of a non-zero spin-splitting field
 the behavior of $k_{s} (\varepsilon)$  is very similar for
 spin-orbit and spin-flip relaxation mechanisms as shown in  Fig.~\ref{Fig:ScalesCollinear}(d).
 
The second term in the r.h.s. of  Eq.~(\ref{Eq:SolFL}). It  describes an effective increase of the quasiparticle temperature
 associated with the $f_L$ nonequilibrium mode. This mode  can only decay  via
inelastic scattering which is  disregarded in the above analysis. In the following we assume the presence of an electrode at  a distance $L$ much larger than the spin relaxation length, and  assume that  the energy mode relaxes there.

\subsubsection{Kinetic equations with supercurrent}
{
 The expressions for the current in Eq.~(\ref{Eq:CurrentsNoPhaseGradient}) and hence the kinetic equations can be modified by driving a supercurrent through the superconducting wire. Its effect can be included by assuming a constant phase gradient $\nabla \phi$ of the order parameter \cite{aikebaier2018}, i.e., $\Delta({\bf r})=|\Delta|e^{i\phi({\bf r})}$. As a result, two new spectral coefficients appear: the spectral supercurrent $j_E$ and the spectral spin supercurrent $j_{Es}$, defined as $j_E = D {\rm Tr}[\tau_3 (\hat g^R \nabla \hat g^R-\hat g^A \nabla \hat g^A)]/(8\nabla \phi)$ and $j_{Es} = D {\rm Tr}[\tau_3 \sigma_3 (\hat g^R \nabla \hat g^R-\hat g^A \nabla \hat g^A)]/(8\nabla \phi)$, respectively. In their presence all distribution function components are coupled so that the spectral currents appearing in Eqs.~(\ref{Eq:fTKinC},\ref{Eq:fTKinCS},\ref{Eq:fLKinE},\ref{Eq:fLKinS}) read 
\begin{equation}
    \begin{pmatrix} {\bm j_e} \\ {\bm j_s} \\ {\bm j_c} \\ {\bm j_{se}} \end{pmatrix} = \begin{pmatrix} D_L \nabla & D_{T3}
 \nabla & j_E \nabla \phi & j_{Es} \nabla \phi \\ D_{T3} \nabla & D_L \nabla & j_{Es} \nabla \phi & j_E \nabla \phi \\ j_E \nabla \phi & j_{Es} \nabla \phi & D_T \nabla & D_{L3} \nabla \\
 j_{Es} \nabla \phi & j_E \nabla \phi & D_{L3} \nabla & D_T \nabla \end{pmatrix} \begin{pmatrix} f_L \\ f_{T3} \\ f_{T} \\ f_{L3} \end{pmatrix}.\end{equation}
Hence in this case the presence of spin-splitting and supercurrent leads to a coupling between all nonequilibrium modes. 
This is in contrast to the case without supercurrent when the matrix in Eq.~(\ref{Eq:CurrentsNoPhaseGradient}) has a block-diagonal form so that the modes are coupled only pairwise. 

In the absence of spin splitting, this leads to a coupling between charge and energy modes, and hence for example to the possibility of creating charge imbalance from temperature gradients \cite{schmid1979,clarke79,pethick79}. For a spin-split superconductor, the supercurrent couples charge and spin modes \cite{aikebaier2018} so that one can be converted into another one. In practice this conversion can be measured by analysing the different symmetry components of the non-local conductance with respect to the injection voltage and the spin polarization of the detector. 

Strictly speaking, the assumption of a constant phase gradient together with the excitation of the nonequilibrium modes leads to a non-conserved total (energy integrated) charge current in the setup. Therefore, the non-linear problem of analysing the dynamics of modes for supercurrent needs to be coupled with the self-consistency equation \eqref{eq:gapeq}, where the presence of the modes $f_T$ and $f_{L3}$ affects the phase of the self-consistent order parameter and therefore the phase gradient. }

\begin{figure}
    \centering
    \includegraphics[width=8cm]{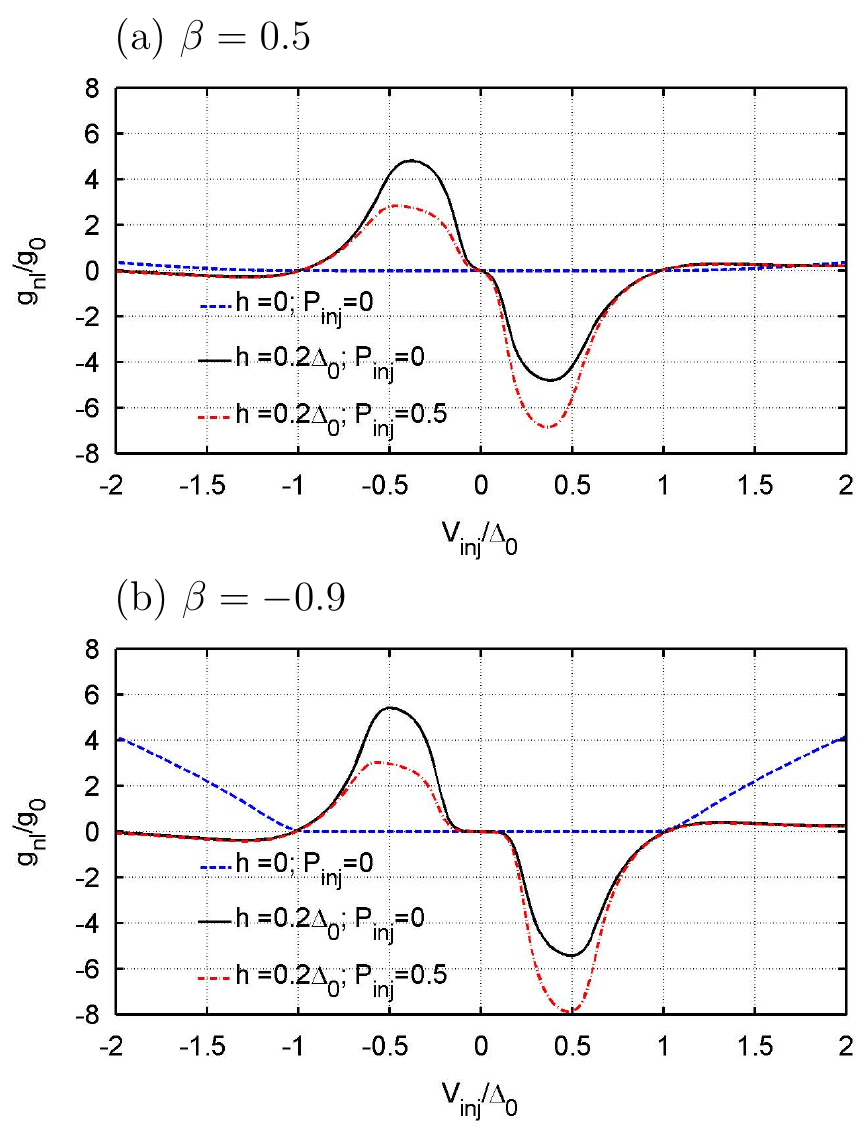}
 \caption{\label{Fig:Cond-nl}  
  Nonlocal conductance as a function of the injecting voltage,
 $g_{nl}(V_{\rm inj})$ for $\alpha_{orb}=1.33$, spin relaxation rate 
 $(\tau_{sn} T_{c0})^{-1} =0.2$ and  $T=0.05\; T_{c0}$,
 effective inelastic relaxation length $L=20\lambda_{sn}$, $L_{\rm det}=5\lambda_{sn}$ and detector polarization $P_{\rm det}=-0.5$.
 Spin relaxation mechanism is 
  (a) spin-flip dominated $\beta=0.5$; (b) spin-orbit dominated
  $\beta=-0.9$. The conductance is normalized to 
  $g_0 =  R_\xi/(R_{\rm inj}R_{\rm det})$,
  where $R_\xi = \xi/(A_s \sigma_N)$ is the normal-state resistance of the wire with length $\xi$ and cross section $A_s$. 
 }
 \end{figure}

\subsubsection{ Non-local conductance measurements}

 In a typical experiment   quasiparticles are injected from a  normal or a ferromagnetic  metal electrode  by applying a voltage $V_{\rm inj}$ between  the electrode and the superconductor.  In this way 
the   nonequilibrium states described by Eqs.~(\ref{Eq:SolFT},\ref{Eq:SolFL})
can be excited.  We determine the coefficients 
  $A_{1,2}$ and $B_{1,2}$  
 using the general boundary conditions (\ref{Eq:bcDistrFunc}) for the currents,
 where the voltage-biased normal electrodes are described by the distribution functions
 $f^{(N)}_T(\varepsilon) = [f_{\rm eq}(\varepsilon +V) - f_{\rm eq}(\varepsilon -V)]/2 $
 and $f^{(N)}_L(\varepsilon) =  [f_{\rm eq}(\varepsilon +V) + f_{\rm eq}(\varepsilon +V)]/2$
 while the remaining components are zero, $f^{(N)}_{L3} = 0$ and $f^{(N)}_{T3} = 0$.
 
 The current at the detector, Eq.~(\ref{Eq:ZeroCurrentYGen}), is  obtained by substitution of Eqs.~(\ref{Eq:SolFT},\ref{Eq:SolFL}) into the 
   expressions (\ref{Eq:ChPot0},\ref{Eq:ChPotZ}) for the chemical potentials.    
Finally we  obtain the  non-local conductance from Eq.~(\ref{eq:gnl0}). The result for $g_{nl}$ as a function of the injecting voltage $V_{\rm inj}$ is shown  in    Fig.~\ref{Fig:Cond-nl}. It reproduces the main features observed in  experiments in Refs.~\cite{Quay2013,Hubler2012a,wolf2014charge}.

We plot the non-local conductance in three different cases (the three curves in Fig.~\ref{Fig:Cond-nl}). 
%
%
The blue curve is for a vanishing spin-splitting field and injector polarization. In that case the signal is solely due to charge imbalance. The black curve is with a non-zero spin-splitting field, but vanishing injector polarization, and in the red curve both the spin-splitting field and the injector polarization are non-zero. As discussed below, the former shows the thermally created spin accumulation, whereas the latter shows a combination of the thermally and electrically created spin accumulations. 

In an experiment where the spin-splitting field is controlled by an external magnetic field, we also need to take into account the orbital depairing effect [Eq.~\eqref{eq:SE-ORB}] of the magnetic field in addition to the Zeeman effect. Whereas the latter leads to the coupling of the energy and spin modes, the former affects especially the relaxation of charge imbalance. For the results in Fig.~\ref{Fig:Cond-nl}, we describe the relative strength of the orbital depairing and the spin-splitting field by setting $(\tau_{\rm orb} T_{c0})^{-1}=\alpha_{\rm orb} (\mu_B B/T_{c0})^2$ and choosing for the dimensionless quantity $\alpha_{\rm orb}$ a value close to the experiments in \cite{Hubler2012a}.

First, in the absence of a Zeeman field, $N_-=0$,
and  according to Eqs.~(\ref{Eq:ChPotZ},\ref{eq:gnl0}) 
only the modes $f_T$ and $f_{T3}$ contribute to $g_{nl}$. In such a case 
the contribution stemming from the spin accumulation is nonzero only if
$P_{\rm inj}\neq 0$, which is the condition to obtain a nonvanishing
$f_{T3}$. However, this function decays over the  spin
diffusion length and therefore is negligibly small at the
distances  $L_{\rm det}>\lambda_{sn}=\sqrt{D\tau_{sn}}$ from the injector. 
Thus, the
detected signal in this case is mostly determined  by the charge imbalance $\mu$.
 This explains the approximate symmetry with respect to
the injecting voltage: $g_{nl} (V_{\rm inj})= g_{nl} (- V_{\rm inj})$ (blue curves in Fig.~\ref{Fig:Cond-nl}). The
charge imbalance contribution to  $g_{nl}$ grows monotonically
when $|eV_{\rm inj}|>\Delta_g$. This behavior is determined by the
increase of the charge relaxation scale at high energies, 
$k^{-1}_{2}\rightarrow\infty$ shown in Fig.~\ref{Fig:ScalesCollinear}(c).

 In the presence of an applied  magnetic field the 
 charge relaxation is strongly enhanced due to the orbital depairing.
 As a result, the charge imbalance background signal is strongly suppressed by an increased $h$. On the other hand the spin imbalance
 contribution stemming from the $f_L$ mode is large. This term describes
 heat injection in the presence of a finite spin-splitting field $h$
 and has a long-range behavior. This contribution leads to the
 large  peaks in $g_{nl} (V_{\rm inj})$  shown in Fig.~\ref{Fig:Cond-nl}.
 The non-linear heating produced by quasiparticles injected at voltages exceeding the energy gap explains the large electric signal observed in the experiments \cite{Quay2013,Hubler2012a,wolf2014charge}.
 Notice that the peaks do not have exactly the same form so that $g_{nl}(V_{\rm inj})\neq -g_{nl}(-V_{\rm inj})$.  
 The deviation from the perfect antisymmetric form is due to the small but finite injector polarization $P_{\rm inj}$,  which modifies the boundary conditions for the spin current.
  
  One important feature shown in  Fig.~\ref{Fig:Cond-nl} is that the spin polarization peaks  exist always in the presence of spin splitting, $h\neq 0$, even if the injector electrode is not ferromagnetic,  $P_{\rm inj}=0$.  This  explains the non-local conductance measurements of Ref.~\cite{PhysRevB.87.024517} by using a normal metal as injector.

The results summarized in  Fig.~\ref{Fig:Cond-nl} can have  direct  applications  in the  field of spintronics.  On the one hand  spin accumulation can be created without using ferromagnets as injectors.  On the other hand, due to the coupling between the spin and energy mode, the spin  signal can be controlled over a very long range.  The only requirement for these two features to occur is  to use a superconductor  with a spin-split  DOS due to a Zeeman or an exchange field.

\subsection{Spin Hanle effect in normal metals and superconductors}
\label{sec:spinhanle}

In the previous section we study the spin injection and spin accumulation 
 in a superconductor assuming that the magnetization  of the 
 injector and the applied field are collinear. We now lift this assumption and consider non-collinear field and magnetizations.
This situation  has been widely
studied in the normal state. The non-collinearity between the external field and the spin of the injected electrons 
leads to a precession of the latter. This is  the spin Hanle effect \cite{PhysRevLett.55.1790,jedema:345,jedema2003spin,Villamor2015,yang2008giant}. 
 This precession can  be measured via the
non-local conductance in a multi-terminal sample as a function of the applied field. 

In order to understand the spin Hanle effect in superconductors it is instructive to first discuss
 the spin precession in a normal metal  within the formalism  described above. 
 We then  demonstrate that in the superconducting state  the spin precession can be  either  enhanced or suppressed, 
 depending on the spin relaxation mechanism.  

\subsubsection{Normal-metal non-local spin valve} \label{Sec:Normal_Metal_Spin_Valve}

In the normal state the Usadel equation can be  drastically simplified because  the retarded and advanced Green's functions do not depend  on energy and have the  simple form
$\hat{g}^{R(A)}=\pm{\tau}_3$. Thus the reduced density of states equals to unity according to  Eqs.~\eqref{eq:dosplus}--\eqref{eq:dosminus}.
It follows  from Eq.~(\ref{eq:K_param}) that  the Keldysh component is then proportional to the distribution matrix: $\hat{g}^{K}=2\tau_3\hat{f}$. 

 It is then  useful to integrate
Eq.~(\ref{eq:Usadel}) over energies and take the trace after multiplication
with the Pauli matrices. This results in a diffusion equation for
the nonequilibrium spin density $\bm{S}$ 
 \[
 \bm S=-\frac{\nu_F}{8}\int dE {\rm Tr} [\tau_{3} \bm\sigma \check{g}^{K} ]
 \]
 given by 
 \begin{eqnarray}
 \nonumber \\
 D\nabla^{2}S_{a}({\bf r})-i\left[{\bm\sigma}{\bm h} ,  {\bm \sigma} {\bm {S}} ({\bm r}) \right]^{a} 
 & = & \frac{S_{a}}{\tau_{sn}} \quad. \label{eq:Hanle_normal} 
 \end{eqnarray}
 
 This is the  spin diffusion equation \cite{zhang2000spin,van1987boundary,dyakonov2008spin} widely used in spintronics. Generalization of this equation in the presence of spin-orbit coupling has been done in several works \cite{mishchenko2004spin,mal2005spin,stanescu2007spin,duckheim2009dynamic}.
 Its right hand side describes the effective spin relaxation time defined as  $\tau^{-1}_{sn} = \tau^{-1}_{so} + \tau^{-1}_{sf}$. 
 The second term on the left hand side describes the  torque induced by the external field via 
 the spin splitting field. This torque leads to the spin Hanle effect, studied in the language of quasiclassics in Ref.~\cite{hernando2000conductance}.
 
 Let us consider a lateral spin valve as the one sketched in Fig.~\ref{Fig:SpinValveSketch}(b).  We assume that  a field is applied in the $x$-direction whereas the injector and detectors are polarized in the 
 $z$-direction. If the width and thickness of the N wire are smaller than the spin relaxation length the problem reduces to a quasi 1D geometry along the length of the wire.  In such a case Eq.~(\ref{eq:Hanle_normal}) leads to two linear coupled equations for the components $S_{y,z}$,
 \[
 \left(\begin{array}{cc}
 \partial_{xx}^{2}-\lambda_{sn}^{-2} & l_{h}^{2}\\
 -l_{h}^{2} & \partial_{xx}^{2}-\lambda_{sn}^{-2}
 \end{array}\right)
 \left(\begin{array}{c}
 S_{z}\\
 S_{y}
 \end{array}\right)=0\;,
 \]
with solutions 
\begin{eqnarray*}
S_{z} & = & Ae^{-\kappa x}+Be^{-\kappa^{*}x}\\
S_{y} & = & -iAe^{-\kappa x}+iBe^{-\kappa^{*}x}\;.
\end{eqnarray*}
 We have defined the spin diffusion length $\lambda_{sn}^{2}=D\tau_{sn}$,
the magnetic length $l_{h}^{2}=D/h$ and ${{\kappa}}=\lambda_{sn}^{-1}\sqrt{1+i(\lambda_{sn}/l_{h})^{2}}$.
The coefficients $A$ and $B$ have to be determined from the boundary condition, 
Eq.~(\ref{eq:bc_gamma}), which in this particular case has the simple
form 
\[
\left.\partial_{x}S_{z}\right|{}_{x=0}=\frac{\nu_FPV_{I}}{2R_{I}\sigma_{N}}.
\]

Here $V_{I}$ is the voltage across the injector/N interface and
$R_{I}$ the resistance per unit area of the barrier. The resulting
spin accumulation in the N wire is
\begin{eqnarray}
S_{z}(x) & = & -\frac{\nu_FPV_{I}}{2R_{I}\sigma_{N}}{\rm Re}\left[\frac{e^{-\kappa x}}{\kappa}\right]\label{eq:muz_normal}\\
S_{y}(x) & = & -\frac{\nu_FPV_{I}}{2R_{I}\sigma_{N}}{\rm Im}\left[\frac{e^{-\kappa x}}{\kappa}\right],\nonumber 
\end{eqnarray}
where $V_{I}$ and $R_{I}$ are the voltage drop and the resistance
of the injector, respectively. A ferromagnetic voltage detector at a distance $x=L$
from the injector detects the spin potential $\mu_{z}=S_{z}(L)/\nu_F$.
The resulting   $\mu_z(B)$ dependence,  described by Eq.~(\ref{eq:muz_normal}), coincides with the Hanle-shaped curves  measured  experimentally \cite{jedema:345,jedema2003spin}.  

\subsubsection{Hanle effect in  spin-split superconductors}
\label{sec:hanle-super}

If the  wire in Fig. 
\ref{Fig:SpinValveSketch}  is in the superconducting state
the non-local signal can change drastically as discussed in Ref.~\cite{silaev2015spin}.    In  contrast
to the normal-metal case,  the retarded and advanced GFs have  non-trivial  energy dependencies and therefore one cannot integrate straightforwardly the diffusion equation
over the energy.

As in previous sections, we assume that the retarded and advanced GFs are position independent and the density of states is the one shown in Fig.~\ref{fig:Calculated-density-of}. 
 In contrast to the normal case the non-local Hanle signal in a superconducting wire depends on the dominating spin relaxation mechanism.  We consider again the two mechanisms: magnetic impurities, Eq.~(\ref{eq:SE-SF}), and extrinsic spin-orbit coupling, 
Eq.~(\ref{eq:SE-SO}), and introduce again the parameter $\beta$ 
describing the relative strength between them 
$ \beta=(\tau_{so}-\tau_{sf})/(\tau_{so}+\tau_{sf})$.

In principle the  Hanle signal in the superconductor would include contributions from the long-range energy mode  
discussed in Sec.~\ref{sec:noneq} for the case of collinear field and magnetizations.  To separate that contribution from the  bare Hanle effect, we concentrate on the linear response regime of low injector and detector voltages, where the effect of the energy mode can be disregarded ({\it cf.} Fig.~\ref{Fig:Cond-nl}).   In such a case the non-local conductance is exponentially small {($\sim \exp(-\Delta/T)$)} and therefore we focus on the non-local resistance $V_{\rm det}/I_{\rm inj}$, where $V_{\rm det}$ is the voltage at the detector in the absence of a current and  $I_{\rm inj}$ is the current at the injector.
To uncover the Hanle effect, we furthermore study the difference $V_S$ between the voltages for the parallel and anti-parallel  orientations of the detector and injector  polarizations. The non-local resistance of interest is thus $R_S=V_S/I_{\rm inj}$. 

If we assume that the external field is applied in the $z$-direction
and disregard the orbital effects, the retarded quasiclassical
Green's function is described by only four nonzero components in Eq.~\eqref{eq:general_GR}. The
energy dependence of these components is shown in Fig.~\ref{fig:gR_Components_Hanle}.

Whereas the spectral terms remain constant in the  S region, the spin-polarized
current $I_{\rm inj}$ that flows through the interface with the injector
causes a spin accumulation characterized by the vector $\vec{\mu}_{s}$.
The current measured at the detector is then given by  \cite{silaev2015spin}  \begin{equation}
 I_{\rm det}=-\frac{V_{\rm det}Y}{R_{\rm det}}+\frac{\bm{\mu}_{s} \cdot \bm{P}_{\rm det}}{R_{\rm det}}\;,\label{eq:ID}
 \end{equation}
where
$R_{\rm det}=R_\square/A$ is the normal-state barrier resistance between the superconducting wire and the detector with cross sectional area $A$ and spin polarization $\bm{P}_{\rm det}$, and $V_{\rm det}$ is the voltage measured at the detector with respect to the superconductor. 
$Y=\int_{0}^{\infty}d\varepsilon N_{+}\partial_{\varepsilon}f_{\rm eq}$ describes the effect of the density of states in Eq.~\eqref{Eq:condG}. Its low-temperature analytical estimate for weak spin relaxation is given in Eq.~\eqref{eq:NFISconductance} below.

To find the voltage induced in the electrically open detector circuit we set $I_{\rm det}=0$ in Eq.~(\ref{eq:ID}). The spin-dependent part of the non-local resistance $R_S = [V_{\rm det}(\bm P_{\rm det}) - V_{\rm det}(-\bm P_{\rm det})]/I_{\rm inj}$ can be found  using the linear response relation for the injector current $I_{\rm inj}=V_{\rm inj}Y/R_{\rm inj}$ with the normal-state resistance $R_{\rm inj}$ of the injector:
 \begin{equation}
 R_{S}=\frac{2R_{\rm inj}}{V_{\rm inj}Y^{2}}\bm{\mu}_{s}\cdot \bm{P}_{\rm det}\;,\label{hanle_rs}
 \end{equation}
 where $V_{\rm inj}$ is the voltage applied at the injector in order to inject
the current. 

We then need to determine the spin accumulation $\vec{\mu}_{s}(x)$
induced along the S wire which can be written in terms of the Keldysh
component of the GF as
 \begin{equation}
 \bm{\mu_{s}}(x)=\int_{0}^{{{\infty}}}\bm{m}(\varepsilon,x)d\varepsilon\;,\label{eq:mu_S_hanle}
 \end{equation}
 with $\bm{m}(\varepsilon,x)={\rm Tr} (\tau_{3}\bm{S}g^{K})/8$. 
 For simplicity we assume that ${\bm P}_{\rm inj}\perp \bm h$ so that  the $f_L$ mode is not induced by the thermoelectric coupling at linear response, see Sec.~\ref{subs:noncollinear}. Therefore the spin accumulation does not contain the  long-range contribution discussed in Sec.~\ref{sec:noneq}.  
 The remaining part of  
 $\bm {m}$ in Eq.~(\ref{eq:mu_S_hanle}) is given by:
 \begin{equation} \label{Eq:SpectrumMagnetization}
 \bm m=N_{+}\bm{f}_{T}+\frac{{\rm Im}g_{33}}{h}(\bm{f}_{T}\times\bm{h}),
 \end{equation}  
where the vector $\bm f_T = (f_{T1},f_{T2},f_{T3})$ contains the   spin
 components of the general distribution function (\ref{eq:distribution_function}).
The first term in this expression is the usual quasiparticle contribution which  also appears in the normal state. It is determined by the nonequilibrium distribution function at  energies larger than the superconducting gap, that is where the total density of states is non-zero.  In contrast,  the second term in Eq.~(\ref{Eq:SpectrumMagnetization}) only appears in the superconducting case and is finite for subgap  energies due to the prefactor ${\rm Im} g_{33}$ ({\it cf.} Fig.~\ref{fig:gR_Components_Hanle}). 
    This term can be related to the  difference $N_-$ between the DOS in spin subbands using the Kramers-Kronig relation for the retarded GF ${\rm Im} g_{33} (\varepsilon)= - \mathcal{P} \int_{-\infty}^{\infty} \frac{N_-(\varepsilon^\prime)}{\varepsilon^\prime - \varepsilon} \frac{d\varepsilon^\prime}{\pi}$.
   

Let us assume that the spin-splitting field in the superconductor points in the 
 $z$-direction,  $ \bm {h} = h \bm {e}_{z} $.
 Then the transverse components $f_{T1},f_{T2}$ 
 satisfy the kinetic equations obtained from Eq.~(\ref{eq:Usadel}),
\begin{subequations}
\label{eq:transversekinetic}
 \begin{eqnarray} \label{Eq:KinfT12-1}
 {\mathcal{D}}_{T1}\nabla^2 f_{T1}+{\mathcal{D}}_{T2}\nabla^2 f_{T2} &=&  X_1 f_{T1} + X_2 f_{T2}
 \\ \label{Eq:KinfT12-2}
 {\mathcal{D}}_{T1}\nabla^2 f_{T2}-{\mathcal{D}}_{T2}\nabla^2 f_{T1} &=&  X_1 f_{T2} -X_2 f_{T1}. 
 \end{eqnarray}
\end{subequations}
 In contrast to the normal case, Eq.~(\ref{eq:Hanle_normal}), the diffusion coefficient is now a tensor with components that depend on energy,
 \begin{eqnarray} \label{Eq:Diffusion-1}
 {\mathcal{D}}_{T1} &=& D(1+|g_{03}|^2-|g_{01}|^2+|g_{31}|^2-|g_{33}|^2)
 \\ \label{Eq:Diffusion-2}
 {\mathcal{D}}_{T2} &=& 2D \; {\rm Im}\;\left(  g_{33}g^*_{03} - g_{31}g^*_{01}  \right)\;.
 \end{eqnarray}
The terms on the right hand side of Eq.~\eqref{eq:transversekinetic} are   $X_1=(S_{T1} - H_1)$,  $X_2=(S_{T2} + H_2)$, with
 \begin{eqnarray}\label{Eq:Hanle}
 H_1 &=& 4 h{\rm Im} g_{33},\;\;\; H_2= 4 h N_+ , \\ \label{Eq:S1}
 S_{T1} &=& 2\tau_{sn}^{-1}[ ({\rm Re} g_{03})^2 +  \beta({\rm Im} g_{01})^2],
 \\\label{Eq:S2}
 S_{T2} &=& 2\tau_{sn}^{-1} ( {\rm Im} g_{33}{\rm Re} g_{03} - \beta{\rm Im} g_{01}{\rm Re} g_{31}  ).
 \end{eqnarray}
 The $H_{1,2}$ terms are the "Hanle" terms that describe the coherent spin rotation and relaxation due to the action of the external field. 
 In the  normal case only $H_2$  and the first term of $S_{T1}$ are non-zero and we recover Eq.~(\ref{eq:Hanle_normal}).

 In order to solve Eqs. (\ref{Eq:KinfT12-1}-\ref{Eq:KinfT12-2}) we need the boundary conditions at the interface with the injector. 
They  can be   obtained from Eq.~(\ref{eq:bc_gamma}). By keeping terms to leading order in the interface parameter $\kappa_I=1/(R_{I\square}\sigma_N)$ we obtain \cite{silaev2015spin}
\begin{subequations}
 \begin{eqnarray} \label{Eq:BCfT121}
 {\mathcal{D}}_{T1}\nabla f_{T1}+ {\mathcal{D}}_{T2} \nabla f_{T2} &=& - 2\kappa_I {\rm Im} g_{33} P_In_-  \\
 \label{Eq:BCfT122}
 {\mathcal{D}}_{T1} \nabla f_{T2} -  {\mathcal{D}}_{T2}\nabla f_{T1} &=& - 2\kappa_I  N_+ P_I n_-  \;.
 \end{eqnarray}
\end{subequations}
These are evaluated at the position of the interface. 
 
\begin{figure}
 \centering
 \includegraphics[scale=0.75]{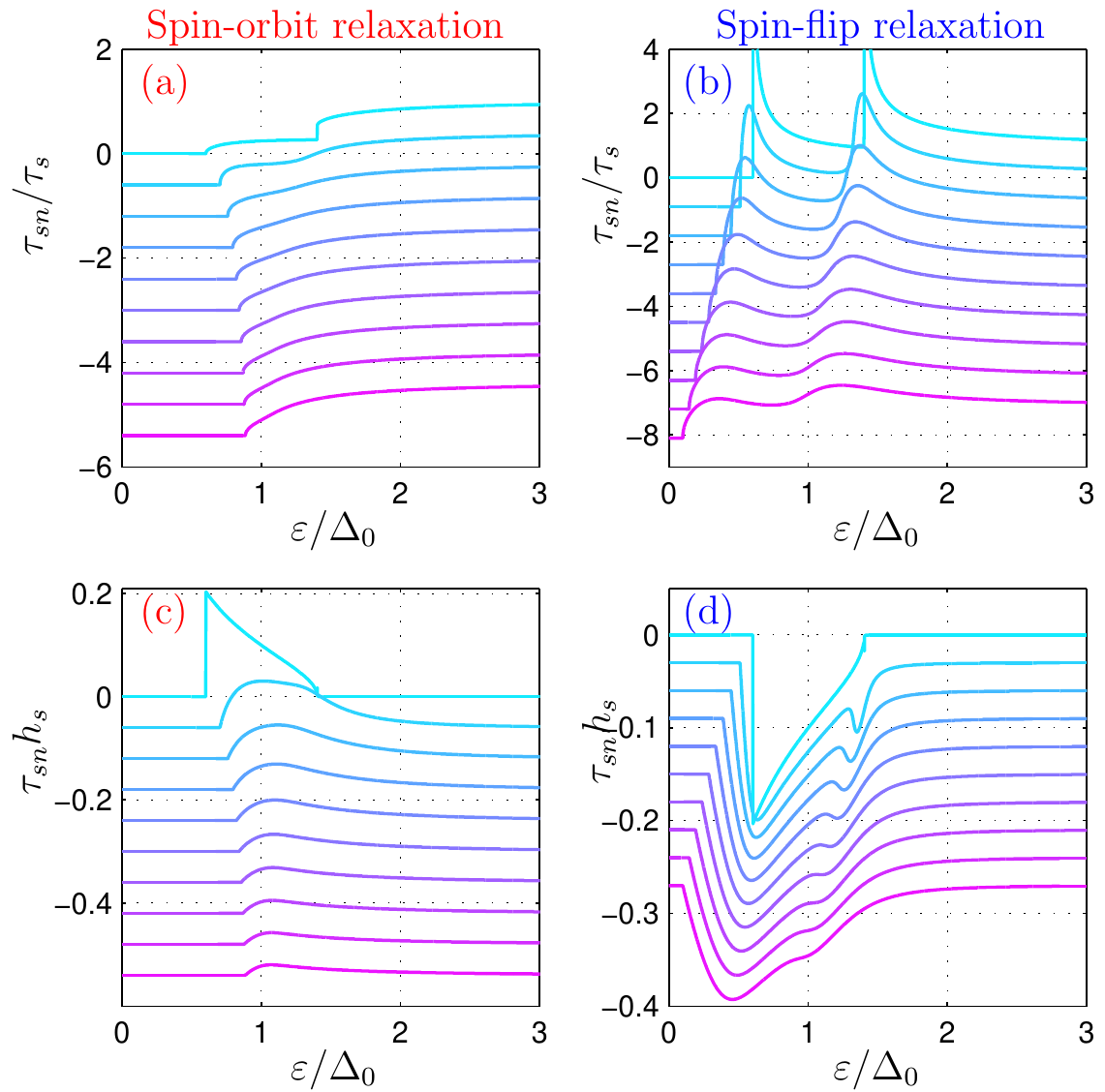}
 \caption{
  (a,b)  Transverse relaxation time of magnetic precession  Eq.~(\ref{eq:hanle-taus-hs}a) and (c,d) shift of the field determining the precession frequency, Eq.~(\ref{eq:hanle-taus-hs}b). For spin-orbit mediated spin relaxation (a,c) $\beta=-1$ the curves from top to bottom are for 
 $(T_{c0}\tau_{sn})^{-1}=0.001 - 30.001$  with steps of $3$.
 For spin-flip 
 relaxation $\beta=1$
 (b,d) the curves from top to bottom are for 
 $(T_{c0}\tau_{sn})^{-1}= 0.001 - 0.601$  with steps of $0.06$. The spin-splitting field is $h=0.4\Delta_0$ in all panels. 
  \label{fig:RelTime}}
 \end{figure}

 Equations (\ref{Eq:KinfT12-1},\ref{Eq:KinfT12-2}) 
 can be re-written for the spectral density of spin polarization, Eq.~\eqref{Eq:SpectrumMagnetization},  in a more familiar  form,  similar to the  Bloch-Torrey  transport equation \cite{Torrey1956}  for the magnetic moment:  
   \begin{align}
   \label{LL}
     \nabla_k\cdot \bm j_{sk} = \gamma{\bm m}\times  ({\bm h}+{\bm h}_s ) - {\bm m}/\tau_S, \\ \label{Eq:SpinCurrentNC}
    \bm j_{sk} = -{\mathcal{D}}_{\parallel} \nabla_k {\bm m} -  {\mathcal{D}}_{\perp} \nabla_k {\bm m} \times {\bm h}.
   \end{align}
  Here $\gamma=-2$ is the electron gyromagnetic ratio
  and $\bm j_{sk}$ is the spin current density in the $k$-th spatial direction.  
    The diffusion coefficients are defined as
   \begin{eqnarray}
   {\mathcal{D}}_{\parallel} &=& \frac{ {\mathcal{D}}_{T1}N_+ + {\mathcal{D}}_{T2} {\rm Im }g_{33} }{ N_+^2 + {\rm Im }g_{33}^2 } \\
   {\mathcal{D}}_{\perp} &=& \frac{ {\mathcal{D}}_{T2}N_+ - {\mathcal{D}}_{T1} {\rm Im }g_{33} }{ N_+^2 + {\rm Im }g_{33}^2 }.
   \end{eqnarray}
  Equations (\ref{LL}-\ref{Eq:SpinCurrentNC}) are written in the steady-state limit, which is enough  for the description of the spin Hanle effect. In the dynamical case, the magnetization depends explicitly on time ${\bm m} = {\bm m} (\varepsilon, t) $, and the  l.h.s. of Eq.~(\ref{LL}) contains  additional terms. When the spin dynamics is slow enough,  with a characteristic  frequency much smaller  than the superconducting energy gap, $\omega \ll \Delta$,
 the Bloch-Torrey equation in the superconducting  state can be written as\footnote{For the general case of arbitrary  large frequencies in the absence of gradients, see Eq.~\eqref{eq:usadeltd-llg}.} 
  %
  %
 \begin{equation}
   \label{LL-1}
   \frac{\partial \bm m}{\partial t} +  \nabla_k\cdot \bm j_{sk} = 
    \gamma{\bm m}\times 
    ({\bm h}+{\bm h}_s ) - {\bm m}/\tau_S.  \end{equation}
    In {Eqs.~(\ref{LL},\ref{LL-1}), $\tau_S$ and ${\bm h}_s$} are the transverse spin relaxation time and  a correction to the effective Zeeman field appearing in the presence of spin-relaxation processes. The latter determines the (Larmor) precession frequency of the spins. Specifically, 
    \begin{subequations}
    \label{eq:hanle-taus-hs}
  \begin{align}
  \tau^{-1}_S= \frac{ 2h(H_2S_{T1} + H_1S_{T2})}{H_1^2+H_2^2}, \\
  {\bm h}_s = {\bm h}\frac{H_2S_{T2} - H_1S_{T1}}{H_1^2+H_2^2}.
  \end{align}
  \end{subequations}
Both the transverse relaxation rate $\tau_S^{-1}$ and the effective field 
shift $h_s= |{\bm h}_s|$ are proportional to the 
normal-state spin relaxation rate 
$\tau_{sn}^{-1}$. In the superconducting state $\tau_S^{-1}$ and  $h_s$ depend differently on energy for different spin-relaxation mechanisms. The typical dependencies are shown in Fig.~\ref{fig:RelTime} for $\beta= \pm 1$.
One can see that $\tau_S^{-1} $
has a step-wise behavior as a function of energy  when the  spin-orbit relaxation dominates [Fig.~\ref{fig:RelTime}(a)]. In contrast, $\tau_S^{-1} (\varepsilon)$
shows  peaks near the spin-split gap edges for  spin-flip relaxation [Fig.~\ref{fig:RelTime}(b)]. The effective field shift $h_s$ has different signs for $\beta =\mp 1$ as shown in Fig.~\ref{fig:RelTime}(c,d).  
Interestingly, the transverse spin relaxation $\tau^{-1}_S$ vanishes at the subgap energies.

Equations (\ref{Eq:KinfT12-1}-\ref{Eq:KinfT12-2}) imply that the transverse components of  ${\bm f}_T$ have the general form 
  \begin{align}
 f_{T1} = -  {\rm Im}( A e^{- k_{T} z } ) \\
 f_{T2} =   {\rm Re}( A e^{- k_{T}z } ),
 \end{align}  
 where $A$ is an    integration constant determined by the boundary conditions, and
 \begin{equation}\label{Eq:kT}
 k_T= \left[\frac{(S_{T1}-H_1)-i(S_{T2}+H_2)}{D({\mathcal{D}}_{T1}-i{\mathcal{D}}_{T2})}\right]^{1/2}
 \end{equation}
   with ${\rm Re} k_T>0$. 
   { In contrast to the normal metal case considered in Sec.~\ref{Sec:Normal_Metal_Spin_Valve}, there is no straightforward relation between $k_T$ and the parameters $\tau_S$ and $h_s$. This is because the diffusion coefficients in the superconducting state entering Eq.~(\ref{Eq:SpinCurrentNC}) are energy dependent and hence the spectral spin current  has a complicated expression that  couples the different  components of the spin polarization. }
   
   The real part 
   ${\rm Re} k_T$ determines the spin relaxation length and the imaginary part ${\rm Im} k_T$ gives the quasiparticle spin precession. In the superconducting state both these scales are  energy dependent.  
 \begin{figure}
  \centering
 \includegraphics[scale=0.24]{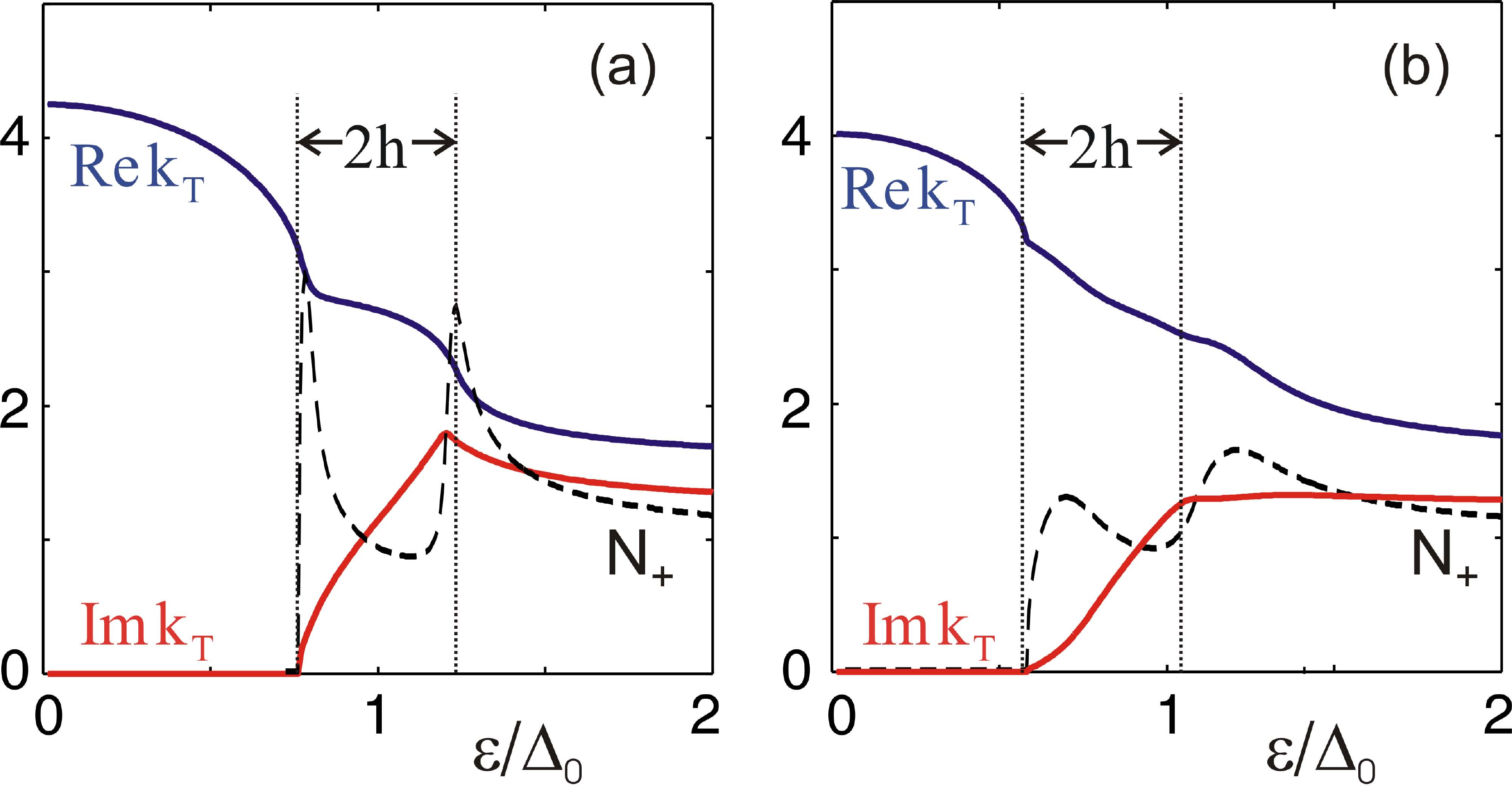}
 \caption{
 Energy dependencies of the inverse length scales that determine spin relaxation ${\rm Re} k_T$ (blue solid curve) and spin precession 
 ${\rm Im} k_T$ (red solid curve). The length scales are normalized by $\lambda_{sn}$. Dashed curves show the total DOS $N_{+}$.  (a) $\beta=-0.9$ (b) $\beta=0.5$. 
 Exchange field $h=0.22 \Delta_{0}$ and spin relaxation rate $(\tau_{sn}T_{c0})^{-1} = 0.2$.
 The precession ${\rm Im} k_T$ vanishes at subgap energies and $k_T (\varepsilon =0) \sim \xi_s^{-1}$ is given by the superconducting coherence length $\xi_s = \sqrt{D/\Delta_0}$.   \label{fig:kT}}
 \end{figure}
 In Fig.~\ref{fig:kT} we show how the precession and relaxation scales depend on energy and on the parameter $\beta$. 

It follows that at low energies ${\rm Re}\,k_T$ is larger than in the {normal-metal} case which corresponds to the limit $\varepsilon\gg \Delta$.  At $\varepsilon =0$ the characteristic relaxation length   $k^{-1}_T (0) $ is determined by the coherence length $\xi_s = \sqrt{D/\Delta_0}$ rather than by the normal-state spin relaxation length $\lambda_{sn}$.
 In contrast, the imaginary part of $k_T$ vanishes at the energies below the spectral gap where the DOS shown by the dashed lined in Fig.~\ref{fig:kT} is absent, i.e., $N_+ =0$. In the case of dominating spin-orbit scattering ($\beta<0$), the precession ${\rm Im} k_T$
  has a shallow peak at $\varepsilon\approx\Delta+h$ [see Fig.~\ref{fig:kT}(a)].
 If the spin-flip mechanism dominates  ($\beta>0$),
 this peak  of ${\rm Im}\, k_T$ is suppressed, Fig.~\ref{fig:kT}(b). 

\begin{figure}
    \centering
 \includegraphics[width=8cm]{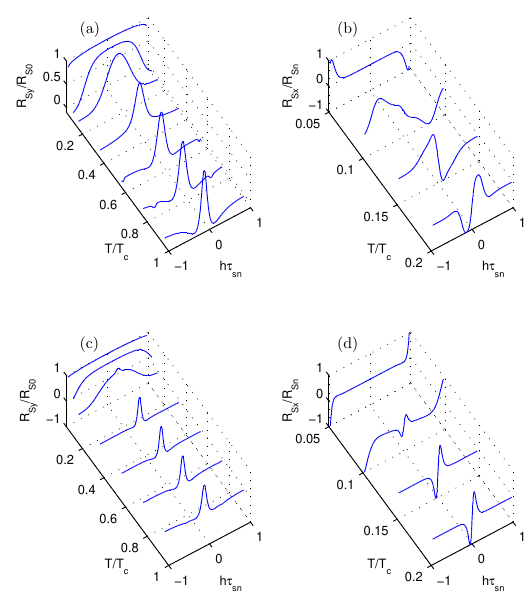}
 \caption{
   Hanle signals $R_{sy}$ and $R_{sx}$ at the distance $L_D=2\lambda_{sn}$ from the injector for 
 (a,b)  $\beta =0.5$ and (c,d) $\beta = - 0.9$. 
 The temperatures are $T/T_c=  0.05,  0.1,  0.15, 0.35, 0.55, 0.75, 0.95$ in panels (a,c) and
  $T/T_c=  0.05,  0.1,  0.15, 0.2$ in panels (b,d). 
  The resistance $R_{sy}$ is normalized by its maximum value obtained at $h=0$
 and $R_{sx}$ is normalized to the local normal state resistance.
 \label{fig:Hanle}}
 \end{figure}

Putting all this together allows us to evaluate the non-local resistance $R_S$ exhibiting the Hanle curves in the superconducting state as shown in Fig.~\ref{fig:Hanle}.  The data at $T=T_c$ corresponds to the normal-metal Hanle signal. The result depends on the relative directions of the injector and detector polarizations. 
 Panels a and c show  the non-local signal $R_{Sy}$ obtained from Eq.~(\ref{hanle_rs})
 when the detector polarization ${\bm P}_{\rm det}=P_{\rm det} {\bm e_y}$ so that 
 ${\bm P}_{\rm det} \parallel {\bm P}_{\rm inj} \perp {\bm h}$.  One can see that 
 both, the precession and decay of the nonlocal signal,  disappear at $T \rightarrow 0$, whereas the shape of the curves at 
intermediate temperatures  depends on the type  of  spin relaxation. 

If instead of the above  configuration we assume that
the three vectors $({\bm P}_{\rm det}, {\bm P}_{\rm inj},{\bm h} )$ are
perpendicular to each other (e.g., ${\bm P}_{\rm det} = P_{\rm det} {\bm e_x}$,
${\bm P}_{\rm inj} = P_{\rm inj} {\bm e_y}$, ${\bm h} = h {\bm e_z}$), the subgap
current is absent in the detector circuit and the corresponding
spin signal $R_{sx}$ has a strong dependence on $h$ even at the temperatures well below $T_c$. This is shown in  Figs.~\ref{fig:Hanle}(c,d).

{We are not aware of the measurement of the spin Hanle signals in the superconducting state. We emphasize that the above results are obtained  in the linear response regime of small voltages. They  disregard the thermal effects coming into play at higher voltages, especially at those of the order of the energy gap $\Delta \pm h$. In the linear response regime the transport quantities in superconductors are exponentially suppressed due to the gap in the density of states, and hence the above theory is rather made for the thermally activated quasiparticles. Because of this, the best way to uncover the superconducting effects on the Hanle response would be to study it rather close but below the critical temperature $T_c$.}

To summarize: The kinetic equations  for the nonequilibrium  modes constitute a rather generic theoretical  tool that we have used in this section to provide  a quantitative description of some  transport properties, such as spin injection and precession,  occurring in superconducting  structures in the presence of a spin-splitting field.  The predictive capability of the derived kinetic equations  has been proven in this section by contrasting  the results  with existing  experiments especially in the case of collinear magnetizations.  This  makes these equations  an  ideal tool to study further effects  that involve the coupling between  different   nonequilibrium modes in superconductors, as for example, the possible thermal spin Hanle {effect} and  other  non-linear effects taking place in systems with non-collinear magnetizations. 

In  section \ref{thermoel}  we  focus  on  thermoelectric effects  that are also direct manifestations of the spin-splitting fields in the superconductor. However, in contrast to the  section here  where the focus is on the diffusion of the nonequilibrium modes within the superconducting wire, section  \ref{thermoel} deals mainly with  interface effects, and disregards the position dependence of the nonequilibrium modes within the superconductors.
The next section discusses applications of the quasiclassical method to time-dependent nonequilibrium problems {and for example generalizes the Bloch-Torrey equation \eqref{LL} to large frequencies.
}

section{Non-equilibrium quasiparticle dynamics}

\label{sec:acdynamics}

The conduction electrons of a superconductor in an oscillating
electromagnetic field can absorb energy via excitation of the existing
quasiparticles, and by creation of new quasiparticles from breaking of
Cooper pairs.  In a stationary state, the excitation is balanced by
corresponding relaxation processes.  These  are sensitive to the
spectrum of states available in a spin-split superconductor.

\begin{figure}
  \centering
  \includegraphics{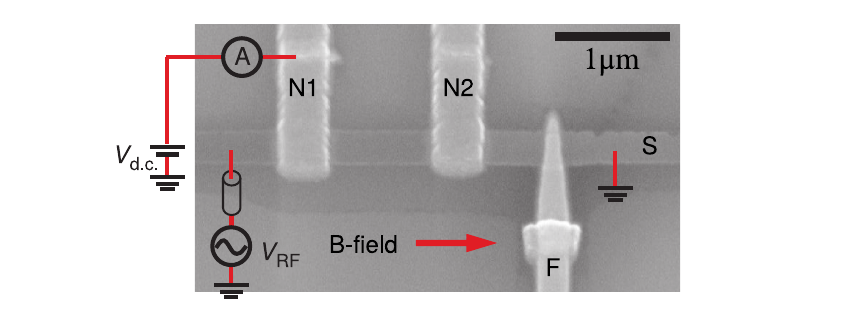}
  \caption{\label{fig:quay-expt}
    Experimental setup from Ref.~\cite{quay2015-qsr}. An applied
    RF field excites quasiparticle spin resonance in a thin Al film (S),
    which is indirectly probed by observing modification of the $I(V)$ curve
    of probe N1.
    Figure adapted from Ref.~\cite{quay2015-qsr}
    (Creative Commons Attribution 4.0 International License).
  }
\end{figure}

Electrons couple to the electromagnetic field via the Zeeman and the orbital terms.  The Zeeman coupling  is the  source for the conduction electron spin
resonance (CESR), which was theoretically considered in early works
\cite{aoi1970-tes,kaplan1965-ces,degennes1966-nsl,tsallis1977-ces,yafet1983-ces}. The corresponding
dynamical susceptibility for the spin-split case was discussed in Refs.~\cite{maki1973-tes,kosov1978-epr,tagirov1987-ske}.
The linewidth of the resonant absorption peak is determined by  the  spin relaxation time in the normal  state, 
$\tau_{\mathrm{sn}}$, and is not affected by the slow relaxation of
the thermal long-range spin imbalance discussed in Sec.~\ref{sec:noneq}. 
The magnetic
screening complicates the experimental observation of the resonance.  Results
however exist in the mixed state of type II superconductors as reported in Refs.~\cite{vier1983-oce,yafet1984-ces,nemes2000-ces}, and more recently,
also in spin-split Al thin films in Ref.~\cite{quay2015-qsr} [see Fig.~\ref{fig:quay-expt}]. For the Al
films, linewidths of $\unit[100]{ps}$ were observed, consistent with
the expected magnitude of $\tau_{sn}$ in aluminum.

The CESR physics is also related to pumping effects, where an
externally driven spin precession drives a nonequilibrium state or
currents in the superconductor, or currents across a junction. Effects of this type have been considered in different ferromagnet/superconductor structures in several works, in adiabatic \cite{brataas2004-scp,skadsem2011} and nonadiabatic cases \cite{houzet2008,richard2012-aci,trif2013-dme,holmqvist2011}.

As mentioned above, a rf field also couples to the orbital degree of freedom of the electrons. This coupling drives ac currents that also excite quasiparticles.  In the absence of spin splitting, the linear and nonlinear response of superconductors has been extensively  studied \cite{TinkhamBook}. In the linear response regime, the complex impedance for
spin-split superconducting films was considered theoretically and
experimentally in Ref.~\cite{bentum1986-fia} adding Zeeman energy shifts to BCS theory results \cite{mattis1958-toa,abrikosov1959-shf}.  In the nonlinear regime, strong driving can modify the observed quasiparticle spectrum \cite{linder2016-mcs}.  Moreover, {at temperatures close to $T_c$, exciting}
quasiparticles can lead to an increase of the superconducting
gap $\Delta$ \cite{eliashberg1970-fss,Eliashberg_enhacement,Klapwijk_enhancement}. In spin-split superconductors, this effect is modified by the spin-split
density of states and spin-flip scattering
\cite{virtanen2016stimulated}.

In this section  we discuss  the dynamic  response of spin-split superconductors  in terms of the time-dependent quasiclassical  equations, both in the linear and nonlinear regimes.  In addition to these effects, oscillating
fields can excite magnetic impurities and nuclear moments. Discussion
of such extrinsic resonance effects can be found in the articles \cite{barnes1981-tes,taylor1975-esr,baberschke1976-esr,garifullin2015-epr},
and are beyond the scope of this review. 
We also do not discuss
magnetization dynamics and how superconductivity affects it; reviews
on this active topic can be found in other
works \cite{tserkovnyak2005,hellman2017}.

\subsection{Linear response of a spin-split superconductor to a rf field}
\label{subs:linrespac}

The dynamic nonequilibrium response of spin-split dirty superconducting
thin films can be studied on the basis of the Usadel equation, Eq. (\ref{eq:Usadel}), which in  time-dependent  representation \cite{kopnin2001-ton} we write in a compact  form
\begin{equation}
  D\hat{\nabla}(\check{g}\circ\hat{\nabla}\check{g})
  =
  [\check{X}\overset{\circ}{,}\check{g}]
  \; ,\label{eq:UsadelT}
  \end{equation}
  where 
  \begin{equation}
  \check{X}
  =
  -i\epsilon\hat{\tau}_3 + i\phi + i\vec{h}\cdot\vec{\sigma}\hat{\tau}_3 + \check{\Delta} + \check{\Sigma}
 \; .
\end{equation}
Now $\check{g}(\omega,\omega')=\int_{-\infty}^\infty\dd{t}\dd{t'}e^{i\omega{}t-i\omega't'}\check{g}(t,t')$
depends on both time indices and matrix products involve
energy convolutions,
$(a\circ{}b)(\omega,\omega')=\int\frac{\dd{\omega_1}}{2\pi}a(\omega,\omega_1)b(\omega_1,\omega')$.
Above, $\epsilon(\omega,\omega')=2\pi\delta(\omega-\omega')\omega$,
and
$\hat{\nabla}X=\nabla X-i[\vec{A}\tau_3\overset{\circ}{,}X]$. External
classical electromagnetic fields couple to the electrons
via the exchange field $\vec{h}(\omega-\omega')$ and the vector
potential $\vec{A}(\omega-\omega')$. We choose the gauge such that the local charge neutrality \cite{gorkov1975-vmr,artemenko1979-efc} fixes the scalar
potential to $\phi(t)=\frac{\pi}{4}\Tr\check{g}^K(t,t)$, and in the
spatially uniform cases discussed below  $\phi=0$.  As above, we consider
thin films, and neglect magnetic screening.

The kinetic equation for the distribution function $\hat{f}$ follows
from Eq. (\ref{eq:UsadelT}) and the above assumptions,
\begin{equation}
  \label{eq:usadeltdkin}
  \frac{1}{8}D\hat{\nabla}[\hat{\nabla}\hat{f} - \hat{g}^R\circ(\hat{\nabla}f)\circ\hat{g}^A]
  +
  \frac{1}{8}\hat{j}^R\circ\hat{\nabla}f - \frac{1}{8}\hat{\nabla}f\circ\hat{j}^A
  =
  \hat{\mathcal{I}}
  \;,
  \end{equation}
where $\hat{j}^{R/A}\equiv{}D\hat{g}^{R/A}\circ\hat{\nabla}\hat{g}^{R/A}$ and the collision integral   is defined as
\begin{equation}
   \hat{\mathcal{I}}=\frac{1}{8}\left[\hat{g}^R\circ\hat{Z} - \hat{Z}\circ\hat{g}^A\right]
\end{equation}
and
\begin{equation}
  \label{eq:usadeltdZ}
  \hat{Z}
  =
  \hat{X}^R\circ\hat{f} - \hat{f}\circ\hat{X}^A - \hat{X}^K
  \,.
\end{equation}
Note that the definition of collision integrals in Eq.~\eqref{eq:usadeltdkin} differs from
Eq.~\eqref{eq:kineticeq}, as several terms in the kinetic equation have here been canceled
by making use of the R/A components of the Usadel equation.  In a spatially uniform case with
$\vec{A}=0$, the l.h.s of Eq.~\eqref{eq:usadeltdkin} vanishes and the
kinetic equation reduces to the condition $\hat {\mathcal{I}}=0$.

The CESR emerges when considering a Zeeman field of the form
$\vec{h}(\omega)=h_0\hat{z} + h_1(\omega)\hat{x}$, where an ac field
$h_1$ perpendicular to a static exchange field $h_0$ excites the
electron spins. The spin dynamics follows a
Bloch-like equation including spin relaxation, similar to the spin
Hanle effect (see Sec.~\ref{sec:spinhanle} and Eq.~\eqref{LL}). Such
equations can be derived from the Usadel equation in linear response
to $h_1$, and from other approaches
\cite{maki1973-tes,kosov1978-epr,tagirov1987-ske,inoue2017}.  Here we briefly
discuss the physics within the Usadel framework.

It is useful to separate the corrections to  the Keldysh  
Green's function in Eq.~\eqref{eq:UsadelT} 
due to the distribution function from those due to the  modification of the
spectral functions: $\delta\hat{g}^K=\delta\hat{g}^K_{\mathrm{reg}} +
\delta\hat{g}^K_{\mathrm{an}}$, defining
$\delta\hat{g}^K_{\mathrm{reg}}=\delta\hat{g}^R\circ{}f_{\mathrm{eq}}-f_{\mathrm{eq}}\circ\delta\hat{g}^A$.
We split $\delta\hat{X}^K$ similarly. The equation
$[\check{X}\overset{\circ}{,}\check{g}]^K=0$
then reduces in linear response to
\begin{gather}
  \label{eq:usadeltdankin}
  \hat{X}_{\mathrm{eq}}^R
  \circ
  \delta\hat{g}_{\mathrm{an}}^K
  -
  \delta\hat{g}_{\mathrm{an}}^K
  \circ
  \hat{X}_{\mathrm{eq}}^A
  =
  \hat{g}_{\mathrm{eq}}^R
  \circ
  \delta\hat{X}_{\mathrm{an}}^K
  -
  \delta\hat{X}_{\mathrm{an}}^K
  \circ
  \hat{g}_{\mathrm{eq}}^A
  \,,
  \\
  \begin{split}
  \delta\hat{X}_{\mathrm{an}}^K
  &=
  -i\sigma_x\tau_3[h_1\overset{\circ}{,}f_{\mathrm{eq}}]
  +
  \frac{1}{8\tau_{so}}\vec{\sigma}\delta\hat{g}^K_{an}\cdot\vec{\sigma}
  \\
  &+
  \frac{1}{8\tau_{sf}}\vec{\sigma}\hat{\tau_3}\delta\hat{g}^K_{an}\cdot\vec{\sigma}\hat{\tau_3}
  +
  \frac{1}{\tau_{orb}}\hat{\tau_3}\delta\hat{g}^K_{an}\hat{\tau_3}
  \,.
  \end{split}
\end{gather}
The solution to Eq.~\eqref{eq:usadeltdankin}
can be obtained with the Ansatz
$\delta\hat{g}_{\mathrm{an}}^K=\hat{g}_{\mathrm{eq}}^R\circ\delta{f}-\delta{f}\circ\hat{g}_{\mathrm{eq}}^A$, where 
$\delta{}f=f_{Tx}\sigma_x+f_{Ty}\sigma_y$. To determine the solution components,
we can take the trace $\frac{1}{8}\Tr[(\ldots)\vec{\sigma}]$ of the
$(\varepsilon,\varepsilon-\omega)$ frequency component of
Eq.~\eqref{eq:usadeltdankin}. This results  in
\begin{align}
  -i\omega\vec{m}
  &=
  -2\vec{m}\times{}h_0\hat{z}
  -
  \frac{N_+'\vec{m}+\beta{}F_+'\vec{m}'}{\tau_{sn}}
  +
  \vec{I}'
  \notag
  \\
  \label{eq:usadeltd-llg}
  &=
  -
  2\vec{m}\times(\vec{h}_0+\vec{h}_s)
  -
  \frac{\vec{m}}{\tau_S}
  +
  \vec{I}'
  \,,
\end{align}
where
$\vec{m}=\frac{1}{8}\Tr[\tau_3\vec{\sigma}\delta\hat{g}_{\mathrm{an}}^K(\varepsilon,\varepsilon-\omega)]$
describes the nonequilibrium magnetization (spin accumulation), 
$\vec{m}'=\frac{1}{8i}\Tr[\tau_1\vec{\sigma}\delta\hat{g}_{\mathrm{an}}^K(\varepsilon,\varepsilon-\omega)]$
describes a correction to spin scattering by superconductivity,
and $N_+'=[g_{03}(\varepsilon)+g_{03}(\varepsilon-\omega)^*]/2$,
$F_+'=[g_{01}(\varepsilon)-g_{01}(\varepsilon-\omega)^*]/(2i)$ are
finite-frequency generalizations of the spin-averaged density of
states and anomalous functions, respectively.  The term
$\vec{I}'=ih_1(\omega)[N_+'\hat{x}-N_z'\hat{y}][f_{\mathrm{eq}}(\varepsilon)-f_{\mathrm{eq}}(\varepsilon-\omega)]$
describes the exciting field. There is redundancy in the
representation, which enables writing $\vec{m}'$ in terms of
$\vec{m}$, resulting in the final Bloch equation. 
Analogously to the spin Hanle effect discussed in Sec.~\ref{sec:hanle-super}, the spin
relaxation time $\tau_S$ is renormalized with respect to the one in the normal state, $\tau_{sn}$, and there is also a  correction $\vec{h}_s$ to the  Zeeman field:
\begin{align}
  \frac{1}{\tau_S}
  &=
  \frac{N_+'}{\tau_{\mathrm{sn}}}
  +
  \beta{}\frac{F_+'}{\tau_{\rm sn}}
  \frac{F_+'N_+'-F_z'N_z'}{(N_+')^2+(N_z')^2}
  \\
  \vec{h}_s
  &=
  -\hat{z}\beta{}\frac{F_+'}{2\tau_{\rm sn}}
  \frac{F_+'N_z'+F_z'N_+'}{(N_+')^2+(N_z')^2}
  \,,
\end{align}
where $N_z'=[g_{33}(\varepsilon)-g_{33}(\varepsilon-\omega)^*]/(2i)$ and
$F_z'=[g_{31}(\varepsilon)+g_{31}(\varepsilon-\omega)^*]/2$.  For slow driving,
$\omega\to0$, the spin relaxation time $\tau_S$ and the Zeeman field
correction $h_s$ coincide with those visible in the Hanle effect
(cf. Eq.~\eqref{eq:hanle-taus-hs} and Fig.~\ref{fig:RelTime}), which
also involves precession of the transverse spin component.

The result~\eqref{eq:usadeltd-llg} describes resonant excitation of the transversal modes,
$f_{Tj}$, which correspond to a nonequilibrium contribution to the spin accumulation \eqref{eq:def_mus},
\begin{align}
  \label{eq:usadeltd-llg-mu}
  \delta\vec{\mu}_{s}(t)
  &=
  \delta\vec{\mu}_{s}^{\rm reg}(t)
  +
  \int_{-\infty}^\infty\frac{\dd{\varepsilon}\dd{\omega}}{2\pi}
  e^{-i\omega{}t}
  \vec{m}(\varepsilon,\varepsilon-\omega)
  \,,
\end{align}
where $\delta\vec{\mu}_{s}^{\rm{}reg}$ arises from the 
modification of the spectral functions,  $\delta{}g^{R/A}$. 
The result contains
the conduction electron spin resonance peak,
$\delta\vec{\mu}_s\propto{}h_1(\omega)/[4(h+h_s)^2 - (\omega +
i/\tau_S)^2]$ at frequency $\omega\simeq2|h_0|$.  As discussed in
Sec.~\ref{sec:noneq}, the $f_{Tj}$ modes can relax due to elastic
spin-flip scattering, which determines the peak absorption linewidth
$\frac{1}{\tau_S}\propto{}\tau_{\mathrm{sn}}^{-1}$. The result
in Ref.~\cite{yafet1983-ces} can be obtained from
Eq.~\eqref{eq:usadeltd-llg} in the quasiequilibrium approximation
$\delta{}f\simeq(\partial_Ef_{\mathrm{eq}})\vec{\mu}_s\cdot\vec{\sigma}$
and in the limit $h_0\ll\Delta$.  Equation \eqref{eq:usadeltd-llg}
also coincides with the result in Ref.~\cite{maki1973-tes}.

If we focus on the orbital effects, 
the complex impedance of the superconductor, \emph{i.e.}, the linear response to an
oscillating electric field described by $\vec{A}(\omega)$, can also be
obtained within the Usadel framework.  
In the above chosen (London) gauge, the
perturbation to time-averaged $\check{g}$ is of the order $D|\vec{A}(\omega)|^2/\omega$
\cite{eliashberg1970-fss} and can be neglected in linear response at
nonzero frequency. The charge current response is then given by
\begin{align}
  \notag
  \label{eq:jac1}
  j(\omega)
  &=
  \frac{\sigma_N}{16}\int_{-\infty}^\infty\dd{\varepsilon} \Tr\tau_3(\check{g}\circ[-i\vec{A}\tau_3\overset{\circ}{,}\check{g}])^K(\varepsilon,\varepsilon-\omega)
  \\
  &=
  -i\omega\vec{A}(\omega)
  [\sigma_1(\omega) - i\sigma_2(\omega)]
  \\
  \frac{
    \sigma_1(\omega)
  }{
    \sigma_N
  }
  &=
  \sum_{\sigma}\int_{-\infty}^\infty\dd{\varepsilon}2R_{\sigma}(\varepsilon,\varepsilon-\omega)
  \frac{f_L(\varepsilon-\omega)-f_L(\varepsilon)}{\omega}
  \,,
  \\
  \frac{
    \sigma_2(\omega)
  }{
    \sigma_N
  }
  &=
  \sum_{\sigma}\int_{-\infty}^\infty\dd{\varepsilon}2R'_{\sigma}(\varepsilon,\varepsilon-\omega)
  \frac{2f_L(\varepsilon-\omega) - 1}{\omega}
  \,,\label{eq:s21}
\end{align}
where
\begin{align}
  \label{eq:ac-abs-kernel}
  R_{\sigma}(\varepsilon, \varepsilon')
  &=
  N_\sigma(\varepsilon)N_\sigma(\varepsilon')
  +
  \Im g_{\sigma,1}(\varepsilon)\Im g_{\sigma,1}(\varepsilon')
  \,,
  \\
  R_{\sigma}'(\varepsilon, \varepsilon')
  &=
  \Im{}g_{\sigma,3}(\varepsilon) N_\sigma(\varepsilon')
  -
  \Re g_{\sigma,1}(\varepsilon) \Im g_{\sigma,1}(\varepsilon')
  \,,
\end{align}
and $g_{\uparrow/\downarrow,1}=(g_{01}\pm{}g_{31})/2$,
$N_{\uparrow/\downarrow}=\Re[g_{03}\pm{}g_{33}]/2$.  
The prefactors in
Eqs.~\eqref{eq:jac1}-\eqref{eq:s21} are given by the normal-state
conductivity, $\sigma_N=2e^2\nu_F D$, and the kernels
$R_\sigma$ have the BCS form \cite{mattis1958-toa,abrikosov1959-shf},
and are decoupled for each spin species.  In other words, the electric
field couples to the orbital motion of the electrons and, in our
approach, it conserves spin.
Therefore, the dissipative response is zero at frequencies $\omega<2\Delta$
in the absence of spin flipping.
Phenomenological modification of the above by allowing direct spin flips in
the coupling matrix element was considered in Ref.~\cite{bentum1986-fia}, 
obtaining a reduction in the dissipative pair-breaking threshold frequency
from $2\Delta$ to $2\Delta-2h$ (c.f. Fig.~\ref{fig:sf-schematic}).
Note however that the derivation of the Usadel diffusion equation
discussed above works in leading order in $\tau_{el}$ and neglects spin-orbit
effects when dealing with the vector potential.

\subsection{Nonlinear spin imbalance of  a spin-split superconductor in a rf field}
\label{sec:ninlin_ac}

We now  go beyond the linear response regime and focus on the
stationary quasiparticle distribution which enters the observables.
This  is determined by  disregarding the gradients, in which case
the  collision term in Eq.~\eqref{eq:usadeltdkin} is equal to the contribution
from the time-dependent vector potential. This results in
\begin{align}
  \label{eq:mw-kinetic-eq}
  \hat{\cal I}_{\rm ac}[\hat{f}] 
  + \hat{\cal I}_{\rm sn}[\hat{f}] 
  + \hat{\cal I}_{\rm in}[\hat{f}] 
  = 0
  \,,
\end{align}
This rate equation, similar to that used in Ref.~\cite{grimaldi1997}, describes the balance between
excitation of quasiparticles  induced by ac fields
($\hat{\cal I}_{\rm ac}$), the spin relaxation by the elastic
spin-flip scattering ($\hat{\cal I}_{\rm sn}$), and the inelastic
relaxation ($\hat{\cal I}_{\rm in}$).  Each term in
Eq.~\eqref{eq:mw-kinetic-eq} can be described within the
quasiclassical approach.

The orbital ac term of the collision integral can be obtained by
assuming a time-dependent spatially uniform vector potential
$A(t)=A_0\cos(\omega t)$. Here, it is useful to simplify the problem
by expanding in $DA_0^2/\omega\ll{}1$, and reduce it to a form
that only involves the dc component of the
distribution function. In this case
\cite{eliashberg1970-fss} 
\begin{align}
  \label{eq:Imw}
  {\cal I}_{\mathrm{ac},\sigma}(\varepsilon)
  &=
  DA_0^2
  \sum_\pm
  R_{\sigma}(\varepsilon, \varepsilon \pm \omega)
  [f_{\sigma}(\varepsilon) - f_{\sigma}(\varepsilon\pm\omega)]
  \,,
\end{align}
for the ${\cal I}_{\uparrow/\downarrow}=\Tr[(1\pm\sigma_z)\hat{\cal
    I}/2]$ components.  We assume that all spin dependent fields are collinear and hence we can define  $f_{\sigma=\uparrow/\downarrow}=f_L\pm{}f_{T3}$, cf.
Eq.~\eqref{eq:distribution_function}. Notice that the
charged modes $f_T$, $f_{L3}$ are not excited by a uniform $A(t)$. 

\begin{figure}
  \centering
  \includegraphics{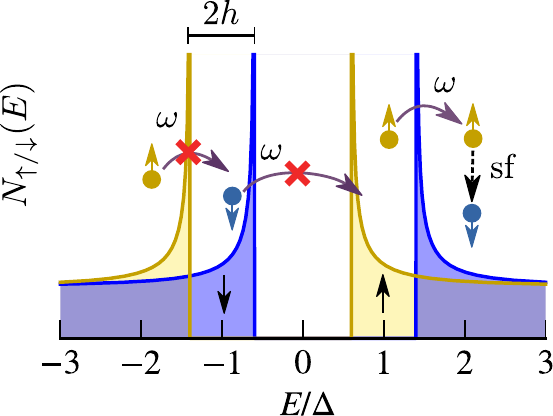}
  \caption{
    \label{fig:sf-schematic}
    Quasiparticle transitions induced by an orbital electromagnetic field.
    The energy dependence of the density of states implies an accumulation
    and depletion of quasiparticles at energy intervals close to gap edges.
    If direct spin-flip transitions are suppressed (red crosses),
    relaxation via
    spin-flip impurity scattering can still transform quasiparticle energy imbalance
    to spin imbalance.
  }
\end{figure}

The absorption kernel $R$ is the one
appearing in the real part of the conductivity,
Eq.~\eqref{eq:ac-abs-kernel}. Qualitatively, this collision integral
results in the depletion
(accumulation) of quasiparticles in an energy band of width $\omega$
above (below) the gap edges $|E|>|\Delta\pm{}h|$ (see
Fig.~\ref{fig:sf-schematic}). At higher energies the electron
distribution corresponds to an increased temperature.

The collision integrals for the relaxation processes can be written as,
cf.~Eq.~\eqref{eq:ST3}, 
\begin{align}
  {\cal I}_{{\rm sn},\sigma} 
  &= 
  \frac{S_{T3}}{8}(f_{\sigma}-f_{\bar{\sigma}})
  \,,
  \;
  {\cal I}_{\mathrm{in},\sigma} = \frac{N_\sigma}{\tau_{\mathrm{in}}}(f_{\sigma}-f_L^{(0)})
  \,,
\end{align}
where we describe inelastic scattering in a relaxation time approximation.

The spin-conserving microwave absorption term does not directly
generate spin imbalance in the superconductor.  However, as
illustrated in Fig.~\ref{fig:sf-schematic}, spin relaxation converts
the accumulation of quasiparticles above the gap to an imbalance in
the number of spin-up and spin-down quasiparticles.  From the kinetic
equation~\eqref{eq:mw-kinetic-eq} we obtain in the limit $1/\tau_{sf}\ll1/\tau_{in}$,
\begin{align}
  f_{\sigma}-f_{\sigma}^{(0)}
  \simeq
  -
  \frac{{\cal I}_{ac,\sigma}^{(0)}\tau_{\mathrm{in}}}{N_\sigma}
  -
  \sigma \frac{S_{T3}\tau_{\mathrm{in}}^2}{8}
  \frac{N_\uparrow{\cal I}_{ac,\downarrow}^{(0)}-N_\downarrow{\cal I}_{ac,\uparrow}^{(0)}}{N_\uparrow{}N_\downarrow}
  \,.
\end{align}
The first term contributes zero total spin imbalance \eqref{Eq:ChPotZ}, but the second gives a nonzero contribution of the order of 
$\tau_{\mathrm{in}}/\tau_{\mathrm{sn}}$. A related effect
in quasiparticle injection was discussed in Ref.~\cite{grimaldi1997}.

The generated spin imbalance can be in principle measured by a
spin-polarized probe junction, as discussed in
Sec.~\ref{sec:nl_detection}.  Note, however, that in this type of
experiments other processes may need to be also considered.  On the
one hand, photoelectric signals can also occur in the absence of spin
splitting as shown in Refs.~\cite{kalenkov2015-dsb,zaitsev1986-pes}.
These processes, however, generally scale with
$\propto{}\tau_{\rm el}$ rather than $\propto\tau_{\rm in}$.  On the
other hand  an ac bias over the S/F junction may
also result in rectification, as discussed in the next section.

\begin{figure}
  \centering
  \includegraphics{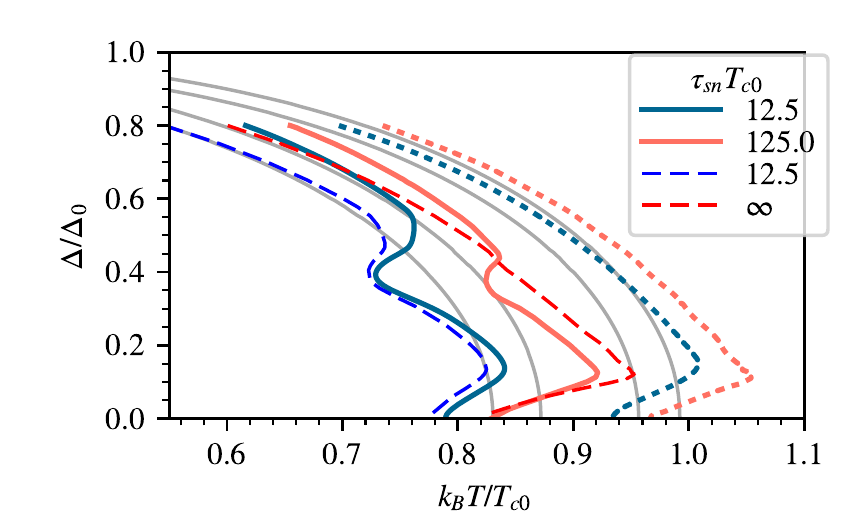}
  \caption{
    \label{fig:Delta-vs-tausn}
    Order parameter $|\Delta|$ for different magnitudes of spin-flip scattering
    $\tau_{sn}$. We take $h/\Delta_0=0.4$, $\tau_{sn}T_{c0}=12.5$,
    $\beta=0.5$ and $\alpha_{\rm orb}=0.01$,
    $\omega/\Delta_0=0.2$, and $DA_0^2/\omega=0.032$.
    The inelastic relaxation is modeled either via a phonon model with
    the collision integral in Eq.~\eqref{eq:ephcolli} (solid lines),
    or relaxation time approximation (dashed) with $\tau_{\text{in}}T_{c0}=100$.
    Results for $h=0$ (dotted) and $A_0=0$ (gray) are also 
    shown.
  }
\end{figure}

It is interesting to notice that the nonequilibrium quasiparticle accumulation generated by the
microwave absorption affects the magnitude of the the superconducting gap $\Delta$, leading to deviations from the results in Ref.~\cite{eliashberg1970-fss}.  Results for the
self-consistent $\Delta(T)$  from numerical calculations \cite{virtanen2016stimulated} are shown  in
Fig.~\ref{fig:Delta-vs-tausn} for $h=0$ and $h>0$.   In the absence of spin-splitting,  $h=0$, a gap
enhancement \cite{eliashberg1970-fss} occurs. For $h>\omega$, an
additional instability develops at $\Delta(T)=h$,
corresponding to a coexistence of two solutions where the spin-averaged
DOS is either gapless ($|\Delta|<h$) or gapped ($|\Delta|>h$).  The
instability requires the presence of strong spin-flip scattering: for $\tau_{sn}\gg\tau_{in}$ the exchange field does not cause
significant qualitative changes, because {in that case} the quasiparticle
accumulations above the gap edges that generate the enhancement, are
approximatively described as two independent copies of the effect at
$h=0$.

\subsection{S/F tunnel junction dynamics}

For adiabatic excitation, $\omega\ll{}\Delta$ dynamics of S/F tunnel
junctions can be described by the dc relations discussed in
Sec.~\ref{sec:quasiclassical_theory} and \ref{thermoel}. That this limit can be reached at time
scales shorter than the spin-relaxation and inelastic times, was used 
in a recent experiment \cite{quay2016-fdm} to probe the
relaxation dynamics in spin-split thin-film superconductors.

At higher frequencies, photoassisted tunneling breaks the adiabatic
description. This can be taken into account via a standard tunneling
Hamiltonian approach
\cite{werthamer1966-nso,larkin1967-teb,barone}.
The result for the tunneling current is
\begin{align}
  \label{eq:current}
  I(t)
  &=
  -2\Re\sum_{\sigma}\sum_{\vec{k}\vec{q}}\int_{-\infty}^t\dd{t'}
  e^{0^+t'}|T_{\vec{k}\vec{q}}^\sigma|^2
  e^{i[\phi(t)-\phi(t')]}
  \\\notag&\qquad\times
  [
    G^<_{\vec{k}\sigma}(t',t)G^>_{\vec{q}\sigma}(t,t')
    -
    G^>_{\vec{k}\sigma}(t',t)G^<_{\vec{q}\sigma}(t,t')
  ]
  \,,
\end{align}
where $T$ is the tunneling matrix element, $G^{>(<)}=\frac{1}{2}G^K\pm\frac{1}{2}(G^R-G^A)$
are (non-quasiclassical) Green functions for
the superconducting ($\vec{k}$) and  non-superconducting ($\vec{q}$) sides,
and the phase $\phi(t)=\frac{e}{\hbar}\int^t\dd{t'}V(t')$ is related
to the time-dependent voltage $V(t)$ across the junction.  The above result applies for
spin-conserving tunneling with collinear magnetizations.  If the
terminals are in an internal equilibrium (no spin or charge
imbalance), 
changing integration variables
yields the time-dependent generalization of Eq.~\eqref{eq:I_tun},
\begin{align}
  I(t)
  &=
  \Re\int_{-\infty}^\infty\frac{\dd{V}}{\pi}
  I_{\rm dc}(V)\int_{-\infty}^{t}\dd{t'}e^{-i(t-t')V}e^{i[\phi(t)-\phi(t')]}
  \\
  &=
  I_{\rm dc}(V(t))
  +
  \Re\int_{-\infty}^{t}\dd{t'} K(t-t')
  [e^{i[\phi(t)-\phi(t')]}
  - e^{iV(t)(t-t')}]
  \,,
\end{align}
where $I_{\rm dc}$ is given by Eq.~\eqref{eq:I_tun}, and the memory
kernel is \cite{marchegiani2016-soq}
\begin{align}
  K(t)
  &=
  \int_{-\infty}^\infty\frac{\dd{V}}{\pi}e^{-itV}
  [I_{\rm dc}(V)-\frac{V}{R_T}-I_{\rm dc}(0)]
  \\\notag
  &=\frac{\pi{}T_F\Delta}{R_T}\frac{J_1(t\Delta)}{\sinh(\pi{}T_Ft)}[i\cos{}ht
  + P_F \sin{}ht]
\,,
\end{align}
and $J_1$ is a Bessel function.

The resulting current-voltage relation is asymmetric,
$I_{\rm{}dc}(-V)\ne{}-I_{\rm{}dc}(V)$ for $h\ne0$. This implies that such junctions
rectify ac signals \cite{quay2016-fdm}. For an ac signal $V(t)=V_{dc} +
V_{ac}\cos(\omega{}t)$, from the above we have the average dc current
\cite{tucker1985-qda} 
\begin{align}
  I_{dc}
  &=
  \sum_{n=-\infty}^\infty{}J_n^2\left(\frac{eV_{\rm ac}}{\hbar\omega}\right)
  I_{\rm dc}(V_{\rm dc} + n\omega)
  \\
  &\simeq
  I_{\rm dc}(V_{\rm dc})
  +
  \frac{V_{\rm ac}^2}{4\omega^2}[I_{\rm dc}(V_{\rm dc} + \omega) + I_{\rm dc}(V_{\rm dc} - \omega)]
  +
  \ldots
  \,.
\end{align}
At $V_{\rm dc}=0$ and small $V_{\rm ac}$ at uniform temperature
$T\gtrsim{}h,\omega$, the rectified current is
$I_{dc}\propto{}P_FV_{\rm{}ac}^2h/(R_T\Delta^2)$, \cite{virtanen2016stimulated} proportional to the
exchange field in S.  Measurements attempting to use F probes for the
nonequilibrium ac effects discussed in the previous sections need to
take into account such rectification, or try to suppress $V_{\rm ac}$
e.g. via a suitable microwave circuit design \cite{horstman1981-gen}.

While dynamical effects in superconductors have been studied
extensively experimentally, we are aware of relatively few studies on 
spin-split  superconductors. Tunnel junction rectification
effects were observed for example  in Ref.~\cite{quay2016-fdm}. However, to our knowledge,   
higher-frequency experiments probing gap enhancement or photoassisted tunneling have not been  
reported so far.

\section{Thermoelectric effects in superconducting structures\label{thermoel}}

A temperature difference across an electric contact typically leads to heat currents aiming to relax this difference. In some cases it may also lead to observable charge currents. Reciprocally, an electrical voltage may drive a heat current not only in nonlinear response (due to Joule heating) but also for small voltages. This connection of charge and heat currents is called the thermoelectric effect. The traditional view of thermoelectric effects in superconductors is that if they exist, they must be very weak. In bulk superconductors, this is partially because any thermoelectrically generated quasiparticle current is screened by a supercurrent \cite{meissner27thermoel}. Ginzburg  suggested \cite{Ginsburg:1944vv} to measure this supercurrent {by using} an additional constraint to the phase of the superconducting order parameter in a multiply connected structure. However, even this thermally created phase gradient tends to be weak, owing to the near-complete electron-hole symmetry in superconductors \cite{galperin74thermoel}.  

Thermoelectric effects typically require strongly energy dependent density of states of the charge carriers. Such energy dependence is present in the BCS density of states of the superconductors. However, typically this density of states is quite symmetric with respect to the Fermi level, and therefore any contribution of the positive energy excitations ("electrons") on thermoelectric effects is cancelled by the negative energy excitations ("holes"). Breaking this electron-hole symmetry would hence allow for the appearance of strong thermoelectric effects. As we discuss in this section, this is
what happens in spin-split superconductors, as an exchange
field breaks the symmetry in each spin sector, but so that the overall
spin-summed energy spectrum remains electron-hole symmetric. Further breaking the spin symmetry in transport through a spin filter can then provide large thermoelectric effects as the two spins are weighed differently \cite{Ozaeta2014a,Machon2013,Machon2014}. This prediction has been recently confirmed experimentally in \cite{Kolenda2016,Kolenda2017}.

In this section we give an overview of the different types of
thermoelectric effects discussed for superconductors, and then
concentrate on the ones obtained in superconductors with a
spin-splitting field. We discuss both the linear response regime and
beyond it, and consider also the limiting features such as the
electron-phonon coupling. We show that under suitable conditions, in
particular for close-to-optimal spin filters, the efficiency of
thermoelectric conversion can become very large and exceed that
obtained for best thermoelectric devices operated at or above room
temperature. Besides the regular quasiparticle current, in devices
coupling two superconductors, one with a spin-splitting field and one
without, the thermoelectric effect can also be converted to a phase
gradient \cite{giazotto2015}. This large thermophase effect is discussed in more detail in 
Ref.~\cite{bergeret2018colloquium}. 

As these effects require low operating temperatures, they obviously
cannot be directly used to improve the efficiency of various everyday
devices. However, {the strong thermoelectric effect may become relevant} in other types of applications, such as sensors, where the measured (wide-band) signal consists of heating one part of the system \cite{heikkila2018}. The thermoelectric conversion can then be used to convert the resulting temperature difference to a charge current.

\subsection{Thermoelectric effects and heat engines}
\label{subs:thermoel}

Biasing a contact with a small voltage $V$ and a small temperature difference $\Delta T$ leads to linear-response charge and heat currents $I$ and $\dot Q$ of the form
\begin{equation}
\label{eq:thermoel}
\left(\begin{array}{c} I \\ \dot Q \end{array}\right) =
\left(\begin{array}{cc} G & \tilde \alpha
  \\ \tilde \alpha & G_{\rm th} T \end{array}\right) \left(\begin{array}{c} V \\ -\Delta
  T/T \end{array}\right).
\end{equation}
The $2\times 2$ conductance matrix in Eq.~(\ref{eq:thermoel})
is a part of the generalized $4\times 4$ Onsager matrix
(\ref{Eq:Onsager}) connecting different interface currents and
potentials as discussed in Sec.~\ref{sec:onsagersection}. This matrix is symmetric in the presence of time-reversal symmetry, and also in all particular cases considered in this section. Here we consider a
conventional situation when the spin-dependent potentials $V_s$ and
$T_s$ are negligibly small. This is relevant {when the spin relaxation rate in the electrodes around the contact exceeds the tunneling rate across the contact.} 
This approximation allows the reduction of the general boundary condition  (\ref{Eq:Onsager}) to the simpler one (\ref{eq:thermoel}) describing thermoelectric response. 

\begin{figure}
\centering
\includegraphics[width=6cm]{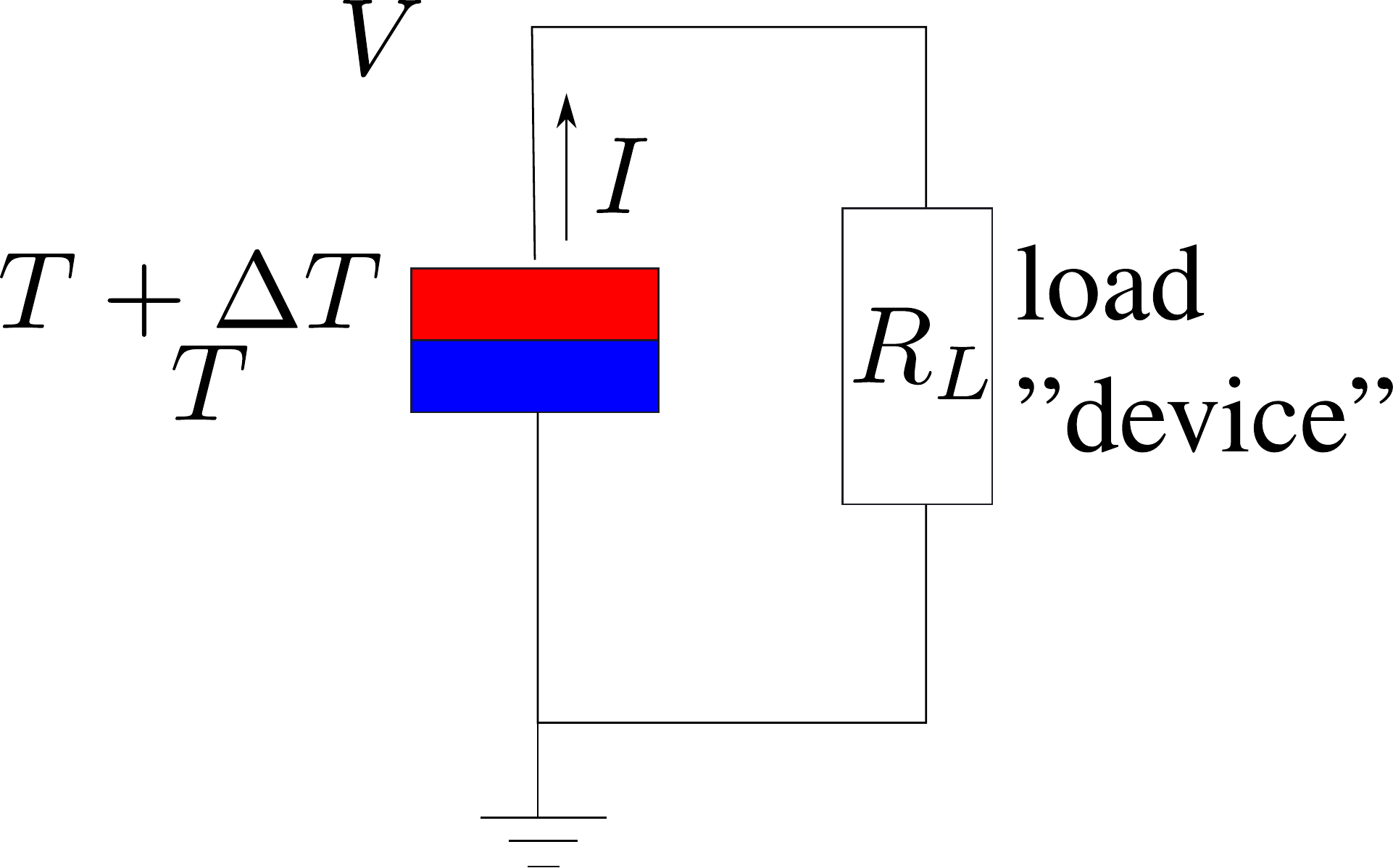}
\caption{Generic electrical heat engine.}
\label{fig:heatengine}
\end{figure}

When $\tilde \alpha$ is not too small, electrical energy may be converted to heat or cooling (Peltier effect), or reciprocally a temperature difference may be converted to electrical power (Seebeck effect). The efficiency of this conversion can be described
by constructing a model for a generic electrical heat engine (see
Fig.~\ref{fig:heatengine}). There, a load with resistance $R_L$ is
driven by the power drawn from the thermoelectric element across which
there is a temperature difference $\Delta T$. The power dissipated on
the load is $P=IV=R_L I^2$, whereas the voltage across the
thermoelectric element is $-I R_L$. Plugging this into
Eq.~\eqref{eq:thermoel} yields 
\begin{equation}
I= - \frac{\tilde\alpha}{T (1+G R_L)} \Delta T.
\end{equation}
The extracted power hence is
\begin{equation}
P=\frac{R_L \tilde \alpha^2 (\Delta T)^2}{T^2 (1+G R_L)^2}.
\end{equation}
On the other hand, the thermoelectric element extracts heat from the
temperature difference, i.e., trying to balance it, with the power
\begin{equation}
\dot Q = \left(G_{\rm th} - \frac{\tilde \alpha^2 R_L}{T (1+G R_L)}
\right)\Delta T.
\end{equation}
The efficiency of thermoelectric conversion is hence
\begin{equation}
\eta = \frac{P}{\dot Q} = \underbrace{\frac{\Delta T}{T_{\rm
      hot}}}_{\eta_{\rm Carnot}} \frac{G
  R_L}{(1+G R_L)^2} \frac{1}{1/N-\frac{G R_L}{1+G R_L}},
\end{equation}
where $N=\tilde \alpha^2/(G_{\rm th} G T) \leq 1$. Alternatively, we can write $N$
in terms of the usual thermoelectric figure of merit,
\begin{equation}
ZT \equiv \frac{N}{1-N} = \frac{\tilde \alpha^2}{G_{\rm th} G T -
  \tilde \alpha^2}=\frac{S^2 GT}{\tilde G_{\rm th}},
\end{equation}
where $S=\tilde \alpha/(GT)$ is the thermopower (Seebeck coefficient) and $\tilde G_{\rm
  th}=G_{\rm th}-\tilde \alpha^2/(GT)$ is the thermal conductance at a
vanishing current. 

Now we should choose $R_L$ to optimize the device. For
example, the maximum efficiency is obtained with $G R_L=1/\sqrt{1-N}=\sqrt{1+ZT}$
yielding 
\begin{equation}
{\rm max} \eta = \eta_{\rm Carnot}
\frac{\sqrt{1+ZT}-1}{\sqrt{1+ZT}+1}.
\end{equation}
This result is consistent with that obtained in Ref.~\cite{snyder2003} in
the linear response limit $T_{\rm cold}/T_{\rm hot} \approx 1$. On
the other hand, optimizing the device to yield a maximum power output
requires $GR_L=1$, corresponding to the limit \cite{novikov1957-eap,curzon1975}
\begin{equation}
\eta=\underbrace{\frac{\Delta T}{2T}}_{\eta_{CA}} \frac{ZT}{2+ZT}.
\end{equation}
Here $\eta_{CA} \equiv 1-\sqrt{T_{\rm cold}/T_{\rm hot}} \approx
\Delta T/(2T)$ is the maximum efficiency obtained when $ZT \rightarrow
\infty$. Both of these efficiencies are maximized when the
thermoelectric figure of merit becomes large. 

Above room temperature, the highest figures of merit are
obtained in certain strongly doped semiconductor structures
\cite{zhao2016,kim2015}. The record values are of the order of $ZT\gtrsim
1\dots 2$. The particular value and the optimal temperature where it is obtained
results from a competition of two generic
temperature dependencies: that of phonon heat conductance, and that of the (typically activated)
process yielding the thermopower. Both of these decrease towards low
temperatures, but the previous decreases as a power law, whereas the
latter decreases exponentially. In an electron system the heat conductance is
typically close to the Wiedemann-Franz limit $G_{\rm th}^{\rm el} \approx {\cal
  L} G T$, where 
  ${\cal L}=\pi^2 k_B^2/(3e^2)$. The total heat
conductance is obtained from the sum of the electronic and phononic
contributions, $G_{\rm th}=G_{\rm th}^{\rm el} +G_{\rm th}^{\rm
  ph}$. Since $ZT \propto GT/G_{\rm th} = 1/[{\cal L}+G_{\rm th}^{\rm
  ph}/(GT)]$, the only way to improve this is to minimize the phononic
contribution. Generally the latter decreases towards low temperatures
faster than linearly. However, the thermopower is typically an
exponential function of temperature, $S \propto \exp(-\Delta/k_B T)$
as the best thermoelectrics contain a gapped dispersion with the gap
$\Delta$. Therefore, the optimal $ZT$  takes place at a
temperature which is some fraction of $\Delta$. As shown in
Sec.~\ref{subs:linearresponseheatengine} (see in particular
Fig.~\ref{fig:ZTsuper1}), the same is true for a normal metal island
coupled to spin-splitted superconductors. In that case, the magnitude
of $\Delta$ is just orders of magnitude lower than in semiconductors
structures.

\subsection{Thermoelectric effects in superconductors} 
\label{subs:sthermoelectrics}

\begin{figure}
\centering
\includegraphics[width=5.5cm]{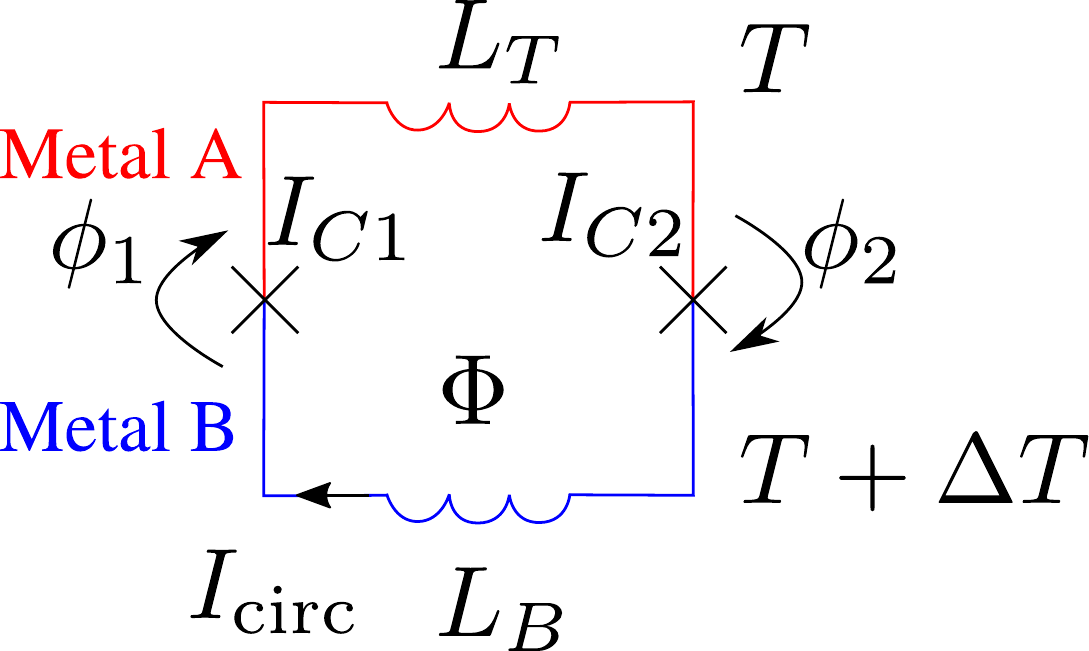}
\caption{Setup for measuring the thermoelectrically induced
  circulating current in a bimetallic loop formed by two different
  superconducting metals A and B. In this case the contact between the
metals becomes a Josephson junction.} 
\label{fig:thermophase}
\end{figure}

Research on thermoelectric effects in superconductors dates back to
the 1920's, when Meissner~\cite{meissner27thermoel} concluded them to be absent because
any thermoelectric current is cancelled by a counterflowing
supercurrent.
Ginzburg~\cite{Ginsburg:1944vv} showed that this is no longer the case in
multiply connected bimetallic superconducting structures. This is because the presence of supercurrent is
linked to the gradient of the phase of the order parameter, and in
multiply connected structures uniqueness of the order parameter
imposes a constraint on the relation between the flux through the ring
and the circulating supercurrent. This situation can be visualized as
in Fig.~\ref{fig:thermophase}. For definiteness, let us consider a
bimetallic superconducting loop, where the thermoelectric effects
mostly take place at the two contacts. The current through contact
$i=1,2$ consists of a sum of the supercurrent and the thermoelectric
current, i.e.,
\begin{equation}
I_{\circ}=I_{ci}\sin(\varphi_i) + I_{\rm th,i}.
\end{equation}
In a closed circuit this current must equal for both junctions, and
thereby it yields a relation between the
phases $\varphi_{1,2}$. A second relation fixing the phases is
obtained in the presence of a flux $\Phi$ through the loop with
inductance $L=L_T+L_B$. Without loss of generality the total loop inductance
can be described with a single quantity. Namely, the phases must be
fixed so that they minimize the energy
\begin{equation}
\frac{\hbar^2}{8e^2 L} \left(n 2\pi + 2\pi \frac{\Phi}{\Phi_0} -
  \varphi_1-\varphi_2\right)^2-\frac{\hbar}{2e} \sum_{i=1,2} I_{ci}
\cos(\varphi_i),
\end{equation}
where $n\in {\mathbb Z}$, $\Phi$ is the magnetic flux through the
junction and $\Phi_0=h/(2e)$ is the flux quantum. In the following we denote
$\Phi/\Phi_0=f + 2 \pi (m-n)$ with $f\in [0,2\pi[$ and $m\in {\mathbb
  Z}$. For $f\ll 1$ and $I_{\rm th,i} \ll I_{c,i}$, the induced phases
are small, and we may linearize the current-phase relations. As a
result, we obtain a circulating current
\begin{equation}
I_{\rm circ}=\frac{f+2m\pi + \frac{I_{\rm th,1}}{I_{c1}}+\frac{I_{\rm
      th,2}}{I_{c2}}}{2e L_{\rm tot}/\hbar},
\end{equation}
where $L_{\rm tot}=L+\hbar(I_{c1}^{-1}+I_{c2}^{-1})/2e$ is the total
inductance of the superconducting loop. 

Let us assume heating the lower superconductor of the bimetallic
loop so that the temperature increases from $T$ to $T+\Delta T$. Such a heating induces thermoelectric currents
\begin{equation}
I_{th,1}=\tilde \alpha_1 \Delta T/T, \quad I_{\rm th,2}=-\tilde \alpha_2 
\Delta T/T.
\end{equation}
This produces a circulating thermoelectric current
\begin{equation}
I_{\rm circ}^{\rm
  th}=\hbar\frac{\frac{\tilde \alpha_1}{I_{c1}}-\frac{\tilde \alpha_2}{I_{c2}}}{2e
  L_{\rm tot}} \frac{\Delta T}{T}.
\end{equation}
It can be measured for example by placing a SQUID on top of the
bimetallic loop, and measuring the induced flux $\Phi_{\rm ind}=MI_{\rm circ}$, where
$M$ is the mutual inductance between the two systems. 

The size of the coefficient $\tilde \alpha$ was calculated in Ref.~\cite{galperin74thermoel} to be 
\begin{equation}
\tilde \alpha = \tilde \alpha_N G(\Delta/T),\quad G(x)=\frac{3}{2\pi^2} \int_x^\infty
\frac{y^2 dy}{\cosh^2(y/2)},
\label{eq:superalpha}
\end{equation}
where the latter form is due to  the reduction of the quasiparticle
density in the superconducting state, and $\tilde \alpha_N$ is the
size of the thermoelectric coefficient in the normal state. The
precise value of $\tilde \alpha_N$ depends on the exact electronic spectrum
of the metals in question and its calculation needs extending the theory beyond the
quasiclassical approximation employed in this review, as within that
approximation $\tilde \alpha_N=0$. For a quadratic
dispersion $\tilde \alpha_N = \frac{\pi^2 G_T k_B^2 T}{6e E_F}$, where $E_F$ is the Fermi energy. At
temperatures $T \ll \Delta/k_B$, $\tilde \alpha$ is thus
 a product of two small coefficients, $\tilde \alpha_N \propto
k_B T/E_F$, and $G(\Delta/T)$. Such a small $\tilde \alpha$ is not easy to
measure quantitatively. Because of many spurious effects in such
measurements, it is not simple to make the experiments agree quantitatively with this theory. 

Nevertheless, at least close to the
critical temperature the thermoelectric flux should be observable, and
the first experiments to measure it were done in the 1970s and early 1980s. One set of experiments \cite{kartsovnik1981,ryazanov1981} followed the idea of Ref.~\cite{galperin74sns} and measured the thermoelectric voltage across a superconductor--normal-metal--superconductor contact as the thermoelectric current exceeded its Josephson critical current. The results of these experiments are in line with the expected magnitude of the thermoelectric signal.  On the other hand, Ref.~\cite{vanharlingen80thermoel} measured the thermoelectrically generated flux in a bimetallic loop formed from superconducting Pb and Ti, close to the critical temperature of the latter. These experiments are also discussed in Ref.~\cite{galperin02}. Surprisingly, the
experiments demonstrated fluxes five orders of magnitude larger than
predicted by theory. This discrepancy annoyed Ginzburg so much that he devoted an
entire chapter on the topic in his Nobel colloquium
\cite{ginzburg04}.

Recently, Ref.~\cite{Shelly16} claim to have solved
this discrepancy with new experiments performed on much smaller
superconducting loops than what was possible in the early 1980s. According
to them, the discrepancy originated from the temperature dependence of
 the inductances
$L_{\rm tot}$ and $M$ as well as the flux $f$ (via the effective area
of the loop), and hence they change as
one end of the bimetallic loop is being heated. This produces 
additional contributions to the thermoelectric flux,
\begin{align}
&\Phi_{\rm ind}=\Phi_{\rm ind}^{\rm th}\\& +\frac{1}{L_{\rm tot}} \left[\left(\frac{dM}{dT}-M
\frac{dL_{\rm tot}}{dT} \right) (f+2m \pi) + M\frac{df}{dT}\right]\Delta T.\nonumber
\end{align}
The additional contributions are in practice much larger than the
pure thermoelectric effect. According to Ref.~\cite{Shelly16}, whereas the effects from $f$ can be
accounted for by measuring the period of oscillations as the external
flux is altered, the effects due to the trapped flux, accounted for by
$m$, is much harder to deduce, and was likely the reason for the
discrepancy, as in the early measurements a geomagnetic field amounted
to $m \sim 10^6$. Using much smaller loops, Ref.~\cite{Shelly16} were able to
control the number of trapped flux quanta, and hence get rid of the
spurious effects. The remaining thermoelectric flux that they observed
is more or less in accord with the value obtained from the above
theory. However, the temperature dependence of the measured flux
depends, besides $G(x)$ above, also on the temperature dependent
inductance, and on the temperature dependence of the main heat contact
between the heated electrons and the phonon bath, so the measurement
could not directly deduce $G(x)$.

The conclusion from these theory and experimental works is that in
conventional superconductors thermoelectric effects can be nonzero, but
they are extremely weak, and therefore difficult to access. 
However, other types of superconducting heterostructures
besides the bimetallic loop do contain relatively strong
thermoelectric effects.  In particular, a supercurrent flowing along a temperature gradient leads to an appearance of charge imbalance \cite{pethick79,clarke80,schmid1979}, which can be measured \cite{clarke79} for example via a non-local geometry similar to those discussed in Sec.~\ref{spininjection}.  A similar type of an effect was found in Andreev interferometers \cite{eom98,chandrasekhar09,parsons03}
composed of a hybrid multiterminal geometry of a normal metal in contact with a superconducting loop. These effects were considered theoretically in Refs.~\cite{virtanen04,virtanen2007,seviour00,titov2008-toi}.
Similar effects are also predicted in ballistic systems as discussed in Refs.~\cite{jacquod10,kalenkov17}. {Nevertheless, these effects either require complicated multiterminal geometries, and otherwise the thermoelectric effects are typically very weak.} 

In what follows we show how this situation can be completely reversed in the presence of the spin-splitting field, provided we add a second
ingredient into the theory: spin filtering. In this case such
superconductor/ferromagnet hybrids can become almost ideal
thermoelectric devices. 
In
particular, Ref.~\cite{Machon2013,Machon2014} showed how three-terminal
proximity-coupled superconductor-ferromagnet devices can show
non-local thermoelectric effects: in this case the density of states (DOS)
in a normal metal coupled both to a superconductor and a ferromagnet
becomes spin dependent, and the spin-resolved DOS is also
electron-hole asymmetric, resulting in the strong thermoelectric
effect. On the other hand, Ref.~\cite{kalenkov14,kalenkov15} showed how a
metallic bilayer consisting of two superconductors or a superconductor
and a normal metal, separated by a spin-active interface (i.e., an interface whose transmission
properties are characterized by a spin-dependent scattering matrix),
can exhibit large thermoelectric response. 
{Around the same time, Ref.~\cite{Ozaeta2014a} showed how  
a superconductor with a spin-splitting field, tunnel coupled to a
normal metal via a spin-polarized interface, exhibits a thermoelectric effect where the figure of merit can become very large. To our knowledge, only this mechanism has been so far accessed experimentally \cite{Kolenda2016,Kolenda2017}. We
explain this mechanism in more detail below.}

Another way to affect the thermoelectric response via magnetism was
discussed in Ref.~\cite{zaitsev1986-pes} and later in
Ref.~\cite{kalenkov12}, who argued how magnetic impurities inside a
superconductor enhance the thermoelectric coefficient by a large
factor  $k_F \ell$ compared to $\tilde \alpha$ in Eq.~\eqref{eq:superalpha}. Here $\ell$ is the elastic mean free path and $k_F$ is the Fermi wavenumber. This effect results from the electron-hole asymmetric Andreev states forming in the vicinity of the magnetic impurities.  

The above discussion concerns metallic structures. In semiconductor quantum dots the thermoelectric effects can be large when the spectrum of the quantum dot is electron-hole asymmetric \cite{beenakker92}, even without the presence of superconductivity or magnetism.  Theoretical studies  showed that the combination of the latter affects the symmetries and may increase the thermoelectric response in quantum dots coupled to superconducting and magnetic electrodes  \cite{hwang16,hwang16b}.

\subsection{Thermoelectric effects at a spin-polarized interface to a
  spin-split superconductor}\label{sec_thermoelectric_sp}

Tunneling into superconductors has been long used to probe the superconducting density of states \cite{giaver1961},  for thermometry \cite{giazotto2006}, and for measuring the nonequilibrium distribution functions \cite{pothier97}. We show below that a {\it spin-polarized} tunnel contact to a superconductor with a spin-splitting field also exhibits a giant thermoelectric effect. Let us denote the spin-dependent normal-state conductance of the tunnel contact by $G_{\uparrow/\downarrow}$ for spin $\uparrow/\downarrow$. In this case the standard tunneling theory yields the spin-resolved charge and heat currents across the tunnel contact from reservoir $R$ to the spin-split superconductor

\begin{align}
I_\sigma = \frac{G_\sigma}{e} \int_{-\infty}^\infty dE &N^R(\varepsilon-\mu_R)
  N_\sigma( \varepsilon-\mu_L )[f_R(\varepsilon)-f_S(\varepsilon)]\label{th_is}\\
\dot Q_\sigma^i = \frac{G_\sigma}{e^2} \int_{-\infty}^\infty d\varepsilon &(\varepsilon-\mu_i)
                N^R(\varepsilon-\mu_R) N_\sigma(\varepsilon-\mu_L)[f_R(\varepsilon)-f_S(\varepsilon)].
\end{align}
The formula for the current is thus the spin-resolved version of
Eq.~\eqref{eq:I_tun}. 
Here $f_{L/R}=n_F(E-\mu_{L/R};T_{L/R})$, $n_F(E;T)=\{\exp[E/(k_B
 T)]+1\}^{-1}$ are the (Fermi) functions of the reservoirs biased at
 potentials $\mu_{L/R}$ and temperatures $T_{L/R}$. We assume the reduced density of states $N^R(\varepsilon)$ (i.e., total density of states at energy $\varepsilon$ divided by the one at the Fermi energy) of reservoir $R$ spin independent for simplicity; the
possible splitting of the Fermi sea in electrode $R$ is included in
$G_\sigma$. The 
reduced density of states in the superconductor for spin $\sigma$ is $N_\sigma(\varepsilon)$.  The heat current
$\dot Q_\sigma^i$ is calculated separately for $i=R,S$, using the
potential $\mu_{R/S}$, because the two heat currents differ by the
Joule power $I(\mu_R-\mu_S)/e$. Let us denote the spin-dependent
reduced density of states via $N_+=N_\uparrow+N_\downarrow$ and
$N_-=N_\uparrow-N_\downarrow$ and define the total tunneling conductance $G_T=G_\uparrow+G_\downarrow$ and spin polarization $P=(G_\uparrow-G_\downarrow)/G_T$. In that case the spin-averaged tunnel
currents $I=I_\uparrow+I_\downarrow$ and $\dot Q^i=\dot Q^i_\uparrow+\dot Q^i_\downarrow$ are
\begin{align}
I&=\frac{G_T}{2e}\int_{-\infty}^\infty d\varepsilon N_R \left(N_++P
  N_-\right) (f_R-f_S)\label{eq:Ith}
  \\
\dot Q_i&=\frac{G_T}{2e}\int_{-\infty}^\infty d\varepsilon (\varepsilon-\mu_i) N_R (N_+ + P N_-) (f_R-f_S).\label{th_qi}
\end{align}
In the following, we first analyze these currents in detail, and then
discuss a method of building a near-optimal heat engine based on such
junctions.  In particular, we use the densities of states described in
Sec.~\ref{sec:quasiclassical_theory}. We first disregard
the spin relaxation effects on the density of states, because this
assumption allows for some analytically treatable limits and because it is a fair approximation for example for Al. This
assumption is lifted in Fig.~\ref{fig:ZT-SpinOrbit}. 

\subsubsection*{Heat current beyond linear response}
Let us consider first the case when the reservoir $R$ is a normal
metal and therefore $N_R(\varepsilon)=1$. The heat current $\dot Q$ describing the cooling of the normal metal is positive at voltages $eV=(\mu_R-\mu_S) \approx \Delta$ even without spin splitting, i.e., the work done by the voltage source can be used to cool the normal metal  \cite{nahum1996,leivo1996,pekola2004}. However, that
heat current is even in the voltage, and therefore it does not
result from the usual Peltier effect (Eq.~\eqref{eq:thermoel} for
$\dot Q$) where the cooling power is linear in  
voltage, and therefore can be reversed by reversing the sign of the voltage. As we show below, in the presence of spin
polarization $P$ and with a non-zero spin-splitting field $h$ in the
superconductor, the cooling power obtains also components that are odd in voltage, in analogy with the usual Peltier coolers. This leads to an improved coefficient of performance of electron refrigeration \cite{kolendaup16,Rouco2017}.

\begin{figure}
\centering
\includegraphics[width=8cm]{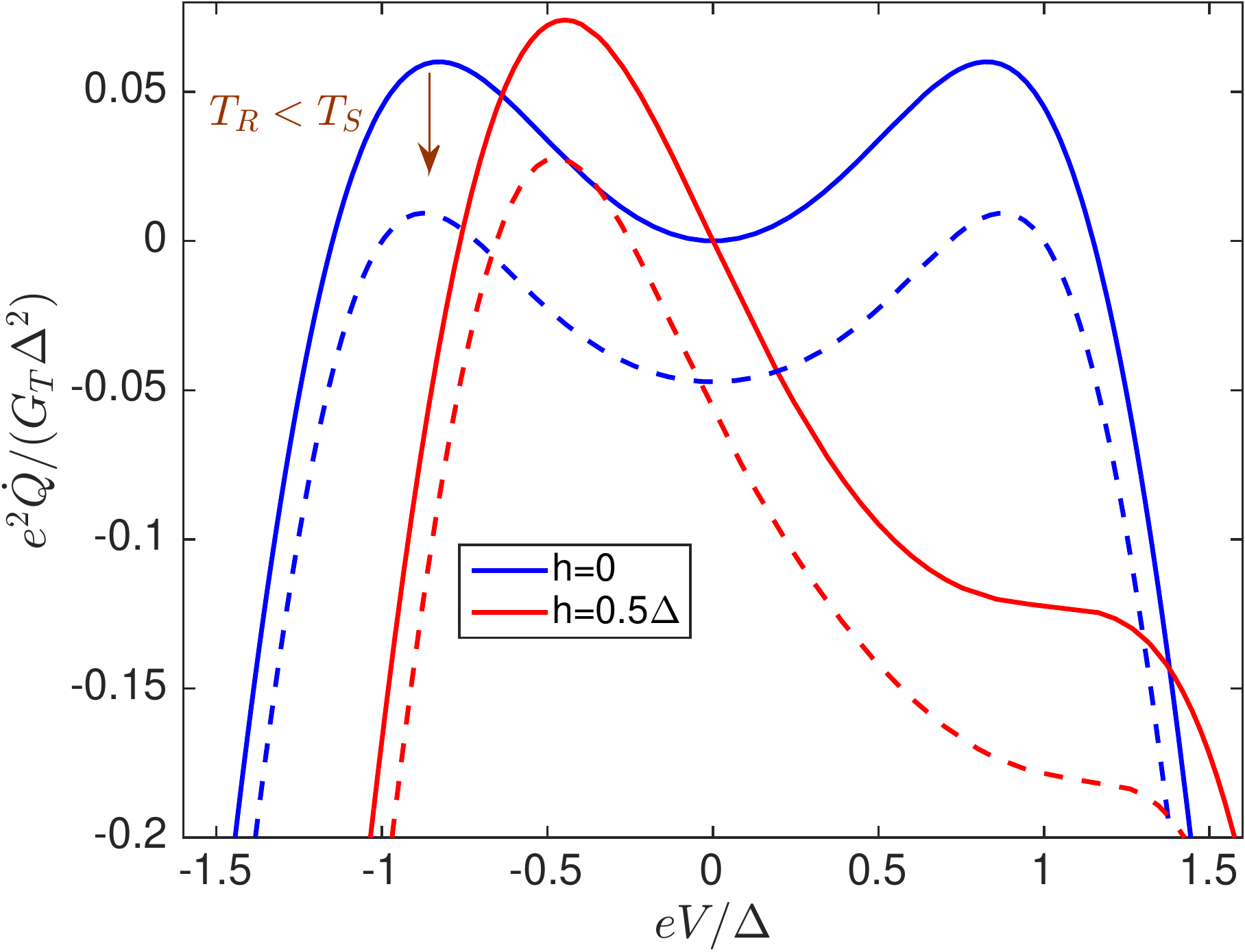}
\caption{\label{fig:coolingpower}
 Cooling power vs.~voltage without (blue line) and with (red) an exchange field in the superconductor, assuming a unit polarization $P=1$. The solid lines are plotted for $T_R=T_S=0.3\Delta/k_B$, and the dashed lines for $k_B T_R=0.22\Delta$ and $k_B T_S=0.3\Delta$.  Changing the sign of $P$ or $h$ 
  inverts the voltage dependence with respect to $V=0$.}
\end{figure}

The cooling power from reservoir $R$ as a function of voltage is shown in Fig.~\ref{fig:coolingpower} with and without an exchange field, assuming the
ideal case of unit spin polarization $P=1$. The figure also shows how decreasing the normal-metal temperature lowers the cooling power. In the absence of extra heating or energy relaxation processes, the normal-metal temperature at the given voltage would then be fixed to the value nullifying the cooling power.

In an electron refrigerator, the cooled element is an island coupled
to two electrodes. Let us consider a normal-metal island playing the
role of the reservoir R, and coupled to two spin-split
superconductors via ferromagnetic insulators with polarizations $P_L$
and $P_R$, respectively. The optimal situation is realized when
$P_L=-P_R=\pm 1$ or when the exchange fields in the two
superconductors are reversed. The total cooling power from the island is $\dot Q_N=\dot
Q_R(V/2,P_L)+\dot Q_R(-V/2,P_R)$. It works against other
relaxation mechanisms, so that the stationary temperature $T$ of 
reservoir $R$ is determined from heat
balance \cite{giazotto2006},
\begin{align}
\label{eq:heatbalance}
\dot Q_N(T,T_S)&=\Sigma_N \Omega_N (T_{\rm bath}^5 - T^5).
\end{align}
Here we assume that the dominant heat relaxation mechanism on the
reservoir $R$ with volume $\Omega_N$ is due to electron-phonon
interaction with strength $\Sigma_N$, described by
Eq.~\eqref{eq:normaleph}. In addition, we assume that the spin accumulation on the island, produced by the nonequilibrium driving, is negligibly small due to spin relaxation. The assumptions relevant for this limit are discussed in Sec.~\ref{subs:spinseebeck} below.

\begin{figure}
\centering
\includegraphics[width=0.325\textwidth]{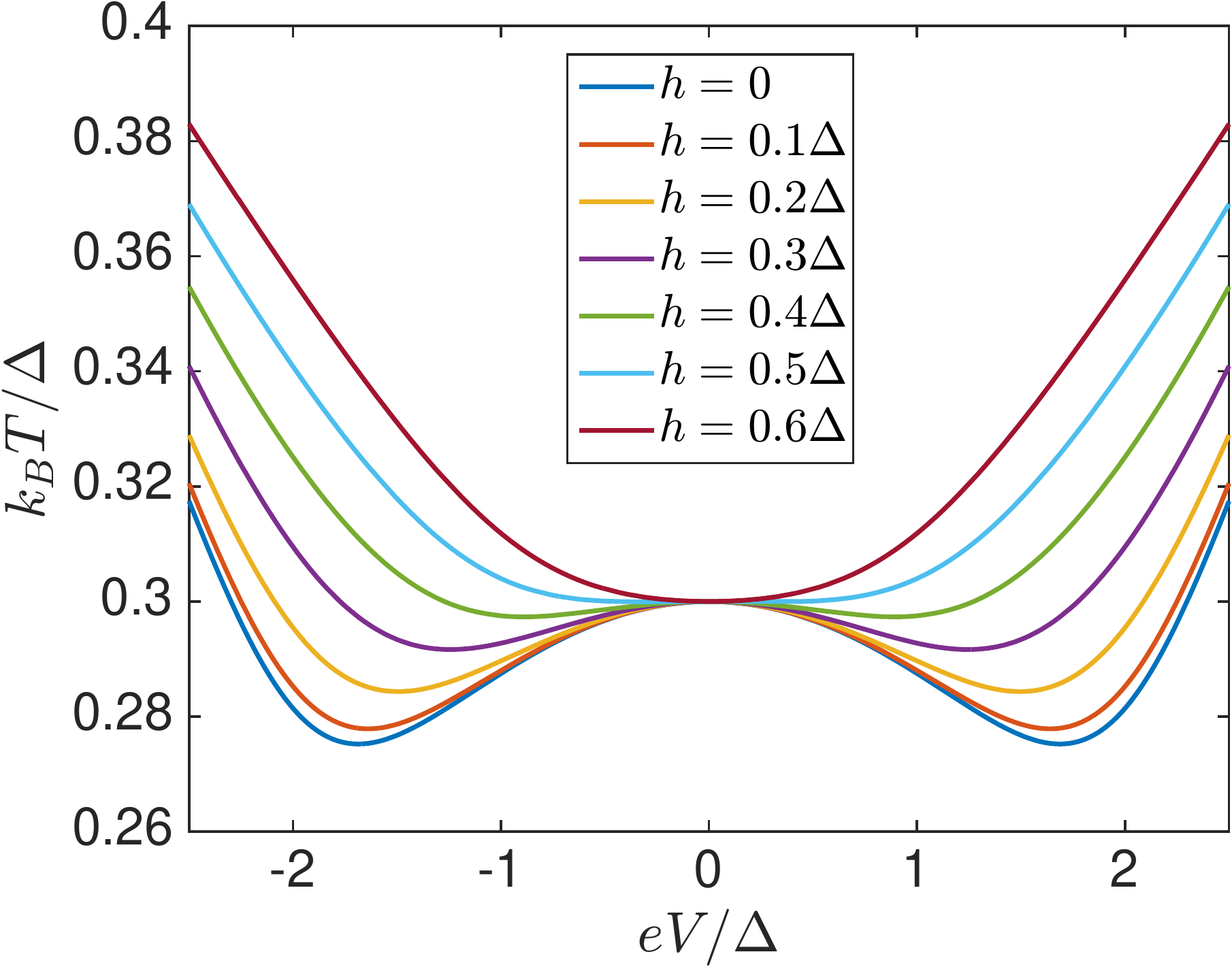}
\includegraphics[width=0.325\textwidth]{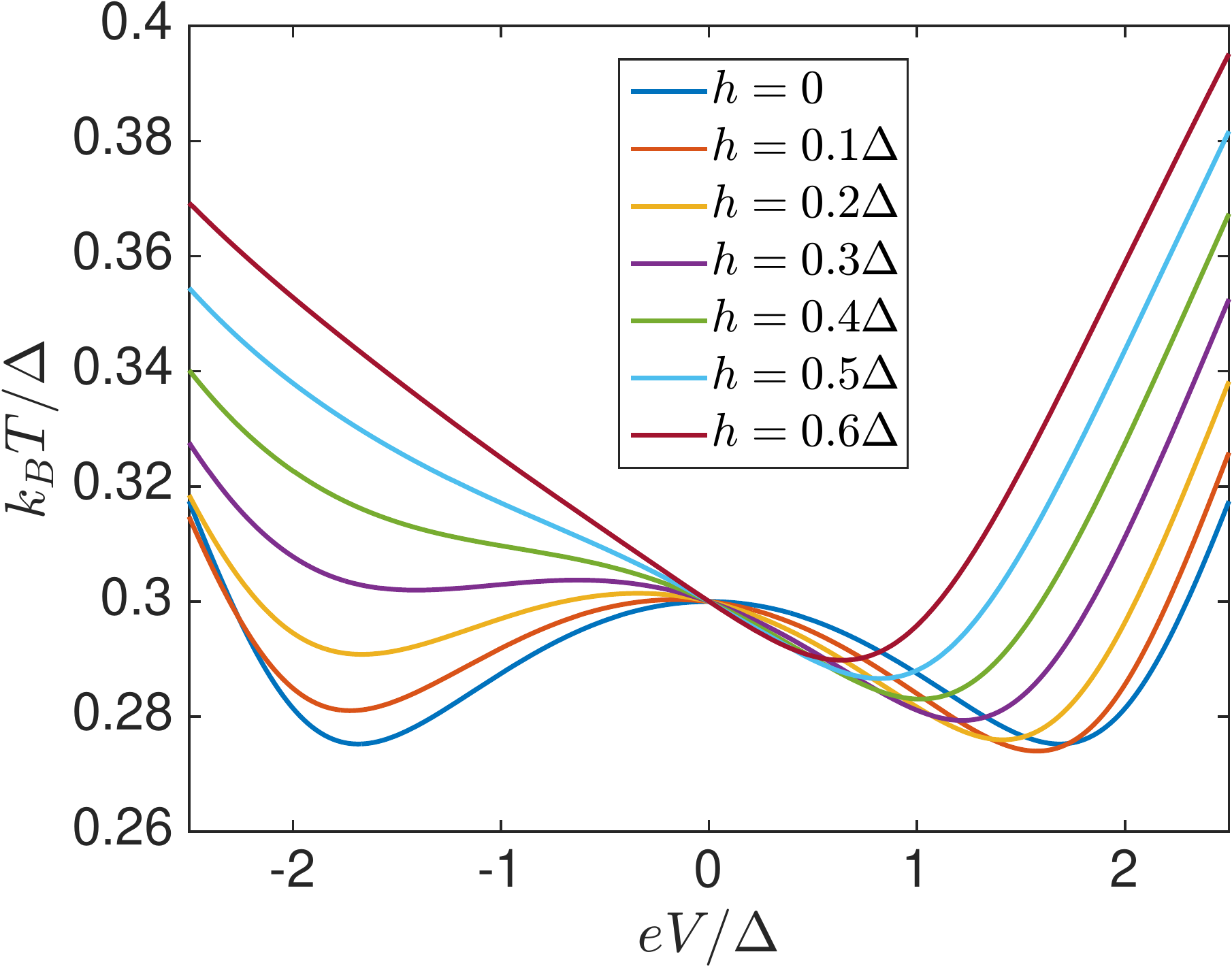}
\includegraphics[width=0.325\textwidth]{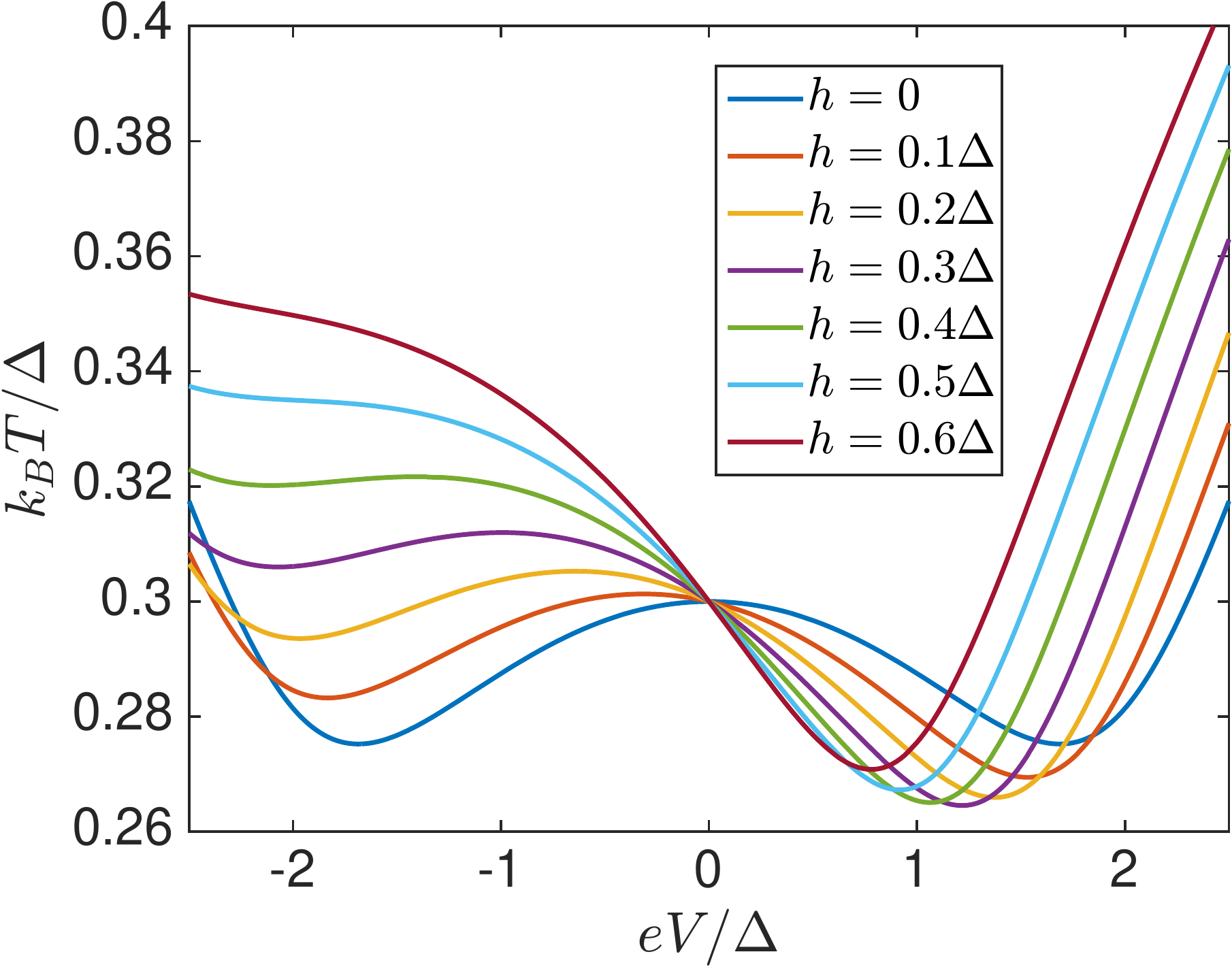}
\includegraphics[width=0.325\textwidth]{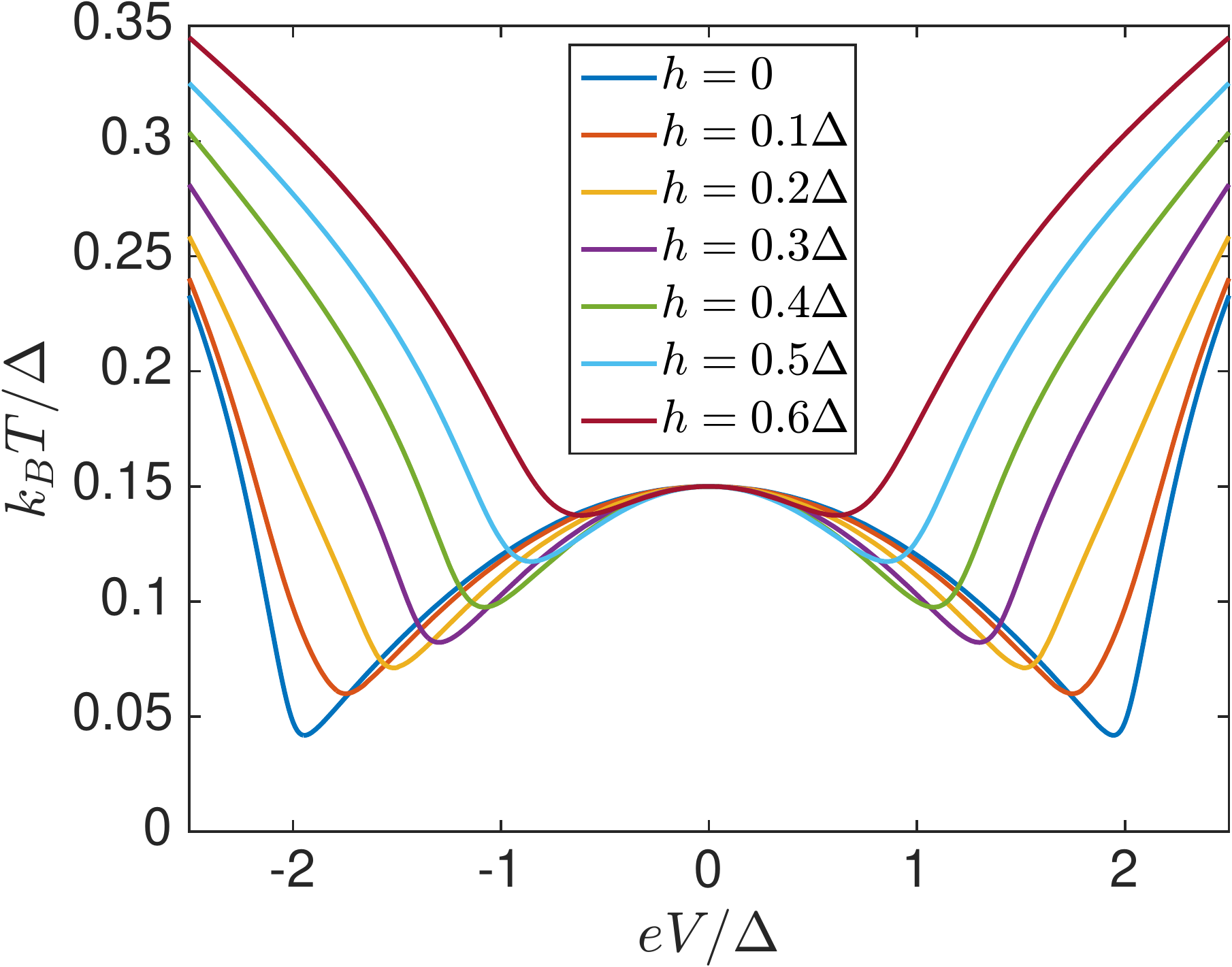}
\includegraphics[width=0.325\textwidth]{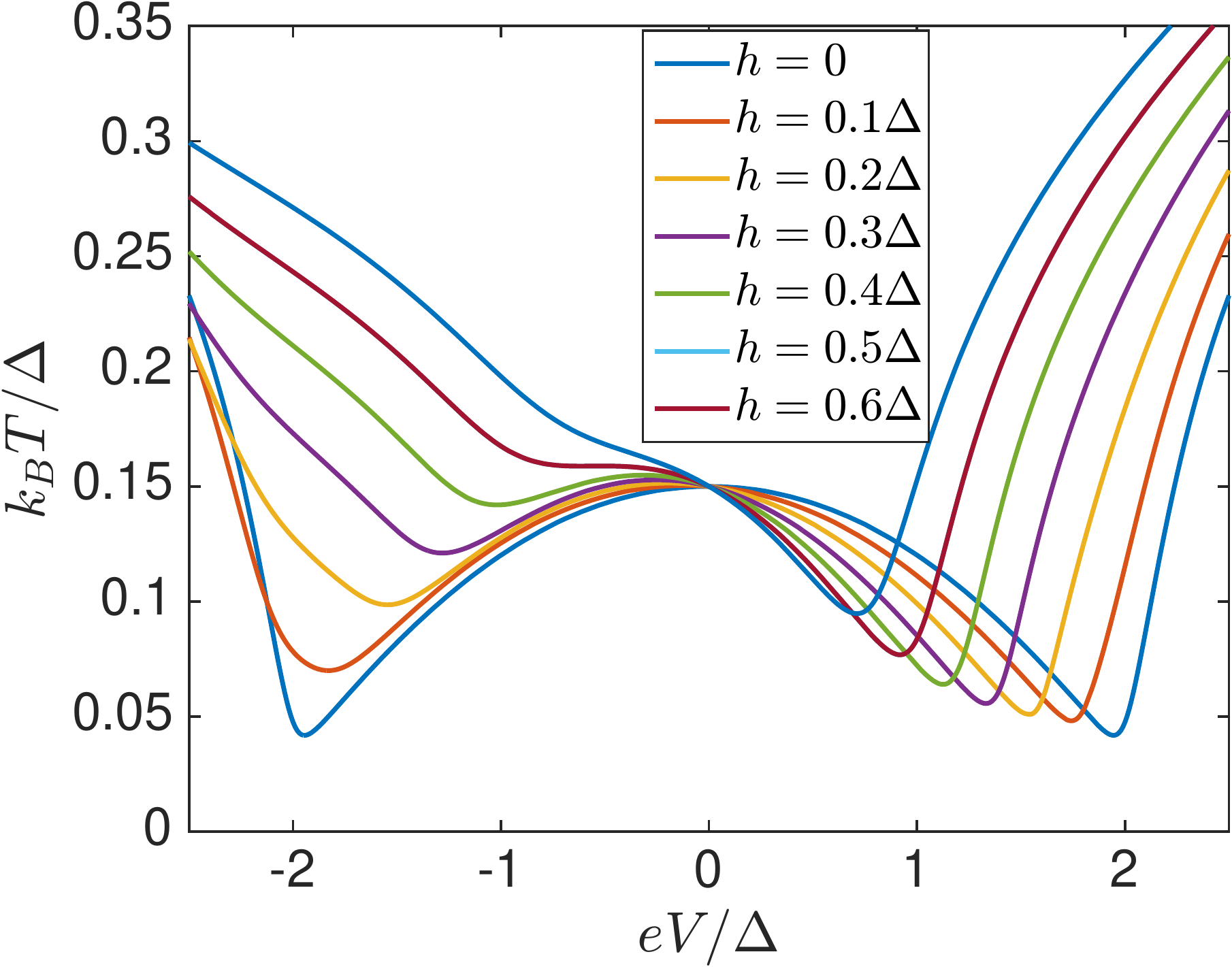}
\includegraphics[width=0.325\textwidth]{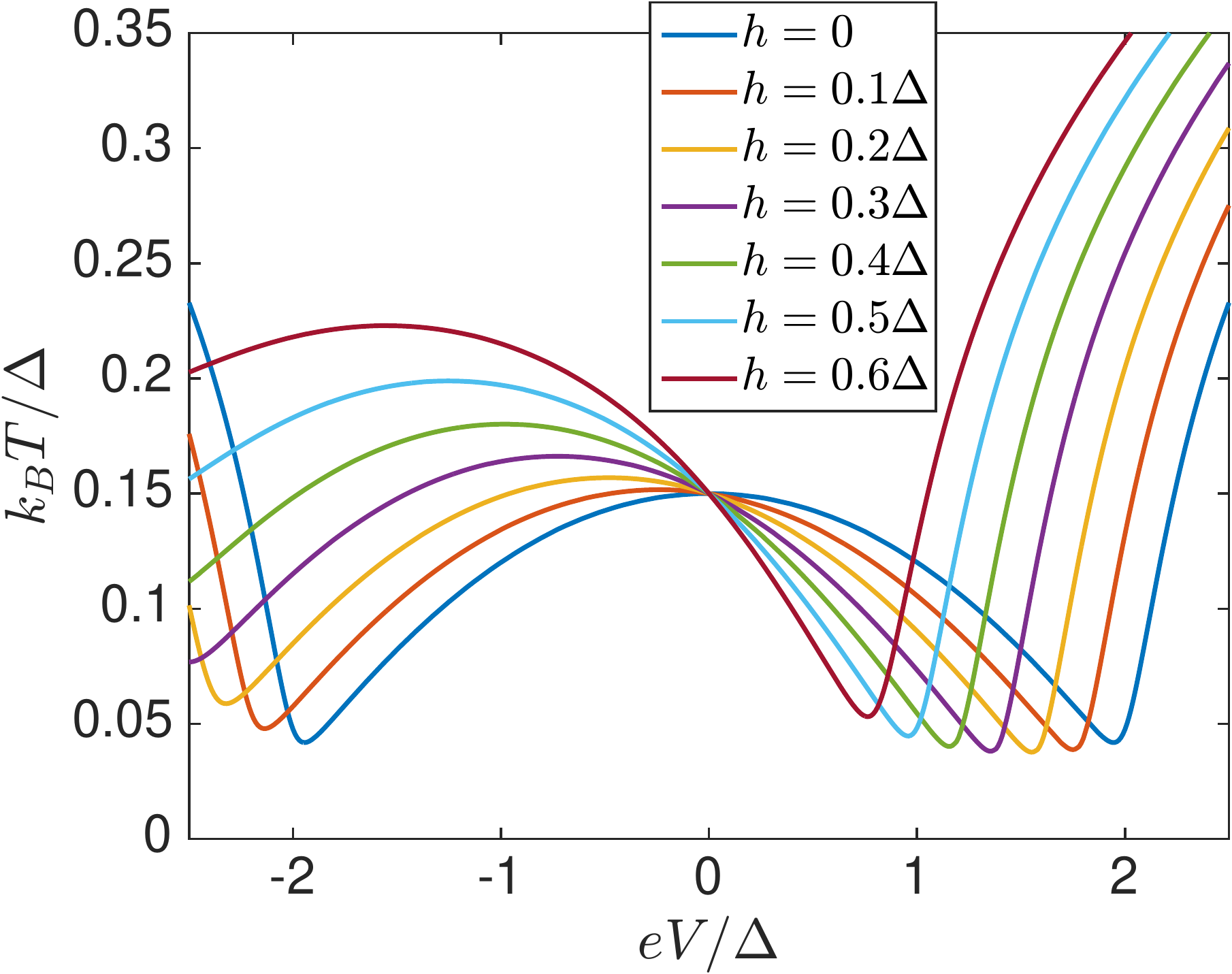}
\caption{\label{fig:islandtemperature}
 Electron temperature obtained with the S-FI-N-FI-S
  electron refrigerator as a function of voltage applied across the
  structure, for different strengths of the exchange field $h$
  inside the superconductor. Top: $k_B T=0.3 \Delta$ and bottom: $k_B
  T=0.15 \Delta$; left: $P=0$, middle: $P=0.5$ and right: $P=1$. The
  electron-phonon coupling strength was chosen to be $\Sigma
  \Omega=100 G_T k_B^5/(e^2 \Delta^3)$.  }
\end{figure}

The resulting island temperatures are shown in
Fig.~\ref{fig:islandtemperature} for different parameters of the
S-FI-N-FI-S junction. We have chosen the parameters so that for
$h=P=0$ they correspond to a typical $T(V)$ curve found experimentally
(see for example \cite{leivo1996}). In the absence of spin filtering, $P=0$,
temperature obtains a minimum at $eV\approx 2(\Delta-h)$, and $T(V)$ is a
symmetric function of voltage. Generally, spin splitting in this case
makes the cooling worse, so that the minimum reached temperature is
higher than in its absence. 

The behavior of $T(V)$ changes in the presence of spin filtering, $P
\neq 0$. In
particular, the curve becomes non-symmetric, and there is cooling even
in the linear response regime, i.e., low voltages. This Peltier effect
is discussed more below. In addition, the minimum reached electron
temperature is generally lowered by an increasing $P$, and for a large
$P \approx 1$, the lowest temperature may be obtained at a non-zero
exchange field. However, this effect seems rather weak for the
considered parameters. 

One possibly relevant aspect of such a magnetic cooler is the fact
that the optimum temperature is obtained at a lower absolute value of
the voltage \cite{kolendaup16}. This translates into a somewhat lowered Joule power
injected to the device. As this heat is dumped into the
superconductor, at the lowest temperatures the heating of the
superconductor becomes the dominant limiting obstacle for cooling
instead of the electron-phonon coupling
\cite{kauppila2013,nguyen2014}. The consequences of this are
  analyzed in \cite{Rouco2017}.

\begin{figure}
\centering
\includegraphics[width=8cm]{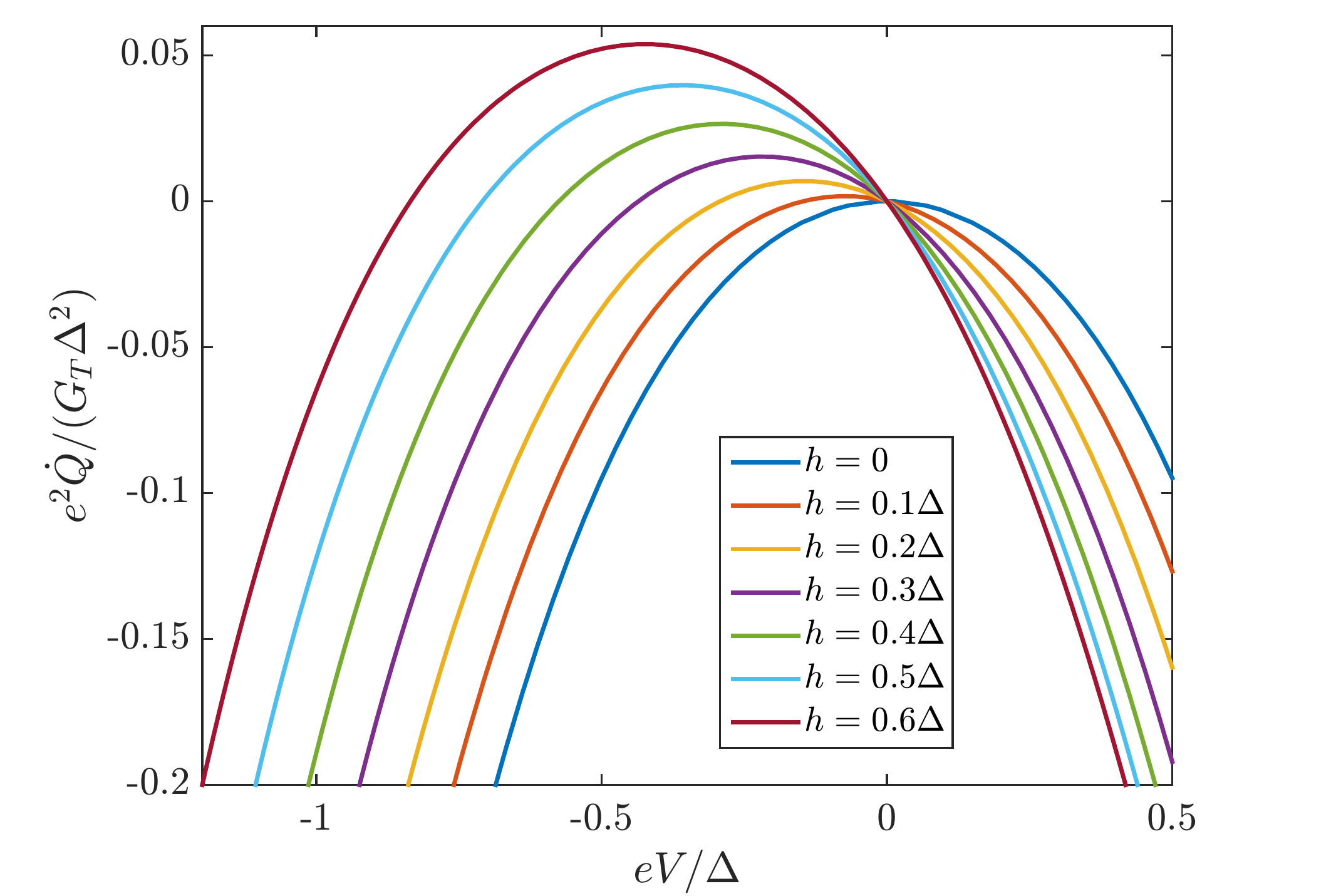}
\caption{Cooling power from the superconductor in a N-FI-S contact
  with a superconductor containing a spin-splitting field $h$. Here
  $P=1$ and $k_B T=0.3\Delta$, close to the maximum cooling
  power. Negative cooling power corresponds to heating.}
\label{fig:Scoolingpower}
\end{figure}

\begin{figure}
\centering
\includegraphics[width=8cm]{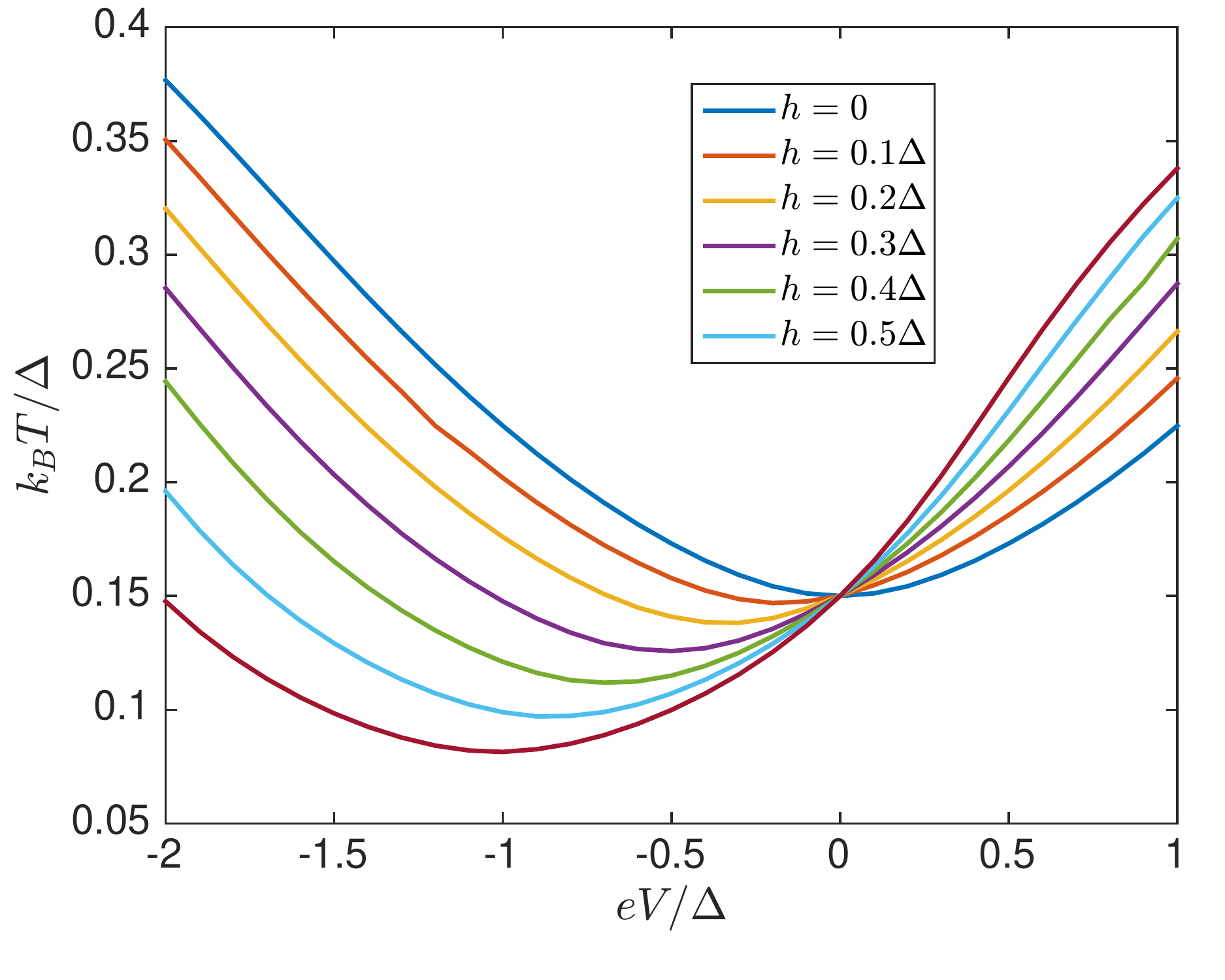}
\caption{Temperature of the spin-split superconducting island in
  contact with two normal or ferromagnetic electrodes via spin filters
of polarizations $P=1$ and $P=-1$, respectively. The magnitude of the
spin-splitting field $h$ is indicated in the legends. The calculation
was done with the electron-phonon coupling strength $\Sigma
  \Omega=100 G_T k_B^5/(e^2 \Delta^3)$. }
\label{fig:Stemperature}
\end{figure}

Besides the normal metal, the magnetic element can be
used to refrigerate the superconductor \cite{Rouco2017,kawabata2013efficient}. We illustrate this by plotting
the cooling power of the superconductor in the case of an N-FI-S
junction in Fig.~\ref{fig:Scoolingpower}. Without spin splitting
and a spin-polarized interface, the cooling power is always negative,
i.e., corresponding to heating. However, since the Peltier cooling
with a non-zero $P$ is
linear in voltage, it has to result to a non-vanishing cooling power
also from the superconductor. The maximum cooling power is less than
in the case of cooling the normal metal
(Fig.~\ref{fig:coolingpower}). However, as indicated in
Fig.~\ref{fig:Stemperature}, the resulting temperature drop is
nevertheless appreciable because the energy gap also weakens the
electron-phonon coupling (see Eq.~\eqref{eq:sephconductance}).

\subsection{Linear response and a heat engine}
\label{subs:linearresponseheatengine}

In the presence of non-vanishing spin polarization $P$ and spin-splitting field $h$, the heat current has a linear component in voltage $V$. This Peltier effect is visible in the $h\neq 0$ curves in Figs.~\ref{fig:coolingpower} and
\ref{fig:Scoolingpower}. In what follows, let us concentrate on low voltages and small temperature differences $eV, k_B \Delta T \ll \Delta$, so that the response of the junction can be described by the linear-response scheme as in Eq.~\eqref{eq:thermoel}. At low temperatures $k_B T \ll \Delta-h$ the thermoelectric coefficients can be evaluated analytically and they are \cite{Ozaeta2014a} 
\begin{subequations}
\label{eq:thermoelcoefsanalytically}
\begin{align}
G&\approx G_T\sqrt{2\pi\tilde\Delta} \cosh(\tilde h)e^{-\tilde\Delta}
\,,\label{eq:NFISconductance}
\\
 G_{\rm th}&\approx \frac{k_B G_T\Delta}{e^2}\sqrt{\frac{\pi}{2\tilde\Delta}}e^{-\tilde\Delta}\left[e^{\tilde h}(\tilde\Delta-\tilde h)^2+e^{-\tilde h}(\tilde\Delta+\tilde h)^2 \right]
\,,
\\
 \tilde \alpha &\approx \frac{G_T P }{e}\sqrt{2\pi\tilde\Delta}e^{-\tilde\Delta}\left[\Delta\sinh(\tilde h)-h\cosh(\tilde h)\right]
\,,\label{eq:alpha}
 \end{align}
 \end{subequations}
with $\tilde \Delta=\Delta/(k_B T)$ and $\tilde h=h/(k_B T)$. In an open-circuit configuration ($I=0$) one rather measures the thermopower 
\begin{equation}
S=\frac{\tilde \alpha}{GT} \approx \frac{P \Delta}{e
T}[\tanh(\tilde{h})-h/\Delta].
\end{equation}
It is maximized for $h=k_B T
{\rm arcosh}[\Delta/(k_B T)]$, where 
\begin{equation}
S_{\rm max} \approx \frac{k_B}{e} P \left[\frac{\Delta}{k_B
    T}-{\rm arcosh}\left(\sqrt{\frac{\Delta}{k_B T}}\right)\right].
    \label{eq:thermopower}
\end{equation}
At low temperatures the thermopower can thus become very large, and within the above scheme it would even diverge for $k_B T \rightarrow 0$. However, in practice this divergence would be cut off via either circuit effects (at $T \rightarrow 0$ the conductance would also tend to zero) or for example the spin relaxation neglected above.  Nevertheless, with proper circuit
design one should be able to measure a thermopower much exceeding
$k_B/e$ in this setup.

\begin{figure}
\centering
\includegraphics[width=8cm]{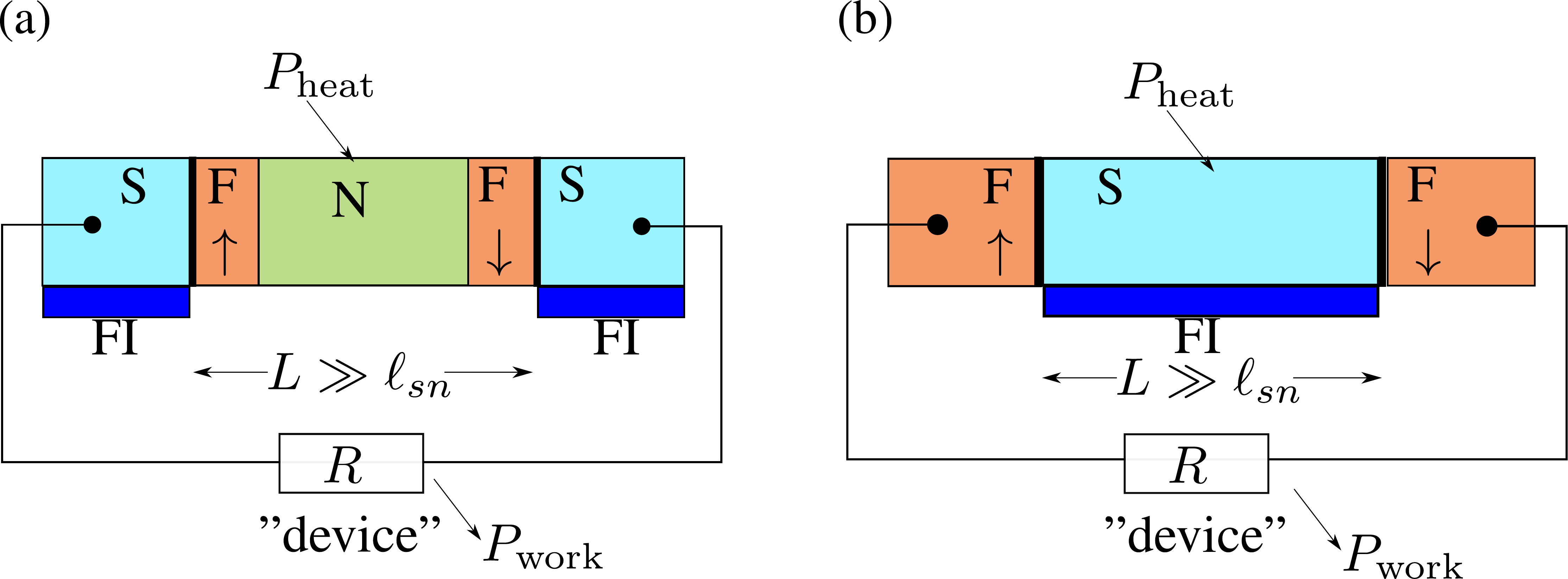}
\caption{\label{fig:sfheatengine} Two realizations of a superconductor/ferromagnet heat engine, where a heating power $P_{\rm heat}$ leading to a temperature increase in the absorber can be converted to electrical power $P_{\rm work}$ and dissipated in the "device" connected to the heat engine. In both cases the setup requires two junctions with antiparallel magnetization directions. The ferromagnetic insulator FI has a magnetization direction along with the two spin-polarized contacts. To disregard spin
  accumulation, the island has to be large compared to the spin
  relaxation length. a) Heat engine with a normal-metal heat absorber. b) Heat engine with a spin-split superconducting
  island. The ferromagnets can also be replaced by a
  normal metal if the interfaces to the superconductor contain a
  ferromagnetic insulator.}
\end{figure}

The prediction for the strong thermoelectric effect was confirmed by a recent experiment by \cite{Kolenda2016} via the measurement of the thermoelectric current $I = -\tilde \alpha \Delta T/T$. This thermoelectric effect also provides a partial explanation \cite{silaev2015long} for the measurements of the long-range non-local spin signal \cite{Hubler2012a,Quay2013}. In what follows, we consider the possibility of realizing a true heat engine with a large figure of merit. We discuss this in the case where either a normal metal or the spin-split superconductor realizes an island and therefore works as the heat absorber, see Fig.~\ref{fig:sfheatengine}.

\begin{figure}
  \centering
\includegraphics[width=8cm]{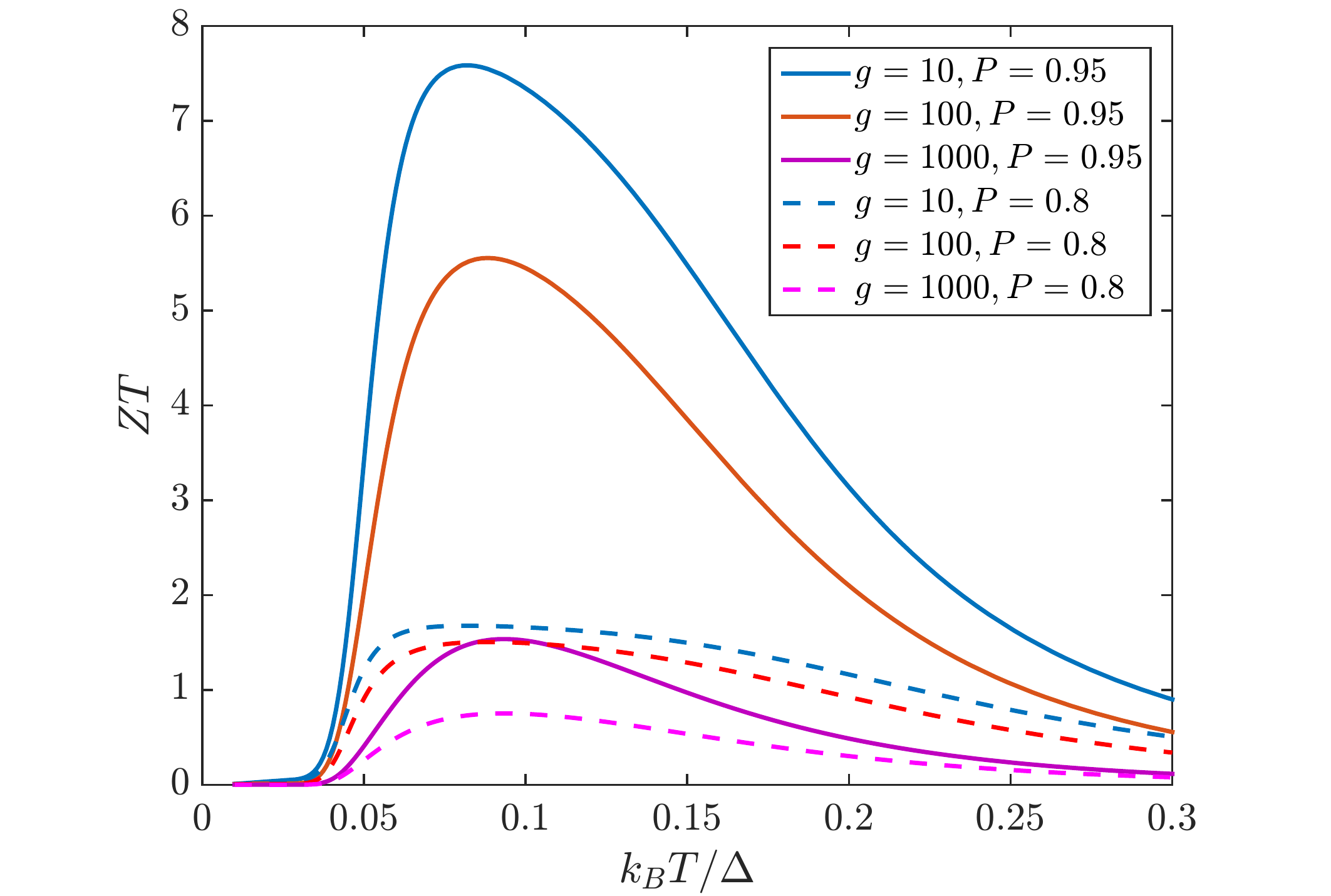}
\caption{\label{fig:ZTsuper1}
 Figure of merit in a S-FI-N-FI-S heat engine as a function of
  temperature for different magnitudes of the electron-phonon heat
  conductance, characterized by the coefficient $g$, and polarizations
  $P$ of the junction. The {curves have} been calculated with
  $h=0.5\Delta$, without calculating $\Delta$ self-consistently. The
  figure of merit is limited from above by $P^2/(1-P^2)$, which for
  $P=0.95$ is 9.3, and for $P=0.8$ it is 1.8.}
\end{figure}

Let us first assume that it is only
the electrons of the ferromagnetic island that are heated, instead of both
the electrons and phonons.\footnote{The latter would require isolating
  the phonons of the island from those of the substrate, which can be
  realized for example by suspending the island.} In this case the figure of merit of the heat engine is limited by the spurious heat conduction due to the
electron-phonon coupling, with heat conductance $G_{\rm e-ph}=5 \Sigma
\Omega T^4$ [see Eq.~\eqref{eq:normaleph}],
and from the heat conductance to the extra superconducting electrodes
$S$, with an exponentially suppressed heat conductance, provided the
heating current is small.  In this case the heat engine has an efficiency described
by a figure of merit that for $k_B T \ll h$ and $P \neq 1$ tends to
\begin{equation}
ZT=\frac{P^2 \pi (\Delta-h)^2 \Delta^4}{g e^{(\Delta-h)/(k_B T)}
  T^{5} \sqrt{k_B T \Delta} +
  (1-P^2) \pi (\Delta-h)^2 \Delta^4 + \pi \Delta^4 (h+\Delta)^2 e^{-2h/(k_B
  T)}},
\end{equation}
where $g=5 k_B^5 \sqrt{2\pi} e^2 \Sigma \Omega\Delta^3/(2 G_T)$ is a
dimensionless quantity characterizing the
strength of electron-phonon coupling. Here we take into account the
fact that the thermoelectric effect takes place across both contacts,
and hence all quantities except the spurious heat conduction channel
are doubled. If we do not describe
half-metals, i.e., $1-P^2 \gg (h+\Delta)^2/(h-\Delta)^2 \exp[-2h/(k_B
T)]$, we can simplify this more by dropping the last term in
the denominator. We plot this as a function of temperature in
Fig.~\ref{fig:ZTsuper1} for some representative parameters. For
example, for $\Omega=0.005$ $\mu$m$^3$, $\Sigma=10^{9}$ W
$\mu$m$^{-3}$K$^{-5}$ and $1/G_T=3$ k$\Omega$, $g=100$. The figure of merit exhibits a peak at around
$k_B T \approx 0.1 \Delta$, and its peak value depends strongly on the
relative strength of the spurious heat conduction channels. As seen in
the figure, reaching high values for the figure of merit is quite
challenging in this way, as besides a high polarization close to one,
it requires quite low normal-state resistance of the junctions, a
combination not very easy to reach with ferromagnetic insulators.

Note that often such spurious heat conduction mechanism that limit the highest available $ZT$ are disregarded from the theoretical analysis of the figure of merit, for example in the case of quantum dots \cite{hwang16}.

\begin{figure}
\centering
\includegraphics[width=8cm]{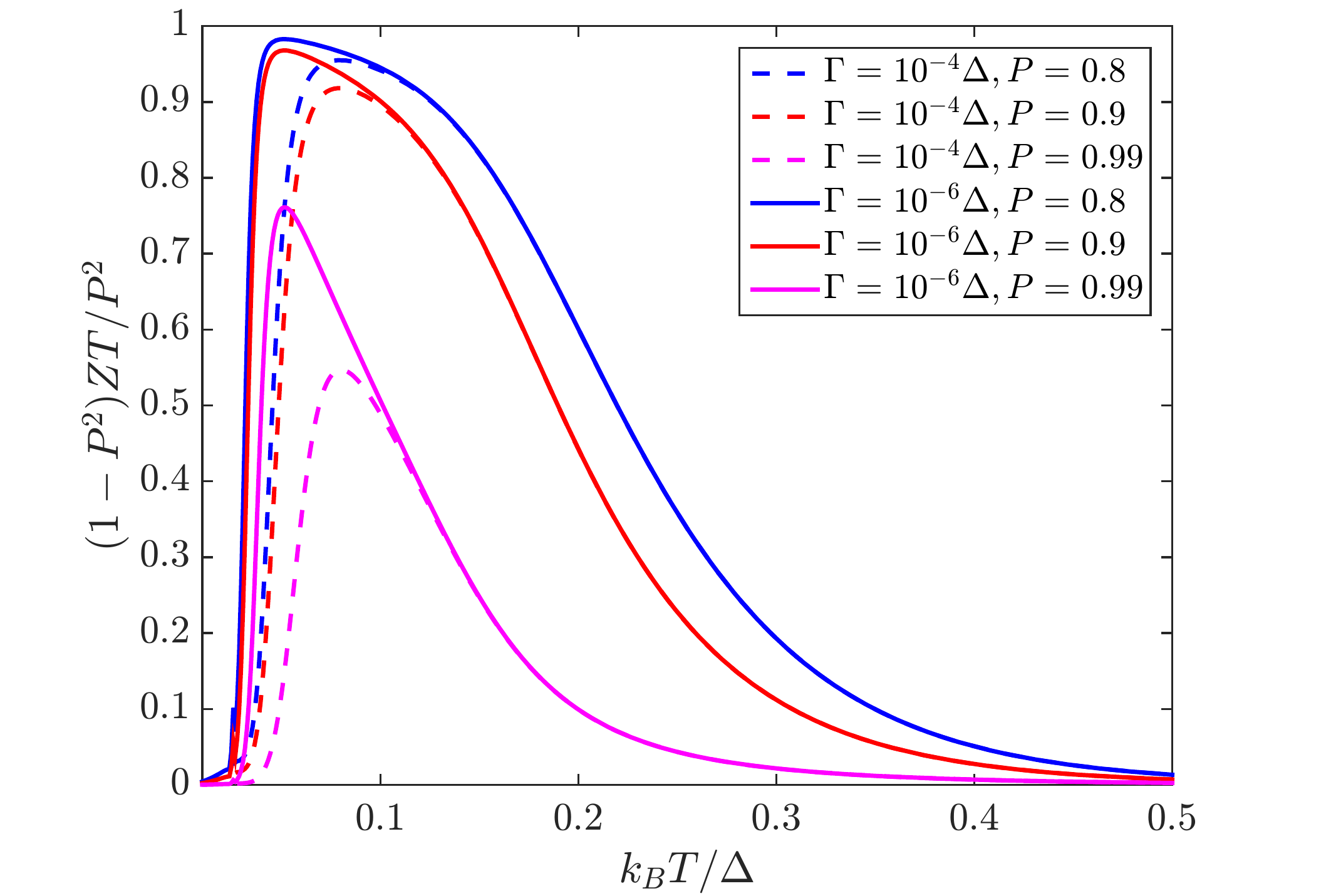}
\caption{\label{fig:ZTsuper2}
  Figure of merit in a N-FI-S-FI-N heat engine as a function of
  temperature for polarizations $P$ of the junction. The figure has been calculated with
  $h=0.5\Delta$ and $g=1000$, without calculating $\Delta$ self-consistently. The
  figure of merit at low temperatures reaches very close to
  $P^2/(1-P^2)$ unless $P$ is very close to unity, but the
  exact temperature scale where this happens depends on the value of
  polarization. At the lowest temperatures $ZT$ is limited by another
  spurious heat conduction process, due to nonzero density of states
  inside the gap, described here by the Dynes $\Gamma$ parameter }
\end{figure}

Another possible setting for the heat engine is the one where the
island is the spin-split superconductor, and it is connected to
normal-metal or ferromagnetic electrodes via spin filters as in
Fig.~\ref{fig:sfheatengine}(b). In the linear response regime this is
otherwise similar to the previous case, except that the
electron-phonon coupling inside the island has an exponentially
suppressed heat conductance, Eq.~\eqref{eq:sephconductance}. In this
case, as long as the pure tunneling limit remains valid, and the
island volume is not overly large, 
 the low-temperature behavior of
$ZT$ is dictated mostly by the deviation of the polarization from
unity as shown in Fig.~\ref{fig:ZTsuper2}. However, because of the
exponentially decaying heat conductance, also in this case other
spurious processes besides electron-phonon coupling start to
limit $ZT$, making it vanish at $T \rightarrow 0$. In the case of
Fig.~\ref{fig:ZTsuper2}, the simplest model for them is due to the small but nonvanishing
density of states assumed for the superconductor \cite{Ozaeta2014a},
and described by the Dynes parameter $\Gamma$ \cite{dynes84}.\footnote{All spectral
  functions, such as the density of states, are calculated here by
  assuming a nonzero imaginary part of size $\Gamma$ in the
  energy. Typically the chosen value for $\Gamma$ is so small that it
  does not affect practical quantities. Figure \ref{fig:ZTsuper2} is
  an exception.}

When the heat input into the heat engine is due to electromagnetic radiation coupled to the device, the conversion of this input power to direct charge energy allows realizing a new type of a superconducting thermoelectric detector of radiation \cite{heikkila2018,chakraborty2018}. In addition, the thermoelectric effects could be used in non-invasive low-temperature thermometry \cite{giazotto2015b} so that the measured thermovoltage would indicate the temperature profile. At room temperature a scanning thermometer based on thermoelectric effects was realized by \cite{menges16}. Utilizing spin-polarized tunneling from spin-split superconductors would allow extending this technique to low temperatures. 

\subsubsection{Effect of spin mixing}
\begin{figure}
\centering
\includegraphics[width=8cm]{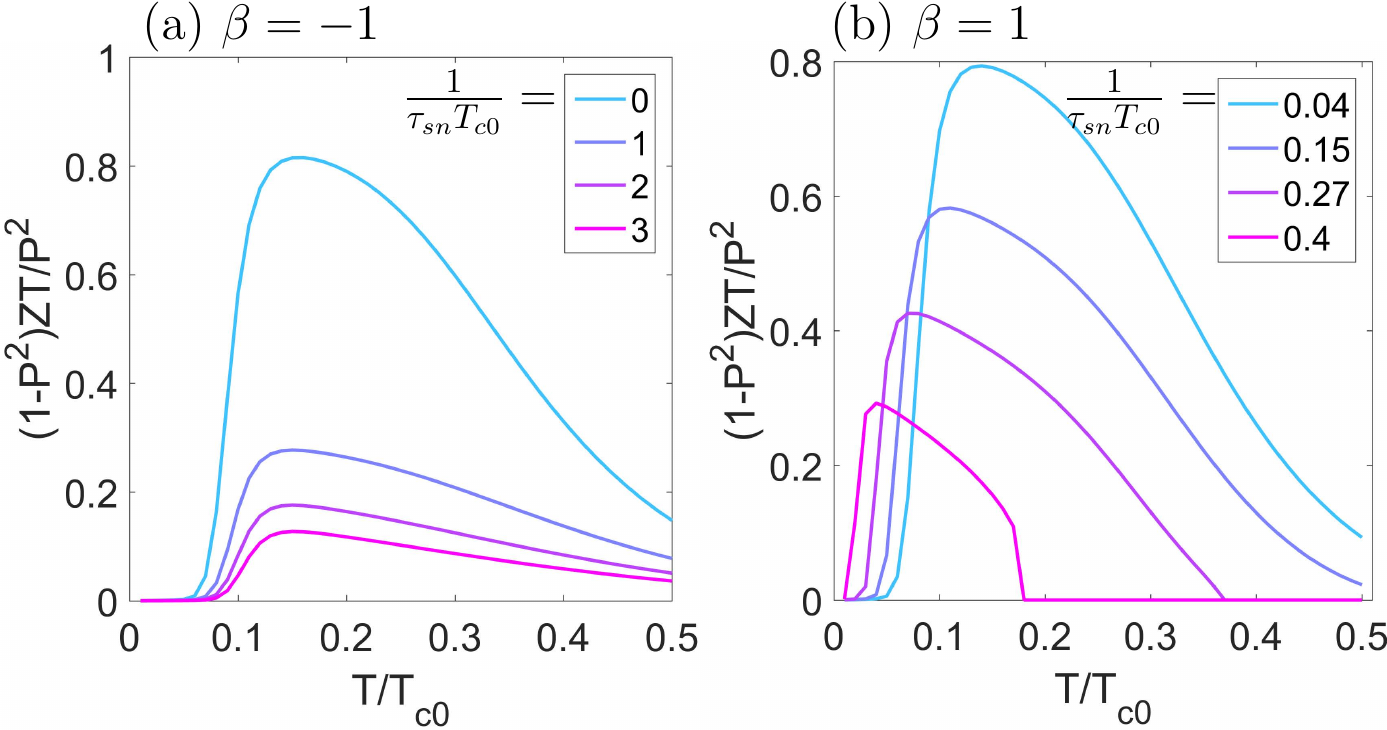}
\caption{
Figure of merit in a N-FI-S-FI-N heat engine as a function of
  temperature in the presence of (a) spin-orbit mechanism of relaxation
  with $\beta=-1$ and (b) spin-flip mechanism of relaxation
  with $\beta=1$. Thy Dynes parameter is $\Gamma = 10^{-4} T_{c0}$, exchange field $h=0.5 \Delta_0$ and $P=0.9$.
  Curves in  (a,b) correspond to
  different spin relaxation rates $1/(\tau_{sn}T_{c0})$. The electron-phonon coupling is disregarded.}
  \label{fig:ZT-SpinOrbita}
\end{figure}

Rather than the Dynes parameter, a more relevant limitation of the  figure of merit in spin-split
superconductors is due to the spin mixing caused by spin-orbit and spin-flip scattering that are disregarded in the above results.  This becomes an issue especially in heavy-metal based superconductors where the spin-orbit scattering is strong. As an example, Fig.~\ref{fig:ZT-SpinOrbita} shows how the figure of merit decays as the  spin-orbit relaxation  becomes stronger.  The high efficiencies are obtained only when the spin relaxation rates are small compared to the  gap energy $\Delta$. Note that for some materials the stronger spin-orbit relaxation due to the increase in the atomic number $Z$ (say, compared to Al) is  partially compensated by the increased critical temperature and therefore the energy gap.

\begin{figure}
\centering
\includegraphics[width=8cm]{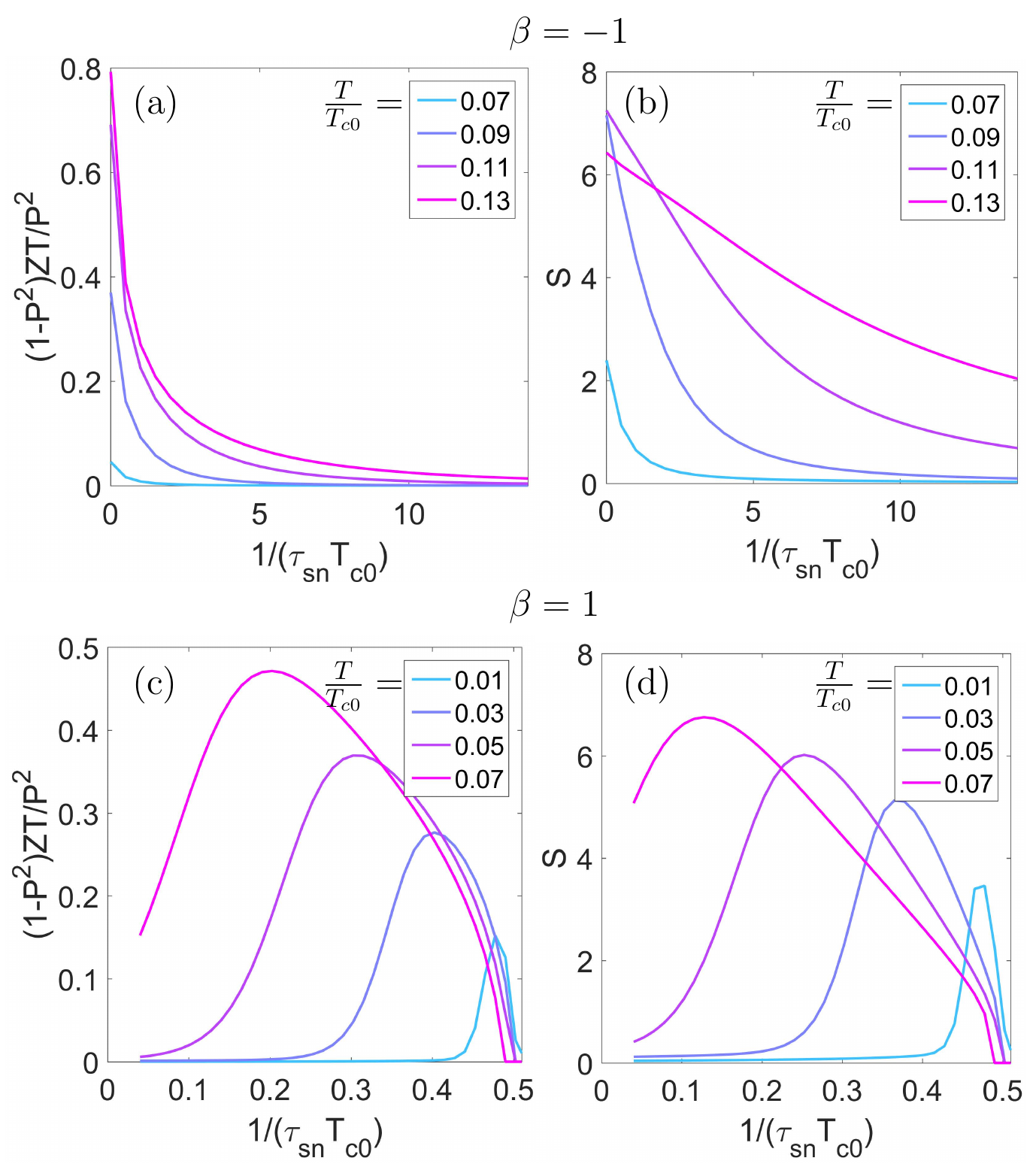}
\caption{
(a,c) Figure of merit in a N-FI-S-FI-N heat engine and 
 (b,d) Seebeck coefficient of the S-FI-N junction as a functions of spin relaxation rate $\tau_{sn}^{-1}$. 
The FI barrier  polarization is $P=0.9$,
exchange field $h=0.5\Delta_0$ and Dynes parameter $\Gamma = 10^{-4} T_{c0}$.
 Panels in the upper row (a,b) correspond to spin-orbit relaxation $\beta=-1$ and in the lower row (c,d) are for the spin-flip relaxation $\beta=1$.
Curves from top to bottom correspond to an increasing temperature.  The electron-phonon coupling is disregarded.   }
  \label{fig:ZT-SpinOrbit}
\end{figure}

As shown in in Fig.~\ref{fig:ZT-SpinOrbit}, the suppression of $ZT$ is monotonous {as a function of an increasing spin-orbit scattering. However, in the presence of a non-vanishing Dynes parameter, the spin-flip scattering may also lead to an increased $ZT$ at low temperatures.}  
For completeness, we also show the effect of spin mixing on the thermopower, i.e., the Seebeck coefficient. The results of the effect of spin-flip scattering are in line with those studied in \cite{rezaei2018}.

\subsubsection{Nonlinear response heat engine}
\begin{figure}
\centering
\includegraphics[width=8cm]{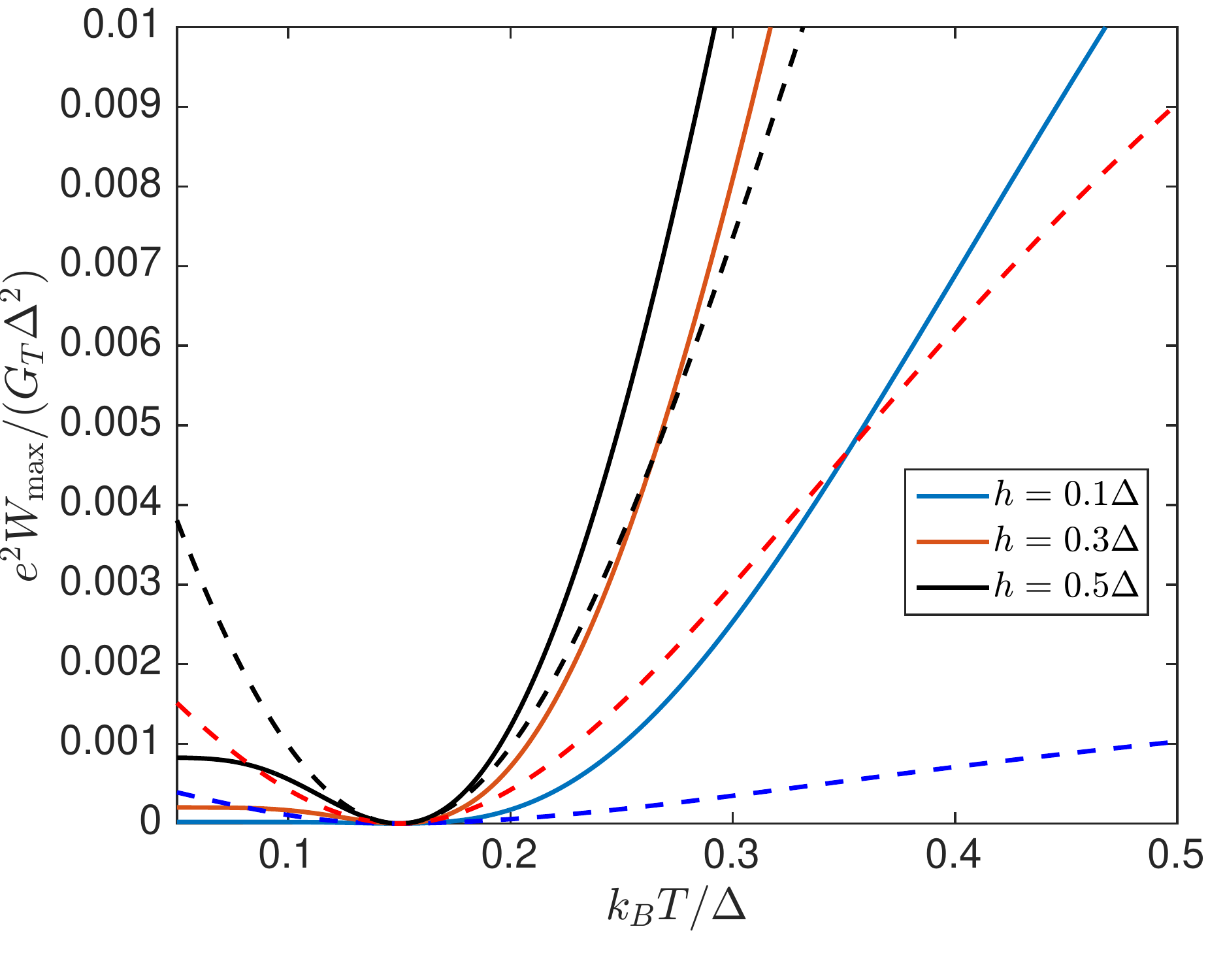}
\caption{Maximum extractable power from a spin-filter contact to a spin-split superconductor. Solid lines depict the situation with $T=T_S$, $k_B T_N=0.15 \Delta$, and dashed lines the one with $T=T_N$, $k_B T_S=0.15\Delta$. The plot has been calculated with $P=0.95$, $\Gamma=10^{-4} \Delta$ and without self-consistency of $\Delta$ (hence everything should be normalized with respect to $\Delta(T_S,h)$). }
\label{fig:thermoelpower}
\end{figure}

\begin{figure}
\centering
\includegraphics[width=8cm]{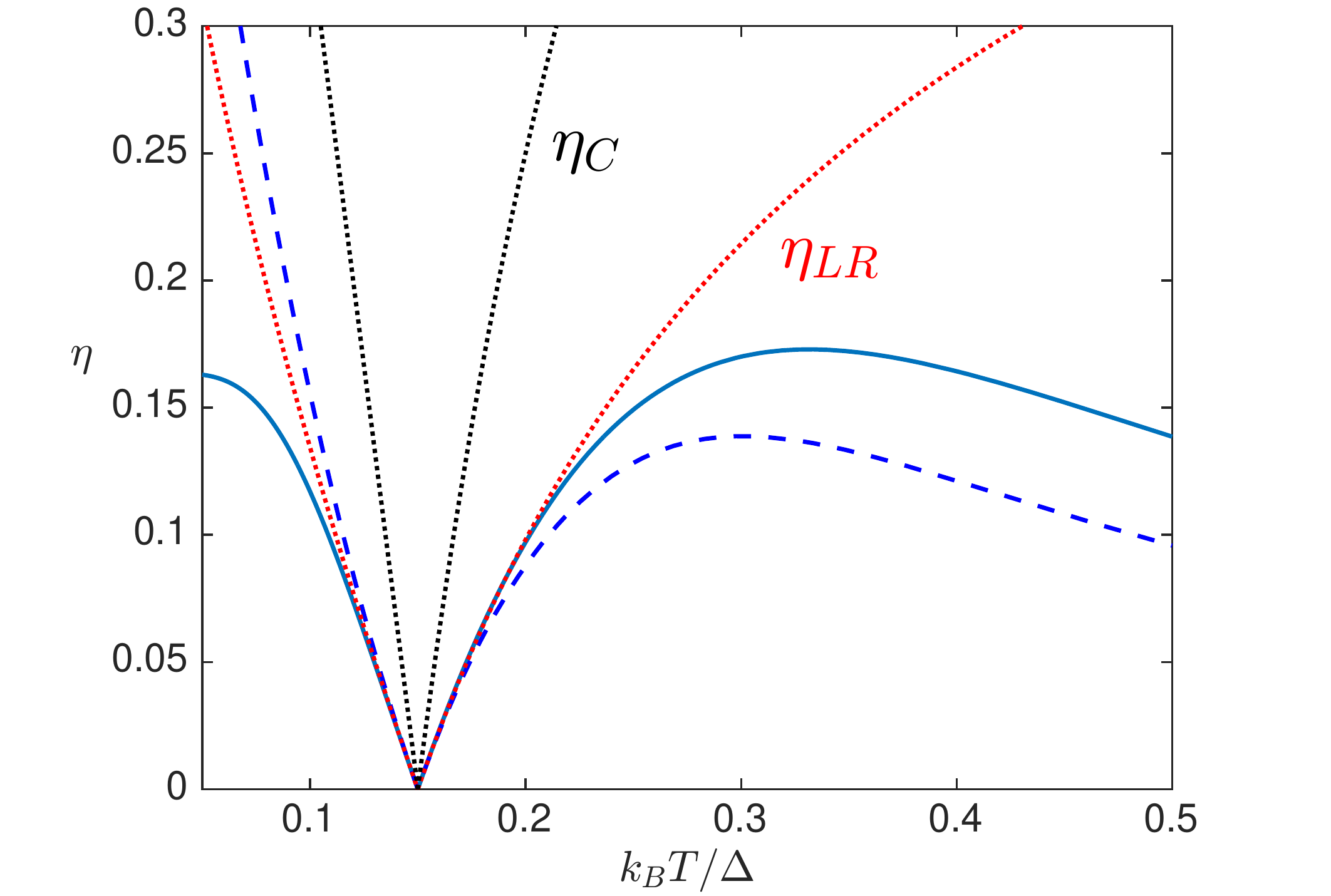}
\caption{Efficiency of the thermoelectric power conversion of a spin-filter junction to a spin-split superconductor, neglecting spurious contributions to the heat current. Solid lines: $T=T_S$, $k_B T_N=0.15 \Delta$, and dashed lines: $T=T_N$, $k_B T_S=0.15\Delta$. The plot has been calculated with $h=0.6\Delta$ and otherwise as in Fig.~\ref{fig:thermoelpower}. Dotted lines show the Carnot efficiency $\eta_C=1-T_{\rm cold}/T_{\rm hot}$ and the linear response result $\eta_{LR}=ZT/(2+ZT) (1-\sqrt{T_{\rm cold}/T_{\rm hot}})$.}
\label{fig:thermoelefficiency}
\end{figure}

The above calculations on the heat engine efficiency have been done in
the strict linear response limit $\Delta T \ll T$. In this limit both the output power and the
Carnot efficiency, proportional to $\Delta T/T$, are vanishingly
small. However, the device can have a rather large output power and efficiency also at
nonlinear response. To illustrate this, we plot the maximum extractable output power $\dot W_{\rm max}={\rm max}_V(-IV)$ from a N-FI-S junction in Fig.~\ref{fig:thermoelpower} at large differences $T_S-T_N$ between the spin-split superconductor and a normal-metal reservoir.  The output power is somewhat larger for $T_S > T_N$ than the opposite situation with the same temperature difference.  The plotted quantity corresponds to the maximum power that is possible to extract from the junction after optimizing the load resistance --- note, however, that this optimized resistance value depends on the exact temperature difference and other parameters. The corresponding efficiency $\eta=\dot W_{\rm max}/\dot Q$ is plotted in Fig.~\ref{fig:thermoelefficiency}. There we take into account only the heat current flowing through the junction itself. The efficiency is reduced if other heat relaxation mechanisms become relevant.  

\subsubsection{Effect of non-collinear magnetizations}
\label{subs:noncollinear}
Above we assume that the polarization ${\mathbf P}=P \hat u_P$ of the tunneling contacts is collinear with the exchange field ${\mathbf h}=h \hat u_h$ in the spin-split superconductor. This is the limit that guarantees the largest thermoelectric current, but it is also interesting to consider how non-collinearity would show up in the thermoelectric response, in particular in the coefficient $\tilde\alpha$ of Eq.~\eqref{eq:alpha}. This coefficient is by definition a scalar quantity. Therefore, it can only depend on a scalar combination of the vectors ${\mathbf P}$ and ${\mathbf h}$, or rather, $\hat u_P$ and $\hat u_h$. This combination is the inner product, $\hat u_P \cdot \hat u_h$. Therefore, the generalization of Eq.~\eqref{eq:alpha} to the non-collinear case is
\begin{equation}
\tilde \alpha_{nc}=\hat u_P \cdot \hat u_h \tilde \alpha_c,
\label{eq:alphanc}
\end{equation}
where $\tilde \alpha_c$ is the thermoelectric coefficient in the collinear case, for example given by Eq.~\eqref{eq:alpha}.

\subsection{Thermally induced spin currents\label{subs:spinseebeck}}

{Besides the large thermoelectric effect discussed above, the
  contact between spin-split superconductors with other conducting
  materials can exhibit a large (longitudinal) spin Seebeck effect,
  where a temperature difference drives spin currents to/from the
  spin-split superconductor \cite{Ozaeta2014a}. The spin current
  $I_S=I_\uparrow-I_\downarrow$ can be obtained from
  Eq.~\eqref{th_is}. In linear response, Eq.~\eqref{Eq:Onsager}, it is
  described by the same coefficient $\tilde \alpha/P = \alpha$ as the thermoelectric
  coefficient, Eq.~\eqref{eq:alpha}, but without the spin polarization
  $P$ of the interface.  When either of the two materials realizes an
  island, the spin current can be converted into a spin accumulation
  $\mu_z$ that is determined from  the balance between thermally
  induced spin currents and spin relaxation within the island. For
  weak spin relaxation, the resulting spin Seebeck coefficient $
  \mu_z/\Delta T$ can be large [{\it i.e.}, not proportional to the number of thermally
  excited quasiparticles $\sim  e^{-(\Delta-h)/(k_B T)}$ as the terms in Eqs.~\eqref{eq:NFISconductance}-\eqref{eq:alpha}], similar to the thermopower in Eq.~\eqref{eq:thermopower}.  The spin currents induced in the case of two spin-split superconductors, and the additional effects of Josephson coupling, magnetization texture and spin-orbit effects are discussed in \cite{linder2016,bathen2017}.

  The above discussion of the thermoelectric cooling and heat engines disregards the spin accumulation caused by the nonequilibrium biases. This assumption is valid if the overall spin relaxation within the island, described by a spin-flip conductance $G_{\rm sf}=e^2\nu_F \Omega/\tau_{sn}$ is much larger than the conductances of the tunnel junctions. Here $\nu_F$ is the normal-state density of states in the Fermi level, $\Omega$ is the volume of the island, and $\tau_{sn}^{-1}$ is the spin relaxation rate within the island.
In the case of a spin-split superconducting island, both the
conductance and the spin relaxation rate contain the same exponential
term $\sim e^{-(\Delta-h)/(k_B T)}$, and for an estimate of the
overall magnitude of the effect, it is hence enough to compare $G_{\rm
  sf}$ in the normal state to the normal-state tunnel conductance
$G_T$. In other words, disregarding the spin accumulation is justified
provided $e^2 \nu_F \Omega/(\tau_{sn} G_T) \gg 1$. On the other hand,
most of the discussion disregards the effect of spin relaxation on the
density of states of the spin-split superconductor. As shown in
Fig.~\ref{fig:ZT-SpinOrbit}, spin relaxation starts to affect the
results when $\hbar/\tau_{sn} \gtrsim k_B T_c$. These two constraints
can be simultaneously satisfied depending on the precise value of
$\tau_{sn}$, provided that $G_T \ll (k_B T_c/\delta_I) e^2/h$, where $\delta_I=1/\nu_F \Omega$ is the average energy level spacing on the island. This constraint is always satisfied in tunnel barriers between metallic systems.

To our knowledge, the spin Seebeck effect has not been directly
measured in spin-split superconductors driven by a temperature
difference. However, outside linear response, this is the effect that
causes the long-range spin signal in spin injection experiments as
discussed in Sec.~\ref{sec:noneq}.  Besides affecting the
nonequilibrium dynamics,  the presence of spin accumulation in a
superconductor with an exchange field couples to the self-consistency
equation for the superconducting order parameter
$\Delta$ as in Sec.~\ref{sec:selfconsistentgap}. Ref.~\cite{bobkova2017} utilized this fact and predicted that
the thermally induced spin accumulation in a spin-split superconductor
can result to changes in the critical temperature: besides only
suppressing it at low temperatures, it can enhance the critical
temperature at some intermediate temperatures, or even lead to a
situation where superconductivity shows up only between two critical temperatures.

This spin Seebeck effect should be contrasted to the analogous phenomenon discussed in non-superconducting materials \cite{Uchida2014}. There, {a} major contribution to the spin Seebeck signal {comes from} the thermally induced spin pumping from ferromagnetic insulators \cite{hoffman2013}. 

}

\section{Summary}\label{sec:outlook}

In this work we review the electronic transport  properties of superconducting hybrid structures with a spin-split density of states. We mainly focus on ferromagnetic insulator-superconductor  systems (FI-S) in which such a splitting field  can be achieved  without the need of an applied magnetic field.
Best FI-S  material combinations {studied so far} are the  europium chalcogenides (EuO, EuS and EuSe) together with Aluminum films.  Interesting of those materials is that thin films of EuO or EuS can also be used as highly efficient  spin filters and  therefore they  are the best candidates for the realization of  the heat engines proposed  in Sec.~\ref{thermoel}. 

One of the goals of this review is to provide a complete theoretical framework to study non-equilibrium effects in superconducting structures with spin-split density of states. 
In this respect, Sec.~\ref{sec:quasiclassical_theory} gives a detailed introduction to the quasiclassical formalism  for diffusive hybrid  systems, and a description of the nonequilibrium modes. The advantages of using this formalism are multiple. On the one hand it provides 
a simple way to identify the quasiparticle nonequilibrium modes.  As shown in Sec.~\ref{sec:noneq-modes}, the combination between superconductivity and magnetism produces, besides the widely studied charge and energy modes, two additional ones: spin and spin energy. The quasiclassical equations show explicitly how all these modes couple to each other in different situations. 

On the other hand, quasiclassical kinetic equations are a powerful tool for the study of hybrid multi-terminal systems, with different materials. 
Effects occurring at hybrid interfaces between different materials can be described by the help of effective boundary conditions.  In Sec.~\ref{sec:hybrid_interfaces} we discuss this issue and present boundary conditions suitable to describe a S-FI interface. 
The inclusion of the interfacial spin-orbit effect and the associated  coupling between charge and spin would be  another interesting further development of the formalism.
The extension of the formalism  to include  additional types of interactions and couplings is in principle  rather straightforward.  For example, 
one can include  magneto-electric effects associated with the charge-spin coupling 
in systems with linear-in-momentum  spin-orbit coupling,  by introducing an effective SU(2) gauge potential in the quasiclassical equations \cite{Bergeret:2013il,bergeret2014spin,bergeret2016manifestation}, or, in the case of an extrinsic spin-orbit coupling, by taking into account higher-order terms in the impurity potential \cite{huang2018extrinsic}. Inclusion of such effects in a non-equilibrium situation 
and with time-dependent fields, is  an interesting further development of the field 
\cite{espedal2017}.

We use  the power of the quasiclassical approach in Sec.~\ref{spininjection}, where we study in detail different transport properties of FI-S structures and spin-split superconductors and contrast the results with experimental findings. Both charge and spin dependent properties  are determined by the behaviour of the nonequilibrium modes of the supercoductor. Specifically we have computed the non-local conductance in a two terminal geometry and the spin Hanle effect in superconductors.

One further perspective of the present work is the extension of the Keldysh quasiclassical  formalism  used in  this review to include magneto-electric effects associated with the spin-orbit coupling (SOC) \cite{Bergeret:2013il,bergeret2014spin,bergeret2016manifestation,huang2018extrinsic, konschelle2015theory,espedal2017}, 
and the coupling   magnetization dynamics 
 to the electronic degrees of freedom via the reciprocal effects of spin transfer torque and spin pumping \cite{tserkovnyak2005}.

In Sec.~\ref{sec:acdynamics} we  discuss  the dynamics of  spin-split superconductors  in rf fields, and review how nonequilibrium effects induced by them manifest in the quasiclassical theory. Historically, magnetic resonance effects in superconductors are well studied, but fewer experiments have probed spin-split thin films.

Another very interesting aspect in studying superconductors with a spin-split density of states are their thermoelectric properties.  In Sec.~\ref{thermoel} we focus on  thermoelectricity in superconducting structures, both in linear and non-linear regimes and discuss the realization of a heat engine based on FI/S structures.  Because of their  extremely  large thermoelectric  figure of merit, using S/FI structures has been proposed for accurate radiation sensing and non-invasive scanning thermometry.  These recently proposed applications add up to the    the rich physics offered by spin-split superconductors, which  have been long used as tools to characterize equilibrium properties of magnets, especially their spin polarization.  
Most likely  there are also many other applications to be uncovered, opened by the possibility for realizing a controlled combination of magnetism and superconductivity. The theoretical framework presented in this review should  help in this task.

The major  challenge in fabricating real devices is to find FI-S combinations 
with a large superconducting critical temperature and simultaneously a large spin splitting and efficient spin-filtering. 
Superconductors with high $T_c$ usually have larger spin-orbit coupling which leads to a broadening of the spin-split peaks in the density of states, as discussed in Sec.~\ref{sec-superwithh}.  Further material research is needed   in this respect in order to find optimal material combinations.

Finally,   the  phenomena discussed in this  review have a direct impact on 
several fields of condensed matter and quantum technologies, as  for example the design of 
hybrid quantum materials and their use in functional devices.  
Transport properties of such  systems can be controlled by the temperature or by external fields, and  to interpret  measurements  it is essential to understand their  nonequilibrium properties, 
which in turn depend on the quality of interface growth and material combination.  
This review provides most of the theoretical tools for such  analysis.

\section*{Acknowledgments}

We thank Faluke Aikebaier, Marco Aprili, Detlef Beckmann,  Wolfgang Belzig, Irina Bobkova, Alexander Bobkov, Giorgio De Simoni,  Matthias Eschrig, Yuri
Galperin, Francesco Giazotto, Vitaly
Golovach, Alexander Mel'nikov, Jagadeesh Moodera,  Risto Ojaj\"arvi,
Asier Ozaeta, Charis Quay, Jason Robinson,  Mikel Rouco, and Elia Strambini   for useful discussions. This work was
supported by the Academy of Finland Center of Excellence (Project No. 284594), Research Fellow  (Project No. 297439) and Key Funding (Project No. 305256) programs and {project number 317118}, the European Research Council (Grant No. 240362-Heattronics), {the
Spanish Ministerio de Ciencia, Innovacion y Universidades (MICINN) (Project No. FIS2017-82804-P),
the European Research
Council under the European Union's Seventh Framework Program
(FP7/2007- 2013)/ERC Grant agreement No. 615187-COMANCHE,  Horizon 2020 and innovation programme under grant agreement No. 800923-SUPERTED.}


\section*{List of Symbols}

\hskip-1.2em
\begin{longtable}{p{10em}p{14em}}
S & superconductor\tabularnewline
FI & ferromagnetic insulator\tabularnewline
$G_{T}$ & tunneling conductance\tabularnewline
$\nu_F$ & normal-state density of states (per spin) at the Fermi level\tabularnewline
$D$ & diffusion constant \tabularnewline
$\sigma_N=2 e^2 \nu_F  D$ & normal-state conductivity in a diffusive wire\tabularnewline
$\check g$ &  8$\times$8 Green's function matrix in Keldysh-Spin-Nambu \tabularnewline
$\hat g$ &  4$\times$4 Green's function matrix in Spin-Nambu \tabularnewline
$N_{\sigma}(\varepsilon)$, $\sigma=\uparrow,\downarrow$ & normalized density of states per spin \tabularnewline
$N_{+}$ & $N_{\uparrow}+N_{\downarrow}=\Re{}g_{03}$ \tabularnewline
$N_{-}$ & $N_{\uparrow}-N_{\downarrow}=\Re{}g_{33}$ \tabularnewline
$n_{F}(\varepsilon)$ & Fermi distribution function\tabularnewline
$f_{eq}=1-2n_{F}$ & quasiclassical equilibrium distribution function\tabularnewline
$f_{T}(\varepsilon)$ & non-equilibrium transversal distribution function \tabularnewline
$f_{L}(\varepsilon)$ & longitudinal distribution \tabularnewline
$f_{Tj}(\varepsilon)$, $j=1,2,3$ & spin transversal function\tabularnewline
$\sigma_{j}$, $j=1,2,3$ & Pauli matrices in spin space\tabularnewline
$\tau_{j}$, $j=1,2,3$ & Pauli matrices in Nambu space\tabularnewline
$g_{ij}$ & $\sigma_i\tau_j$ component of the retarded Green function \tabularnewline
$\tau$ & elastic relaxation time\tabularnewline
$\tau_{sf}$ & spin-flip scattering time\tabularnewline
$\tau_{so}$ & relaxation time at impurities with spin-orbit coupling\tabularnewline
$\tau_{sn}=[\tau_{sf}^{-1}+\tau_{so}^{-1}]^{-1}$ & normal state spin relaxation time\tabularnewline
$R_{\square}$ & barrier resistance per unit area\tabularnewline
$j_{c,e,s,se}$ & charge,energy,spin,spin-energy density currents\tabularnewline
$I$ & charge current\tabularnewline
$\dot{Q}$  & energy current\tabularnewline
$I_{s}$ & spin current \tabularnewline
$\dot{Q}_{s}$  & spin-energy current\tabularnewline
$P$ & polarization or spin-filter efficiency of FI interface\tabularnewline
$\Delta$ & superconducting order parameter\tabularnewline
$\Delta_0$ & value of $\Delta$ at $T=h=0$ and without spin
             relaxation\tabularnewline
$T_{c0}=\Delta_0/1.76$ & BCS critical temperature for $h=0$ and
                         without spin relaxation\tabularnewline
$D$ & Diffusion coefficient \tabularnewline
$G$ & electrical conductance\tabularnewline
$G_{th}$  & thermal conductance\tabularnewline
$\xi_s$ & superconducting coherence length \tabularnewline
\end{longtable}

\section*{References}

\bibliography{superferro}

\appendix

\section{Conservation of spin and charge by the electron-phonon scattering }
\label{app:ephonon}
 The inelastic electron-phonon scattering solely cannot relax charge the spin polarization and therefore
 it should keep $\mu$ and $\mu_z$ constant. 
   The electron-phonon collision integral (CI) reads
   \begin{align}\label{Eq:e-ph} 
   &  I_{qp-ph} (\varepsilon) = \frac{g_{ph}}{4}\int_{-\infty}^{\infty} 
    d\omega |\omega|\omega \hat K(\varepsilon,\varepsilon^\prime)
    \\ \nonumber
   & \hat K =  \left[ 
    \hat\Sigma^R_{ph} \hat g^K (\varepsilon)- \hat g^K (\varepsilon) \hat\Sigma^A_{ph}-
    \hat g^R (\varepsilon) \hat\Sigma^K_{ph} + \hat\Sigma^K_{ph} \hat g^A (\varepsilon)
    \right]
   \end{align}
   where    $\omega=\varepsilon^\prime-\varepsilon$ and the self energies are
   \begin{align} \nonumber
       & \hat\Sigma^K_{ph} (\varepsilon,\varepsilon^\prime) = \coth (\omega/2T) \hat g^K(\varepsilon^\prime) - 
     ( \hat g^R(\varepsilon^\prime) -\hat g^A(\varepsilon^\prime) )  
     \\ \nonumber
     & \hat\Sigma^A_{ph} (\varepsilon,\varepsilon^\prime) = \coth (\omega/2T) \hat g^A(\varepsilon^\prime) + \hat g^K(\varepsilon^\prime)/2 
     \\ \nonumber
     & \hat\Sigma^R_{ph} (\varepsilon,\varepsilon^\prime) = \coth (\omega/2T) \hat g^R(\varepsilon^\prime) - \hat g^K(\varepsilon^\prime)/2 .
   \end{align}

   We are interested in conservation laws for the charge and spin 
    \begin{align} \label{Eq:ChargeRel}
   \dot\mu =   \int_{-\infty}^{\infty} d\varepsilon {\rm Tr} \left[ \tau_3 \hat I_{qp-ph} \right] =0 \\
    \label{Eq:SpinRel}
    \dot\mu_z = \int_{-\infty}^{\infty} d\varepsilon {\rm Tr} \left[ \sigma_3 \hat I_{qp-ph} \right] =0
    \end{align}        
   For this purpose we write the kernel of CI (\ref{Eq:e-ph}) as follows: 
   \begin{align} \label{Eq:Kernel1}
     \hat K  = \coth (\omega/2T) \left[ \hat g^R(\varepsilon^\prime) \hat g^K (\varepsilon) - \hat g^R (\varepsilon) \hat g^K(\varepsilon^\prime) \right] +
     \\ \nonumber
    \coth (\omega/2T) \left[ \hat g^K(\varepsilon^\prime) \hat g^A (\varepsilon) - \hat g^K (\varepsilon) \hat g^A(\varepsilon^\prime) \right] 
     \\   \nonumber
   - \left[ \hat g^K(\varepsilon^\prime) \hat g^K (\varepsilon) + \hat g^K (\varepsilon) \hat g^K(\varepsilon^\prime) \right] /2 
     \\ \nonumber
     - \hat g^R (\varepsilon) \hat g^A (\varepsilon^\prime) - g^R (\varepsilon^\prime) g^A (\varepsilon)  \\ \nonumber
   + \hat g^R (\varepsilon) \hat g^R (\varepsilon^\prime) + \hat g^A (\varepsilon^\prime) \hat g^A (\varepsilon) 
      \end{align}
                 
 To prove the conservation laws it is enough to show that 
 \begin{equation} \label{Eq:aux1}
 \int_{-\infty}^{\infty} d\varepsilon {\rm Tr} 
 \hat\gamma 
 \left[  
 \hat K (\varepsilon,\varepsilon+\omega) - 
 \hat K (\varepsilon,\varepsilon-\omega)
 \right] = 0
 \end{equation}
where $\hat\gamma = \sigma_3$ for spin and 
$\hat\gamma = \tau_3$ for charge. 
If the relation (\ref{Eq:aux1}) holds then 
integral in (\ref{Eq:e-ph}) is identically zero.

At first we note that the CI (\ref{Eq:e-ph}) vanises if the distribution function is the equilibrium one, $\hat f = \tanh (\varepsilon/2T)$. Hence we can rewrite the kernel in the form
  \begin{align} \label{Eq:Kernel2}
     \hat K  = \coth (\omega/2T) \left[ \hat g^R(\varepsilon^\prime) \hat g^K_1 (\varepsilon) - \hat g^R (\varepsilon) \hat g^K_1(\varepsilon^\prime) \right] +
     \\ \nonumber
    \coth (\omega/2T) \left[ \hat g^K_1(\varepsilon^\prime) \hat g^A (\varepsilon) - \hat g^K_1 (\varepsilon) \hat g^A(\varepsilon^\prime) \right] 
     \\   \nonumber
   - \left[ \hat g^K_1(\varepsilon^\prime) \hat g^K_1 (\varepsilon) + \hat g^K_1 (\varepsilon) \hat g^K_1(\varepsilon^\prime) \right] /2 ,
          \end{align}
where the new Keldysh function is $\hat g^K_1 = \hat g^K - \tanh (\varepsilon/2T) \hat g^{RA}$.
Then the relation (\ref{Eq:aux1}) follows from the fact that all terms in the kernel (\ref{Eq:Kernel2})  are symmetric with respect to the interchange of $\varepsilon$ and
   $\varepsilon^\prime$ and the simultaneous sign change of $\omega$.

\section{Pauli matrices in Nambu--spin space}
\label{app:paulimatrices}
In this review, we study the properties of the Green's function that is a matrix in the $4\times 4$ space spanned by the outer product of $2\times 2$ Nambu (electron-hole) and spin spaces. We represent matrices in these spaces in terms of the identity operator and the three Pauli matrices $\tau_i$ and $\sigma_j$ ($i,j=1,2,3$) for the Nambu and spin space separately. Therefore, products of such matrices $\tau_i \sigma_j$  should be understood as outer products. There are hence in total 16 such matrices: the $4\times 4$ identity matrix $\hat 1$, the nine combinations of the pairs of Pauli matrices, and the six Pauli matrices acting on either space alone. In the latter matrices we do not explicitely write the outer product with the $2\times 2$ identity matrix as it would make equations unnecessarily cumbersome. In these cases the individual Pauli matrices show up alone. The exact matrix representation of these operators is often not very relevant, but for completeness we show a few examples:
\begin{align*}
\tau_1 &= \begin{pmatrix} 0 & 0 & 1 & 0\\0 & 0 & 0 & 1\\1 & 0 & 0 & 0\\0 & 1 & 0 & 0 \end{pmatrix}, \quad \sigma_1 = \begin{pmatrix} 0 & 1 & 0 & 0\\1 & 0 & 0 & 0\\0 & 0 & 0 & 1\\0 & 0 & 1 & 0\end{pmatrix}\\\tau_3 &= \begin{pmatrix} 1 & 0 & 0 & 0 \\0 & 1 & 0 & 0 \\0 & 0 & -1 & 0 \\0 & 0 & 0 & -1\end{pmatrix}, \sigma_3 = \begin{pmatrix} 1 & 0 & 0 & 0 \\0 & -1 & 0 & 0 \\0 & 0 & 1 & 0 \\0 & 0 & 0 & -1\end{pmatrix}\\
\tau_3 \sigma_3 &= \begin{pmatrix} 1 & 0 & 0 & 0 \\0 & -1 & 0 & 0 \\0 & 0 & -1 & 0 \\0 & 0 & 0 & 1\end{pmatrix}, \tau_1 \sigma_1 = \begin{pmatrix} 0 & 0 & 0 & 1 \\0 & 0 & 1 & 0 \\0 & 1 & 0 & 0 \\1 & 0 & 0 & 0\end{pmatrix}
\end{align*}
In this representation, the rows thus correspond to the order ``spin up electron'', ``spin down electron'', ``spin up hole'' and ``spin down hole'', respectively.

\end{document}